%% file: _main.tex
\documentclass{pas}

\usepackage{multirow}
\usepackage{setspace}

\input{commands/siunits}

\input{commands/commands}

\input{commands/commands_other}
\input{commands/commands_galfit_generated}

\input{commands/commands_acsgc_generated}

\input{commands/commands_offsets_generated}
\input{commands/commands_lts_generated}

\input{commands/commands_frbpop_generated}
\renewcommand{\HII}{H\,{\sc ii}}

\renewcommand{\ba}{\ensuremath{q}}

\begin{document}

\lefttitle{Host Galaxy Inclination and FRB Dispersion Measure}
\righttitle{L. Marnoch et al}

\jnlPage{1}{4}
\jnlDoiYr{2026}
\doival{10.1017/pasa.xxxx.xx}

\articletitt{Research Paper}

\title{A Relationship Between Fast Radio Burst Dispersion Measure and Host Galaxy Inclination}

\author{
\sn{Lachlan} \gn{Marnoch}$^{1,2,3}$, 
\sn{Stuart~D.} \gn{Ryder}$^{1,2}$,
\sn{Sunil} \gn{Simha}$^{4,5}$,
\sn{Alexa~C.} \gn{Gordon}$^{4}$,
\sn{R.~M.} \gn{Shannon}$^{6}$,
\sn{Nicolas} \gn{Tejos}$^{7}$,
\sn{Marcin} \gn{Glowacki}$^{8,9}$,
\sn{Ziteng} \gn{Wang}$^{10}$,
\sn{Elizabeth~K.} \gn{Mahony}$^{3}$,
\sn{J.~Xavier} \gn{Prochaska}$^{11,12,13}$,
\sn{Wen-fai} \gn{Fong}$^{4}$,
\sn{Richard~M.} \gn{McDermid}$^{1,2}$
}

\affil{
$^1$School of Mathematical and Physical Sciences, Macquarie University, NSW 2109, Australia; 
$^2$Astrophysics and Space Technologies Research Centre, Macquarie University, Sydney, NSW 2109, Australia; and
$^3$CSIRO Space And Astronomy, P.O. Box 76, Epping, NSW 1710, Australia; and
$^4$Center for Interdisciplinary Exploration and Research in Astrophysics (CIERA) and Department of Physics and Astronomy, Northwestern University, Evanston, IL 60208, USA; and
$^5$Department of Astronomy and Astrophysics, University of Chicago, William Eckhardt Research Center, 5640 S Ellis Ave, Chicago, IL 60637, USA
$^6$Centre for Astrophysics and Supercomputing, Swinburne University of Technology, Hawthorn, VIC, 3122, Australia; and
$^7$Instituto de F\'isica, Pontificia Universidad Cat\'olica de Valpara\'iso, Casilla 4059, Valpara\'iso, Chile; and
$^8$Institute for Astronomy, University of Edinburgh, Royal Observatory, Edinburgh, EH9 3HJ, United Kingdom; and
$^9$Inter-University Institute for Data Intensive Astronomy, Department of Astronomy, University of Cape Town, Cape Town, South Africa; and
$^10$International Centre for Radio Astronomy Research, Curtin University, Bentley, WA 6102, Australia; and
$^11$Department of Astronomy and Astrophysics, University of California, Santa Cruz, CA 95064, USA; and
$^12$Kavli Institute for the Physics and Mathematics of the Universe (Kavli IPMU), 5-1-5 Kashiwanoha, Kashiwa, 277-8583, Japan; and
$^13$Division of Science, National Astronomical Observatory of Japan, 2-21-1 Osawa, Mitaka, Tokyo 181-8588, Japan
}

\corresp{Lachlan Marnoch, Email: lachlan.marnoch@gmail.com}

\history{(Received 09 October 2026; revised xx xx xxxx; accepted xx xx xxxx)}

\begin{abstract}
It is commonly thought that fast radio bursts (FRBs) trace star formation, and that they are produced by magnetars. Both hypotheses imply that their sources are hosted primarily (but not exclusively) in galactic disks. Although there is evidence in that direction, it has yet to be shown conclusively. In addition, a tension exists with the fact that at least one FRB source is found in a globular cluster far from the disk of the host.
If FRBs originate primarily in disks, the galaxy orientation could leave an impression in the sample of observed FRBs, either as a selection bias against FRBs in edge-on galaxies or as a correlation with the host contribution to the DM (\DMHost{}). Statistical tests against the inclinations of FRB host galaxies, derived from parametric morphology applied to optical and near-infrared imaging, can search for these effects. 
Such a search requires an adequately large sample of known FRB hosts that have been imaged to a sufficient depth, while estimates of \DMHost{} also require the redshift of each host to be known. These conditions are best met in the Commensal Real-time ASKAP Fast Transient (CRAFT) survey sample of FRB host galaxies, for which the majority have been imaged at optical wavelengths with FORS2 on the Very Large Telescope (VLT). 
Here these data are fitted with the morphology software \galfit{} to derive the elliptical axis ratio $q$ (a proxy for inclination) for \nGALFIT{} hosts from the CRAFT sample, detected with ASKAP and followed up with VLT imaging. We find no bias against FRB detections from edge-on galaxies; instead, we uncover a new relationship between the inclination of an FRB host galaxy and \DMHost{}. This is consistent with a primarily disk-based origin for CRAFT FRBs.
Using this relationship we are able to reduce the scatter of the host sample about the Macquart relation by 40\%, demonstrating a new empirical method for estimating \DMHost{} that requires only deep optical imaging.
\end{abstract}


\maketitle

\section{Introduction}
\label{introduction}

Fast radio bursts are very bright, brief radio transients, and have been detected in galaxies ranging in distance from the Milky Way itself \citep{BochenekSGR1935, CHIMESGR1935} to $z>2$ \citep{CalebFRB20240304B}. 
An FRB is an exceptional tool for probing plasma, equipped with physical properties that precisely measure the integrated column density (dispersion measure), magnetisation (rotation measure) and turbulence (scattering timescale) of free electrons along the line of sight between source and observer. This is useful for cosmology, enabling studies of the distribution of ionised material in the Universe. For maximum utility, the distance that the FRB has crossed must be estimated, requiring the measurement of a cosmological redshift. As FRBs themselves carry no identifiable spectral features, the host galaxy must usually be identified and observed optically or in the infrared. Given that FRBs exist chiefly as a cosmological population, across a very large redshift range, matching precise radio localisation with deep optical imaging is ideal.

Identifying and characterising the host galaxies of FRBs is also important for unlocking the fundamental nature of the transients, including the mechanisms and objects that produce them.
As with gamma-ray bursts (GRBs; e.g. \citealp{FongBerger2013}) before them, a deepening understanding of the population of FRB hosts is shedding light on the nature of the transients themselves \citep{Heintz2020, BhandariFRB20191001A, Bhandari2022, GordonProspector, Sharma2025, GordonSpatial}. The statistical power of these studies grows with the size and homogeneity of the sample; large samples of FRBs with both precise localisation and deep follow-up imaging are required to solve these mysteries. However, the difficulty of detecting and precisely localising such brief transients (on the scale of a millisecond) if they do not repeat (most have not been seen to), combined with the availability of 8~metre-class telescope time to identify and characterise distant host galaxies, has meant that large samples have been slow to materialise. 
The Commensal Realtime ASKAP Fast Transients survey, or CRAFT \citep{CRAFT}, has worked to alleviate this state of affairs since 2018 \citep{BannisterFRB20180924B}, using the Australian SKA Pathfinder \citep[ASKAP;][]{ASKAP} to detect and localise FRBs and the European Southern Observatory's Very Large Telescope (VLT) to follow them up. It has accumulated a respectable sample of FRB host galaxies, along with the deep imaging necessary to study them \citep{ShannonICS}. 

Although the mechanisms by which FRBs are generated remain uncertain, scientific consensus is converging on magnetars (neutron stars with magnetic fields in the regime of $10^{11}$ to $10^{15}$ G; \citealp{Magnetars}) as the origin of most, if not all, FRBs. 
These objects are believed to be young, with ages $\lesssim 10^5$ years \citep{Magnetars}. Like all neutron stars, they are understood to primarily form in core-collapse supernovae (CCSNe; e.g., \citealt{Pardo-Araujo2026}), although alternate channels have been proposed \citep{Kremer2023}. 
CCSNe are produced by massive stars ($\gtrsim\SI{8}{\solarmass}$), which are short-lived ($<\SI{20}{\mega\year}$).
Thus, CCSNe should trace star formation with a relatively small delay; and magnetars, by and large, should trace star formation by proxy, with a delay not much greater. 
Although evidence is accumulating of a connection between FRBs and star formation \citep{GordonProspector}, the presence of some FRBs within older environments demonstrates that this connection is not necessarily universal \citep{GlobularCluster,HorowiczMargalit2025, EftekhariFRB20240902A}. 

Star formation occurs predominantly in galactic disks.
A consequence of the youth of the Milky Way magnetar population is that they have tended not to travel far from the plane of the disk despite powerful kicks from their natal supernovae \citep{Magnetars}. The scale height of the population, assuming an exponential distribution, has been measured at just 20 to 30 pc \citep{MagnetarCat}.
Hence, evidence of a disk connection for FRBs would be consistent with a magnetar connection. The spatial distributions of FRBs indicate that most do originate from the disks of their galaxies, as opposed to the more extended spatial distributions expected for globular clusters  \citep{GordonSpatial}.

\begin{figure}
    \centering
    \includegraphics[trim={4cm 18cm 0.5cm 10cm}, width=\linewidth, clip]{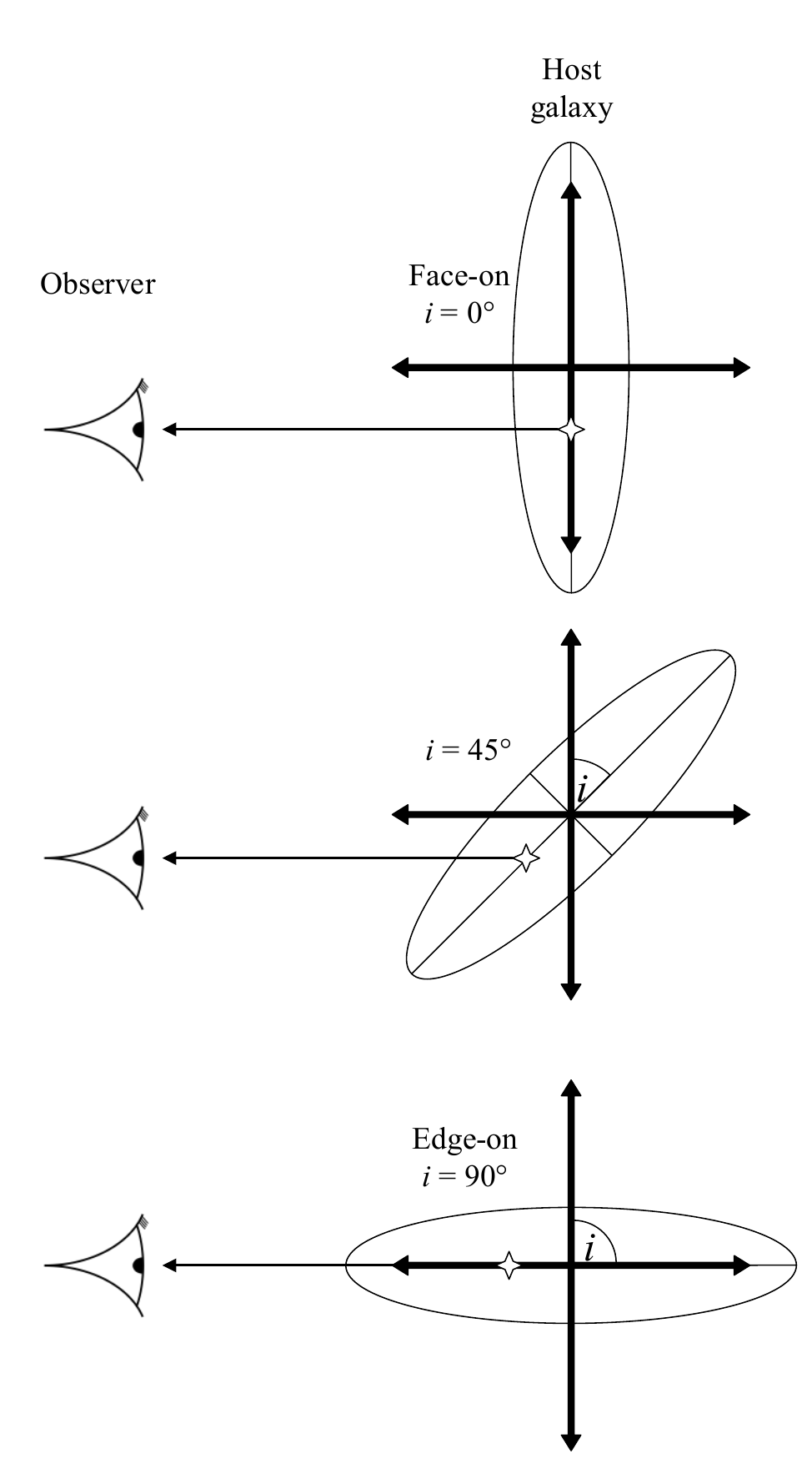}
    
    \caption[Diagram of inclined host galaxy]{A diagram illustrating host galaxy inclination, demonstrating the increasing path length with inclination through a host galaxy disk. The four-pointed star represents an FRB source.}
    \label{fig:inclination-diagram}
\end{figure}



The dispersion measure (DM) of an FRB, measured from the frequency-dependent sweep of its arrival time, provides the integrated free electron density along its line-of-sight. For events at cosmological distances, it is weighted by the redshift at which the DM is contributed, so that $\DM{} = \int \en/(1+z(l)) \dn l$ (where $\en$ is the free electron density). For an FRB, the quantity can be decomposed into $\DM{}=\DMHost{} + \DMCosmic{} + \DMMW{}$, where \DMMW{} is the contribution of material associated with the Milky Way (both the halo and interstellar medium), \DMCosmic{} of the intergalactic medium and intervening structures such as foreground halos, and \DMHost{} of the source's host galaxy. \DMHost{} is perhaps the most difficult DM component to estimate, being strongly affected by the nature of the host galaxy and of the environment in which the FRB source is embedded; attempts to do so are discussed further in \autoref{dm}. These relationships are difficult to disentangle, although work by \cite{Leung2025} has found a correlation between stellar mass and \DMHost{}, with more massive hosts tending to exhibit a lower value for \DMHost{}.
By fitting the ASKAP FRB population, \cite{JamesH0} infer a log-normal \DMHost{} distribution with a median of \(186^{+59}_{-48}\dmunits\), indicating that the contribution is both highly variable and potentially large; some of this variation may be accounted for by orientation of the host galaxies relative to us, the observers. 

An FRB from a disk tilted relative to the observer will have to travel through a greater segment of the disk of its host than if the host was face-on; this is illustrated in \autoref{fig:inclination-diagram}. 
A longer path length through the gaseous content of the disk will result in a larger \DMHost{}. In addition, there is a greater chance of intersection with \HII{} regions, located primarily in galactic disks, which will tend to increase \DMHost{} significantly; \cite{Ocker2024} have found that foreground Galactic \HII{} regions explain the majority of high-DM Milky Way pulsars. The turbulent gas in these structures will also tend to increase the scattering timescale.
Thus, the relative geometry of the host galaxies could affect detected FRBs in two primary ways:
\begin{enumerate}
    \item A correlation of the host contribution to the dispersion measure (\DMHost{}; see \autoref{dm}) with inclination;
    \item A correlation of the scattering with inclination.
\end{enumerate}

However, no correlation has been found between DM and FRB scattering \citep{Day2020}, which may be attributed to selection biases and fluctuations in the cosmological component of DM \citep{Mas-Ribas2026}. \cite{GlowackiScattering} have conducted a search for a correlation between FRB scattering and host inclination, among other quantities, finding no strong correlation.

Recently, studies within cosmological simulations have predicted the former trend; using 1000 sightlines from $16500$ IllustrisTNG galaxies, \cite{Kovacs2024}  predict an exponential trend of \DMHostRest{} with host inclination (we use a $^\prime$ to indicate a quantity in the host rest frame, while its absence implies the observer frame). 

\cite{Orr2024} perform a similar study using the Feedback in Realistic Environments (FIRE-2) simulations.
Interestingly, they predict that such a trend will only exist for `older' ($\gtrsim\SI{10}{Myr}$) stellar populations, as those originating from younger populations will be dominated by the contribution of their local \HII{} region (except for extreme edge-on cases). 
They base this on a sample of just 6 simulated hosts. Nonetheless, their results broadly agree that disk hosts with greater inclinations will exhibit greater \DMHostRest{}.

Apart from being observed directly, a correlation of inclination with \DMHost{} and/or scattering might also manifest in a selection bias in the FRBs that are detected; that is, there could be a lower detection rate of FRBs from galaxies with high inclinations, due to the following effects:

\begin{enumerate}
\item Larger \DMHost{} may:
\begin{enumerate}
    \item Introduce DM smearing, thus reducing signal-to-noise;
    \item Increase the likelihood of \DMFRB{} being pushed beyond the DM search range of an FRB survey (although for most surveys this would require an increase of thousands of \dmunits{}).
\end{enumerate}
\item Greater scattering may:
\begin{enumerate}
    \item Smear FRBs across a larger timescale, and potentially push the burst outside of search width parameters;
    \item Lower the S/N in each spectral channel even for a perfectly matched search width, as more noise is added to the same signal; the detection metric of FRB searches is inversely proportional to the square root of the width.
\end{enumerate}

\end{enumerate} Statistically, the hunt for such a bias can be conducted by comparing the FRB host population inclination distribution to a control sample of background galaxies. \citet[hereafter B+24]{Bhardwaj2024} searched for both a selection bias against FRBs from inclined hosts, and for a correlation of \DMHost{} with inclination, using a sample of 23 hosts from the CRAFT and DSA-110 surveys; they find strong evidence for the former and no evidence for the latter. We conduct similar searches in this work, 
with results that appear to contradict the B+24 results in both respects. We will discuss these discrepancies in \autoref{discussion:bhardwaj}.

\section{Data}

\input{tables/craft_sample}

For this work we studied \nGALFIT{} FRB sources with identified host galaxies, all detected with ASKAP, localised to \hbox{(sub-)arcsecond} level precision by CRAFT and followed up using optical and near-infrared instruments on the VLT. 
The FRB sample and relevant host properties used here are summarised in \autoref{tab:craft_sample}.

The imaging used here was obtained from the FOcal Reducer and low dispersion Spectrograph \citep[FORS2;][]{FORS} in $g$, $I$ and $R$-bands, and processed using the CRAFT Optical Pipeline\footnote{\href{https://github.com/Lachimax/craft-optical-followup}{https://github.com/Lachimax/craft-optical-followup}}, largely as described by \cite{MarnochFRB20210912A}. De-biasing and flat-fielding in the pipeline are performed using \esoreflex{}. Astrometric calibration is carried out on individual frames using a local installation of \astronet{} with index files derived from the \textit{Gaia} DR3 catalogue. The frames are coadded with \montage{} and \ccdproc{} using a sigma-clipped mean pixel stack. \sextractor{} and \psfex{} are deployed on the coadded imaging to measure photometry of field stars, which are cross-referenced with catalogues for photometric calibration, and to determine the empirical PSF, which is used by \galfit{}.
Host identification is confirmed using the Probabilistic Association of Transients to their Hosts (PATH) code \citep{AggarwalPATH}, with \galfit{} then used to parametrise the morphology of the host galaxies (see \autoref{morph}). 

FRB detection and localisation are described by \cite{ShannonICS} and \cite{CRACO1}. \cite{ShannonICS} describe the criteria for imaging follow-up. The \nGALFIT{} hosts used in this study are not specially selected. All CRAFT host galaxies with both FORS2 imaging and a known redshift prior to September 2024 were included initially; four such hosts were removed from the sample due to difficulties fitting the imaging with \galfit{} (see \autoref{morph}).
A sample cut based on PATH posterior was considered, but became unnecessary as all remaining sample members are high confidence (with the least-secure being the host of \FRB{20181112A} at $\POx=0.761$).


\section{Host morphology and inclination}
\label{morph}

\newcommand{\figmod}{0.9}

\begin{figure*}
    \centering
    \includegraphics[trim={0.3cm 0.31cm 0.25cm 0.26cm}, width=\figmod\linewidth]{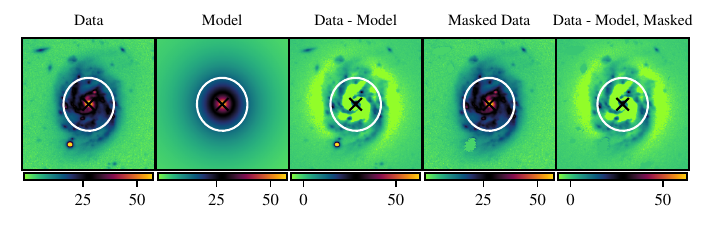}
    
    \caption{A demonstration of \galfit{} model parameters, masking, and subtraction, using the host of \FRB{20211127I}. The white ellipse shows the axis ratio of the model ($\ba=0.953(2)$, resulting in a derived $i=\SI{18}{\degree}$), with the semi-major axis set to $\Reff{}$. The black cross marks the model centroid. Units of pixel brightness are arbitrary.}
\label{fig:galfit-subtracted-20211127I}
\end{figure*}

\begin{figure*}
    \centering
    \includegraphics[trim={0.3cm 0.31cm 0.25cm 0.26cm}, width=\figmod\linewidth]{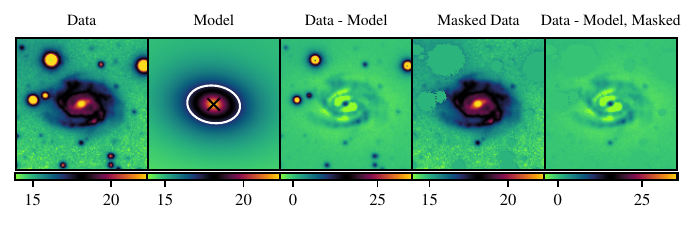}
    
    \caption{\galfit{} model subtraction for the host of \FRB{20210807D}, as in \autoref{fig:galfit-subtracted-20211127I} with $\ba=0.707(4)$, $i=\SI{46}{\degree}$.}
\label{fig:galfit-subtracted-20210807D}
\end{figure*}

\begin{figure*}
    \centering
    \includegraphics[trim={0.3cm 0.31cm 0.25cm 0.26cm}, width=\figmod\linewidth]{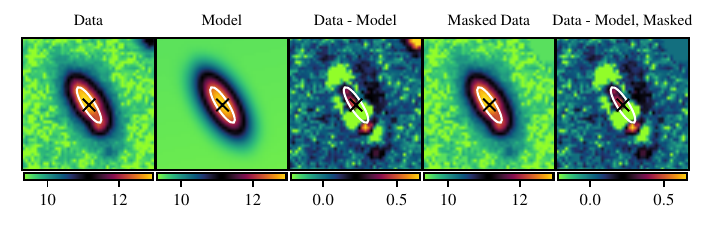}
    \caption{\galfit{} model subtraction for the host of \FRB{20200906A}, as in \autoref{fig:galfit-subtracted-20211127I} with $\ba=0.274(2)$, $i=\SI{78}{\degree}$.}
\label{fig:galfit-subtracted-FRB20200906A}
\end{figure*}

\begin{figure*}
    \centering
    \includegraphics[trim={0.3cm 0.31cm 0.25cm 0.26cm}, width=\figmod\linewidth]{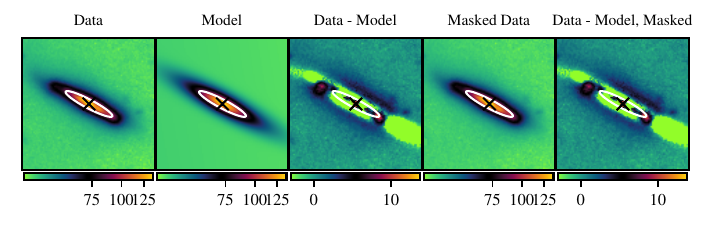}
    \caption{\galfit{} model subtraction for the host of \FRB{20240201A}, as in \autoref{fig:galfit-subtracted-20211127I} with $\ba=0.1666(2)$, $i\sim\SI{90}{\degree}$.}
\label{fig:galfit-subtracted-20240201A}
\end{figure*}

The inclination $i$ of a disk galaxy is the angle between the line-of-sight to the observer, and the normal to the plane of the galaxy disk, as shown in \autoref{fig:inclination-diagram}. A face-on galaxy has $i=\SI{0}{\degree}$ and an edge-on galaxy \SI{90}{\degree}. The image of a galaxy is a two-dimensional projection of a three-dimensional object with an intrinsic shape that is not firmly known, and even disky galaxies vary widely in intrinsic size and shape; thus, it is difficult to accurately measure inclination directly from imaging. We adopt the proxy of the elliptical axis ratio between the semi-major and semi-minor axes of the model, \ba{} = $b/a$.
With some geometrical assumptions, $i$ is related to \ba{} by the following 
equation 
\citep{Hubble1926,Tully2013}:
\begin{align}
\label{equ:inclination}
    \cos^2(i)   = \dfrac{q^2 - q_0^2}{1 - q_0^2}
\end{align}
where \qint{} is the axis ratio of the galaxy when viewed fully edge-on. This formula relies on the following simplifications:
\begin{enumerate}
    \item Disks are assumed to be intrinsically circular, i.e. $\ba=1$ for a fully face-on host.
    \item Any distortion or asymmetry in the disk 
    is neglected. 
    \item A common intrinsic axis ratio \qint{} is assumed to be shared by all galaxies.
    \item Our FRB host galaxies are implicitly assumed to have disks, and not to be of the elliptical type.
\end{enumerate}
\noindent Despite these limitations, this is the best way to extract an inclination angle using only imaging. Other methods, such as those using the 2D velocity field of the galaxy, require optical IFU or radio spectral line cube data. We use the axis ratio \ba{} directly for the important comparisons, in order to avoid downstream effects of these assumptions, providing this formula mainly to demonstrate the value of \ba{} as a proxy variable.

To obtain \ba{}, we use the software \galfit{} \citep{Galfit1,Galfit2}, which models light profiles from galaxy imaging, fitting two-dimensional parametric functions with $\chi^2$ minimisation. We fit \sersic{} profiles \citep{SersicProfile,SersicProfile2} to the VLT/FORS2 imaging of CRAFT FRB host galaxies, for which we obtain the effective radius \Reff{}, \sersic{} index $n$, integrated magnitude \mgalfit{}, centroid positions, and elliptical axis ratio \ba.
We fit a single \sersic{} profile per galaxy; although multiple components may be desirable in some cases, the analysis in \autoref{populations} requires comparison to a large catalogue of galaxies, on which only single-component fits were performed.

We have used \galfit{} to successfully fit the majority of our host galaxy sample; we remove the following, for which it failed. The host of FRB\,20230731A \citep{ShannonICS} is located in the Galactic plane, and the fit was affected by the presence of foreground stars. The host of FRB\,20240210D \citep{TNSCRAFT2026} is not well-resolved, with the galaxy size similar to the PSF FWHM. The centre of the host galaxy of FRB\,20230526A \citep{ShannonICS} was saturated in the VLT/FORS2 imaging. The highest-redshift FRB in the original sample, FRB\,20220610A \citep{RyderFRB20220610A}, was found to originate from a small group of galaxies in HST imaging \citep{GordonFRB20220610A}, which is not well resolved in our VLT/FORS2 observations.
We were able to model the remaining host galaxies using \galfit{}, and obtain the axis ratio \ba{}, defining the elliptical shape of the profile. This is provided in \autoref{tab:craft_sample}. Example figures showing the models and the residuals after subtraction are provided in \autoref{fig:galfit-subtracted-20211127I} to \autoref{fig:galfit-subtracted-20240201A}, with the full set in \autoref{app:residuals}.

While in reality \qint{} will vary from galaxy to galaxy, we adopt the conventional \citep[e.g.,][]{Tully2013} $\qint{}=0.2$.
Although some of our hosts would suggest a small $q_0$ (we measure $\ba{}<0.2$ for four hosts), we also cannot assume that this is true for the entire population, and it cannot be measured for individual galaxies. 
We set the four hosts with $\ba<0.2$ to $i=90^\circ$. These values are presented in \autoref{tab:craft_sample}. Our primary analysis is performed directly with the axis ratio \ba{}, circumventing contamination from mismatched assumptions.

From visual inspection of the residuals after model subtraction, in the cases of three well-resolved face-on galaxies with prominent bulges, \galfit{} has primarily fit the bulge instead of the galactic disk. These are the hosts of FRBs 20190608B, 20220725A, and 20231226A. The bulges of these galaxies have smaller axis ratios than their disks, and so the inclinations will be exaggerated. This will have contaminated our results, but given that the exaggerated inclinations are not extreme, the bias should be minimal; we account for this in \autoref{correlations}. The disk appears to dominate in all other cases. 

\section{Dispersion measure modelling}
\label{dm}

\DMHost{} is the host contribution in the observer frame. As we are attempting to constrain an intrinsic host property we will focus on calculations in the host rest frame; redshift being a linear transformation, the same equations apply.
\DMHostRest{} can be decomposed as:
\begin{equation}
    \DMHostRest{}   = \DMSourceRest + \DMHostISMRest + \DMHostHaloRest
    \label{equ:dm-host}
\end{equation}\DMSource{}, being the contribution of material immediately local to the source, cannot be reliably modelled at the present time.
However, it seems likely that \DMSource{} will only be significant in the case of a young source, still embedded in the ejecta from its natal supernova (as hypothesised for the repeating source \FRB{20121102A}; \citealp{MargalitMetzger2018}). 
This component is unlikely to correlate significantly with host galaxy orientation and is likely more dependent on age.

\DMHostHalo{}, 
the contribution of the host's circumgalactic material, has been estimated in several cases \citep[e.g.][]{SimhaFRB20190608B, SimhaExcess, COBRADR1, Bernales-Cortes2025}
by scaling a baryon halo to the host's stellar mass 
\massstar{},
as estimated through SED fitting. It is likely to be significant in most cases, and has been estimated on the order of several tens of \dmunits{} \citep{COBRADR1}.
However, the relationships used for this scaling \citep[e.g.][]{Behroozi2013, Moster2013, Kravtsov2018} have significant scatter (on the order of 0.3 dex), and we lack 
\massstar{} 
estimates for many of our sample; thus we do not attempt to decompose 
\DMHost{} 
into its constituents for the main sample. Assuming spherical symmetry of the halo, as is usually done, \DMHostHalo{} should not vary greatly with the host orientation. We will thus assume that the majority of any variation of \DMHost{} with host orientation will be explained primarily by \DMHostISM{}.

Efforts have been made to quantify \DMHostISM{} observationally, either using nebular emission lines from the host \citep[e.g.][]{TendulkarR1, ChittidiFRB20190608B, Bernales-Cortes2025} or the FRB scattering timescale \citep[e.g.][]{Cordes2022}.
However, these methods are likely to have large scatter as they depend strongly on assumptions concerning the nature of the host galaxy, and the depth of the FRB source in the observed column of ionised material. How well they represent the actual contribution of the host ISM remains an open question.

While it is difficult to measure \DMHost{} directly, we can employ a proxy in \DMExcess{} (also called $\mathrm{\DM{excess}}$) by modelling and subtracting the other DM components, which are more easily estimated \citep[e.g.][]{Bernales-Cortes2025}.
We calculate it as:
\begin{align}
\DMExcess       &= \DMFRB - \DMMWISM - \DMMWHalo - \tavg{\DMCosmic}\\
\DMExcessRest{} &=\DMExcess{}(1+\zhost)
\end{align}

\noindent\DMMWISM{} is the contribution of gas in the Milky Way interstellar medium. We adopt the NE2001 model for this component \citep{NE2001, NE2001-II}, \DMNE{}. We estimate an uncertainty in \DMMWISM{} as the difference with the prediction of the YMW16 model, \(\Delta\DMMWISM = |\DMNE{}-\DMYMW|\).

\DMMWHalo{} is the contribution from the Galactic halo. Here we adopt the estimate \(\DMMWHalo\sim\DMSI{40}\) derived from the modified Navarro-Frenk-White profile \citep[mNFW;][]{MathewsProchaska2017, ProchaskaZheng2019}. This estimate is broadly consistent with the results of \citet{Cook2023}, \citet{Ravi2025}, and \cite{Hoffman2026}. The mNFW model is spherically 
symmetric, and so varies very little between sightlines; thus, we have adopted a fixed value. Our results are therefore insensitive to the choice of \DMMWHalo{}, as it would affect only the $y$-intercept of our linear fits to DM. 

\DMCosmic{}, on the other hand, will vary significantly by line-of-sight, depending on the structure of the intervening cosmic web and the quantity of foreground halos. 
Surveys such as FLIMFLAM \citep{FLIMFLAM, COBRADR1} are focused on constraining this quantity for particular lines-of-sight, with applications to cosmology; this is a time, data and effort-intensive exercise. However, the expectation value \DMCosmicAvg{}, as a function of redshift, is a well-understood cosmological result based on empirical knowledge of baryon distributions and their ionisation state \citep{Ioka2003, MacquartCosmicDM, Inoue2004, James2022A}. We calculate it using the \textsc{FRB} code\footnote{\url{https://github.com/FRBs/FRB}}. Following \cite{SimhaExcess}, and based on numerical simulations by  \cite{FLIMFLAM}, we adopt a 20\% uncertainty for the IGM contribution to \DMCosmic{}, which is compounded into the \DMExcess{} uncertainty; we caution that this is very likely an underestimate.

\DMExcess{} can be negative for those few hosts with below-average \DMCosmic{} contributions, lying along lines-of-sight that intersect little of the cosmic web (in our sample, we find only FRBs 20190611B and 20210807D to have $\DMExcess{} < 0$). Hence, how closely \DMExcess{} reflects the true \DMHost{} will, once more, vary from source to source. We hence refer to it as a proxy variable. Nonetheless, if we can believe that the variation in \DMHost{} is a significant contributor to the total variation in \DMFRB{}, then \DMExcess{} should certainly correlate with it, enabling us to leverage a large imaging sample.

\section{Search for a selection bias}
\label{populations}

We first investigate whether the CRAFT FRB sample is biased against edge-on hosts, as claimed by B+24, by searching for a deviation in the elliptical axis ratio $\ba{}=b/a$ from the general galaxy population. We do this by comparison with the Advanced Camera for Surveys General Catalogue \citep[ACS-GC;][]{ACSGC}, which is a reprocessing of public data from the Advanced Camera for Surveys (ACS) on the \textit{Hubble Space Telescope} (HST).

We use the two-sample Kolmogorov-Smirnov (K-S; \citealp{KStest}) and $k$-sample Anderson-Darling (A-D; \citealp{ADtest, KsampleAD}) tests, as implemented in \scipy{}, with the `two-sided' null hypothesis; that is, that the two samples are drawn from identical distributions. 
We select the standard $\alpha$ of $0.05$; at $p < 0.05$, we rule out that our sample is drawn from the same distribution as the comparison sample, with a greater {\pval{}} leaving the possibility that the two samples have a common origin. The higher the {\pval{}}, the less unlikely a common origin is.


We aim to reduce systematic biases by comparing our \galfit{} outputs to the ACS-GC, which was also derived using \galfit{}. That catalogue was generated using \sextractor{} and \galfit{}, combined using the \galapagos{} \citep{galapagos} code. 
\galapagos{} uses \galfit{} to fit single-\sersic{} profiles to sources first detected with \sextractor{}, using \sextractor{} outputs for initial guesses (which is similar to how the CRAFT \galfit{} sample has been derived).
ACS-GC has reanalysed all imaging from the All-wavelength Extended Groth strip International Survey \citep[AEGIS;][]{AEGIS}, the Cosmological Evolutionary Survey \citep[COSMOS;][]{COSMOS}, Galaxy Evolution from Morphology and SEDs survey \citep[GEMS;][]{GEMS}, and the Great Observatories Origins Deep Survey \citep[GOODS;][]{GOODS}, amounting to $\sim470,000$ objects with \galfit{} parameters.
It also cross-matches with a number of other HST catalogues to provide photometric and spectroscopic redshifts for 74\% of its catalogued objects \citep{ACSGC}. 

\galfit{'s} use of PSF convolution when fitting its models mitigates the error caused by the difference in PSF between HST and VLT imaging. 
\cite{Dewsnap2023} compare \galfit{} models of a common sample of 4016 galaxies across HST and Subaru imaging.
In contrast to the \sersic{} index, they find that \ba{} was in reasonable agreement between the Subaru and HST samples, with an average difference of $q_\mathrm{Subaru} - q_\mathrm{HST} = 0.02\pm0.15$. 
In the absence of a comparable \galfit{} catalogue derived from a ground-based telescope, we feel that the effect on \ba{} is small enough for our purposes.

We combine the F606W ($\sim V$-band) measurements from the GOODS, AEGIS and GEMS portions of the catalogue; COSMOS has only F814W ($\sim I$-band). while $V$-band lies between the two targeted bands for our VLT imaging, $g$ and $R$ (with 5 objects in $I$ instead due to fitting issues). Ideally, all models would have been derived from imaging in a similar rest-frame bandpass. However, the CRAFT VLT sample is not homogeneous in this way; our hosts have either $g$ or $R$ band, but rarely both. This compromise is made so that we can make full use of our sample. Future large samples such as LSST in a consistent set of bandpasses
will help overcome this shortcoming. 
However, we do not expect this difference in bandpass to affect apparent inclination, which is the primary target of this study, in any appreciable way.

\begin{figure}
    \centering
    \includegraphics[width=\linewidth]{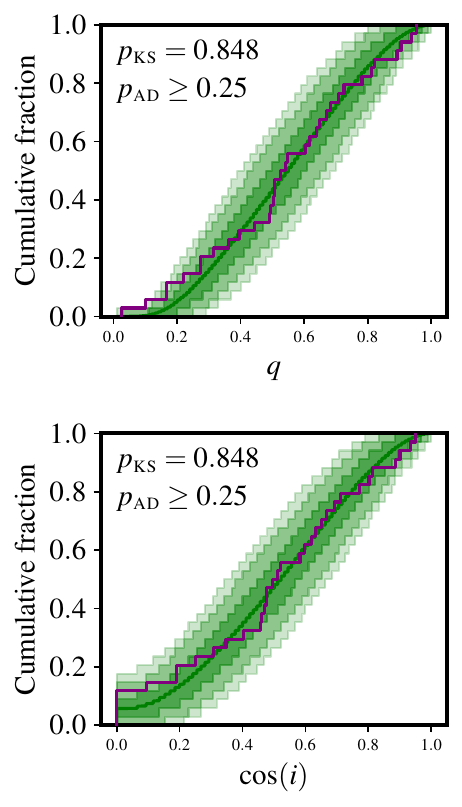}
    
    \caption[Comparison of inclination-related CRAFT host properties with ACS-GC.]{Comparison of the axis ratio (top) and $\cos{i}$ (bottom) of the full sample of CRAFT hosts (purple) with ACS-GC (green). The shading provides the 68\%, 96\% and 99\% bootstrap confidence regions.}
    \label{fig:acsgc-inclinations}
\end{figure}

We additionally cut the ACS-GC sample to objects with a \galfit{} magnitude brighter than \magmax{} to match the brightness range of our sample, and limit the sample to those with \texttt{FLAG\_GALFIT $=0$}, indicating an extended source and a good fit \citep{ACSGC}. A redshift cut was considered instead, but the magnitude cut avoids relying on photometric redshifts, or spectroscopic redshifts that would be subject to selection effects.
After performing these sample cuts, the ACS-GC $V$-band sample is $\sim15000$ objects. The tests with this sample, against the \nGALFIT{} CRAFT hosts,
are shown in 
\autoref{fig:acsgc-inclinations}.

A further sub-sample is carved out with $n<2$, applied to both CRAFT hosts and the ASC-GC, in an effort to limit the analysis to disky galaxies. This results in a sample size of 
\nGALFITn{} 
for CRAFT (the $\diamond$ sample as defined in \autoref{tab:craft_sample}) and $\sim12000$ for ACS-GC. 

For comparison with the results of B+24, which are discussed in detail in \autoref{discussion:bhardwaj}, we derive another subsample by imposing $\mgalfit < 21$ on both the CRAFT (the $\star$ sample as defined in \autoref{tab:craft_sample}) and ACS-GC samples. This is performed separately to the $n$ cut.

\begin{figure}
    \centering
    \includegraphics[width=\linewidth]{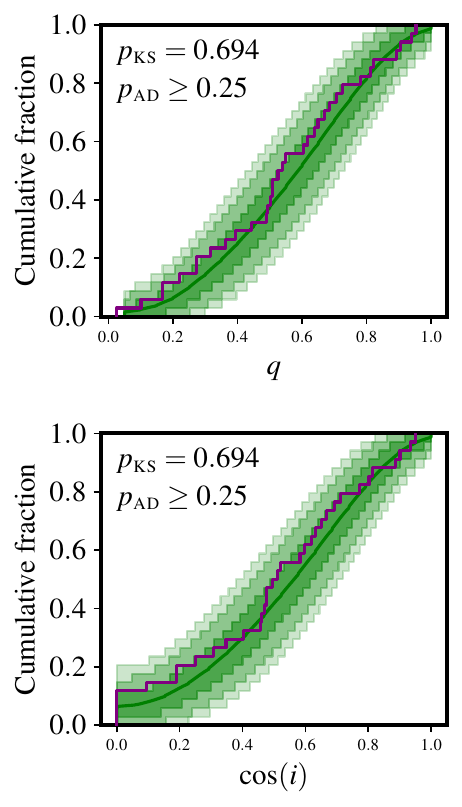}
    
    \caption[Comparison of inclination-related CRAFT host properties with SDSS DR16.]{Comparison of the axis ratio (top) and $\cos{i}$ (bottom) of the full sample of CRAFT hosts (purple) with SDSS DR16 (green). The shading provides the 68\%, 96\% and 99\% bootstrap confidence regions.}
    \label{fig:sdss-inclinations}
\end{figure}

\input{tables/acsgc}

The results of all of the population tests against ACS-GC are provided in \autoref{tab:population-tests:acsgc}.
The K-S test comparing \ba{} from the CRAFT sample to \ba{} from the ACS-GC sample (shown together in \autoref{fig:acsgc-inclinations}) generates a high $\pks = \KSgalfitaxisratioCRAFTvACSGC$ and $\pad \geq \ADgalfitaxisratioCRAFTvACSGC$, providing no evidence that they are sampled from different distributions. This is true for all three sample cuts, to varying \pval{}, but never less than 0.495 for the K-S tests or 0.25 (the maximum value for the \scipy{} implementation) for the A-D tests. This indicates that there is a strong correspondence between the measured inclination distributions of the two populations. That is, no selection bias against FRBs related to host galaxy inclination has been found in the CRAFT FRB sample, contrary to the findings of B+24. 

For comparison, we repeat the full-sample tests against an SDSS DR16 \citep{SDSS16} sample of 500,000 galaxies obeying the same constraints as that used by B+24 ($u-r<2.3$, $m_r < 21$, $\ln({L}_{\exp{},r}) > \ln({L}_{\mathrm{DeV},r})$, mode $=1$, clean $=1$, type $=3$); this is presented in \autoref{fig:sdss-inclinations}. Although the CRAFT sample does not trace the SDSS sample as well as it does the ACS-GC sample (possibly due to the different methodologies with which they were derived), at $\pks=0.694$ and $\pad=0.25$, there is insufficient evidence to claim a significant deviation from the background population. This again contradicts the results of B+24.

\section{Correlations and linear fits}
\label{correlations}

Here, we take measurements of our host galaxies, derived from \galfit{} and otherwise, and search for correlations with the properties of the FRB signals themselves.

When searching for correlations, we examine both the Pearson correlation coefficient \rpearson{} \citep{Pearson} and the Spearman's rank correlation coefficient \rspearman{} \citep{Spearman}, which are non-parametric measurements of correlation between variables in a sample. 
The Pearson coefficient represents only linear correlation, while the Spearman coefficient provides a measure of how well the data would be fit by an unspecified monotonic function. 

The null hypothesis implicit in these coefficients is that the correlation between the two populations is actually zero. 
Once again, we adopt the \pval{} of 0.05 as the threshold of statistical significance; that is, the point below which a correlation can be taken seriously.

To then attempt regression of correlated values, we use the Python package 
\ltsfit{} \citep{LTSFit}, which implements the Least Trimmed Squares technique proposed by \cite{LTS} for robust fitting of data containing severe outliers, while accounting for uncertainty in all dimensions. 
Here we fit only linear functions as a first step.

\begin{figure*}
    \centering
    \includegraphics[width=0.8\linewidth]{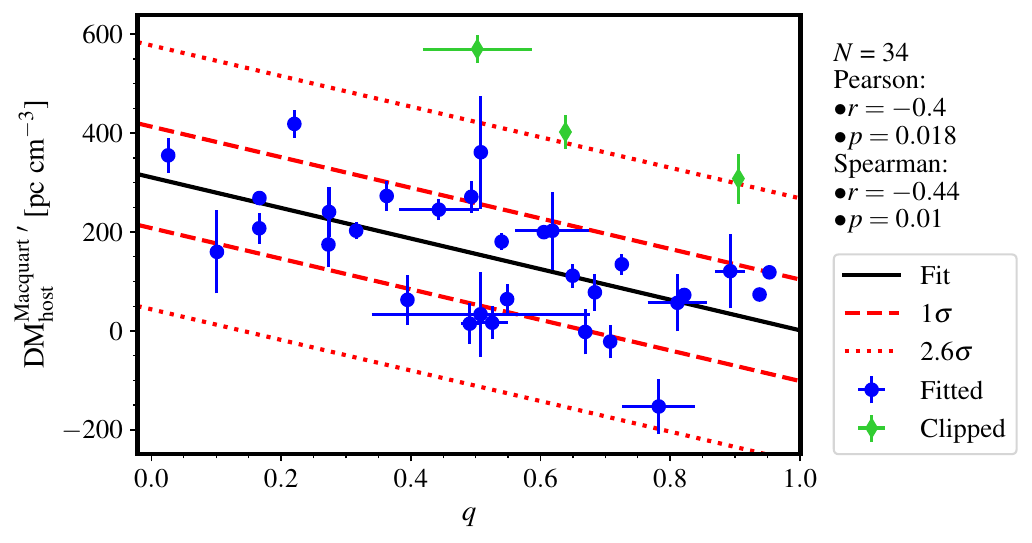}
    
    \caption[Linear fit to \DMExcessRest{} versus axis ratio]{Scatter plot of, and linear fit to, the relationship between the axis ratio $q=b/a$ of the host galaxy and \DMExcessRest{} \textbf{in the host frame}. \xsigma{2.6} lines are drawn as these indicate the 99th percentile for a normal distribution \citep{LTSFit}. The equation of the line of best fit is provided in \autoref{tab:correlations-all}.}
    \label{fig:dmexcess-axisratio}
\end{figure*}

\begin{figure}
    \centering
    \includegraphics[width=\linewidth]{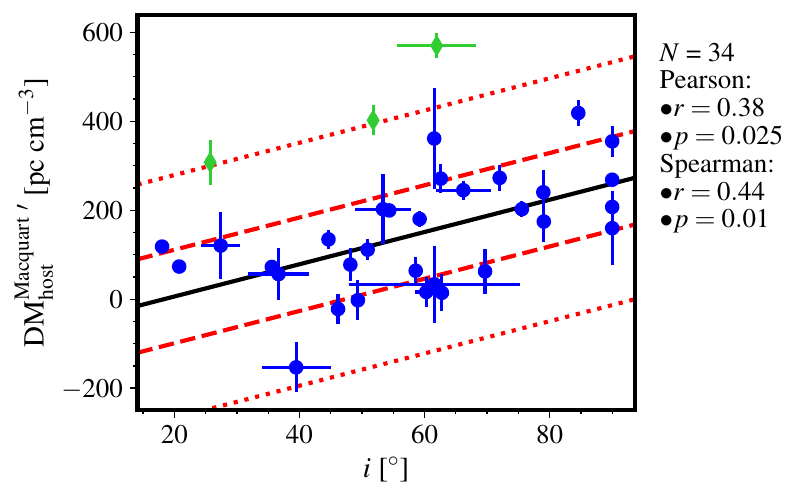}
    
    \caption[Linear fit to \DMExcessRest{} versus inclination]{Scatter plot of, and linear fit to, the relationship between the modelled inclination of the host galaxy (calculated from the axis ratio $q$) and excess DM in the host frame. The equation of the line of best fit is provided in \autoref{tab:correlations-all}.
    }
    \label{fig:dmexcess-inclination}
\end{figure}

\begin{figure}
    \centering
    \includegraphics[width=\linewidth]{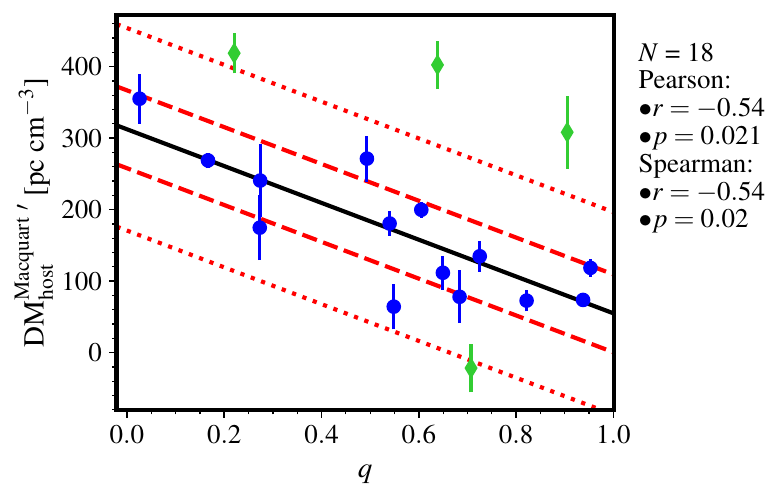}
    
    \caption[Linear fit to \DMExcessRest{} versus axis ratio, with $\mgalfit{} < 21$]{As \autoref{fig:dmexcess-axisratio}, but with a sample ($\star$) limited to $\mgalfit{} < 21$. The equation of the line of best fit is provided in \autoref{tab:correlations-mag}.
    }
    \label{fig:dmexcess-axisratio-mag}
\end{figure}

\begin{figure}
    \centering
    \includegraphics[width=\linewidth]{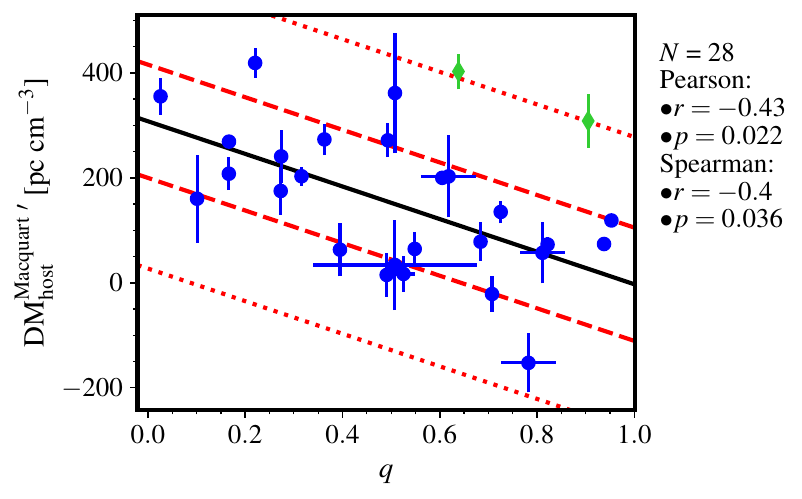}
    
    \caption[Linear fit to \DMExcessRest{} versus axis ratio, with $n<2$]{As \autoref{fig:dmexcess-axisratio}, but with a sample ($\diamond$) limited to $n < 2$. The equation of the line of best fit is provided in \autoref{tab:correlations-n}. 
    }
    \label{fig:dmexcess-axisratio-n}
\end{figure}

\begin{figure}
    \centering
    \includegraphics[width=\linewidth]{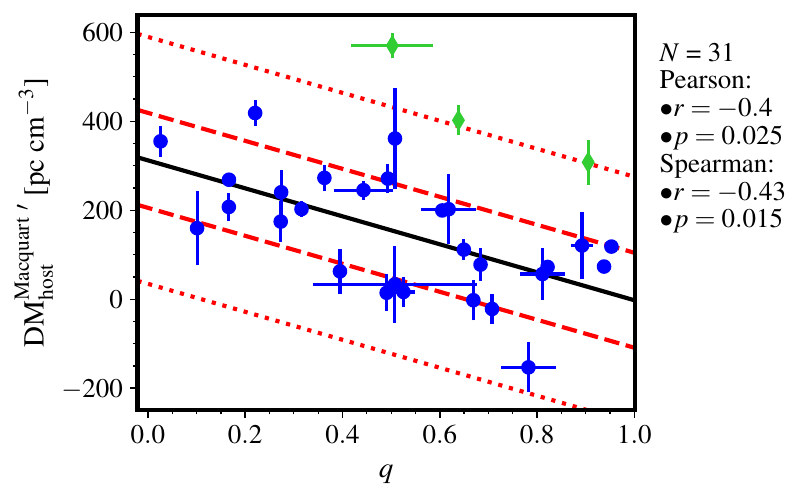}
    
    \caption[Linear fit to \DMExcessRest{} versus axis ratio, with poor \galfit{} fits removed.]{As \autoref{fig:dmexcess-axisratio}, with the poorly-fit (by \galfit{}) hosts of FRBs 20190608B, 20220725A and 20231226A removed from the sample ($\dagger$). The equation of the line of best fit is provided in \autoref{tab:correlations-no-bad-bulges}. 
    }
    \label{fig:dmexcess-axisratio-bulge}
\end{figure}

We begin by checking for any correlation between the observer-frame \DMExcess{} and axis ratio \ba{} in the full sample.
Before clipping, the two variables exhibit correlation coefficients of $\rpearson=\Pearsongalfitaxisratiodmexcesscraftgalfit$ and $\rspearman=\Spearmangalfitaxisratiodmexcesscraftgalfit$. With $\ppearson=\PPearsongalfitaxisratiodmexcesscraftgalfit$ and $\pspearman=\PSpearmangalfitaxisratiodmexcesscraftgalfit$, we consider this moderate correlation to be statistically significant. 

Motivated by the probability that \DMHost{} is driving the correlation with axis ratio, we convert \DMExcess{} to the host frame as \DMExcessRest{}; it is plotted against \ba{} in \autoref{fig:dmexcess-axisratio}. $\rpearson=\Pearsongalfitaxisratiodmexcessrest$ is slightly stronger, and the associated $\ppearson=\PPearsongalfitaxisratiodmexcessrest$ is smaller. $\rspearman$ increases to $\Spearmangalfitaxisratiodmexcessrest$, with $\pspearman$ reduced to $\PSpearmangalfitaxisratiodmexcessrest$.
The strengthening of the correlation (and its statistical significance) in the host frame for both tests lends weight to the case that the effect may be associated with the host galaxy. 
When axial ratios are converted to inclination the correlation is less significant, but still with $p < 0.05$,
as shown in \autoref{fig:dmexcess-inclination}; 
this is likely because of the issues
(discussed in \autoref{morph})
with the toy model used to derive the inclination values. 

Motivated by comparison with the results of B+24, who perform a cut of $\m{r} > 21$ on their FRB host galaxy sample, we remove galaxies with $\mgalfit{} > 21$ for another test (that is, the $\star$ sample). The correlation 
(shown in \autoref{fig:dmexcess-axisratio-mag})
becomes stronger ($\rpearson=\Pearsongalfitaxisratiodmexcessrestmagcut$, $\rspearman=\Spearmangalfitaxisratiodmexcessrestmagcut$) 
with \pval{s} remaining at the 2\% level, despite a halving of the sample size. This could be attributed partially to the increased S/N of the objects resulting in better fits and more reliable axis ratios on average. Dwarf galaxies will have also been disproportionately removed by this cut, increasing the proportion of disks in the sample, which would strengthen any disk-driven correlation. 
We do the same for the $n<2$ sample ($\diamond$), finding the difference from the main sample insignificant ($\rpearson=-0.43$; $\ppearson=0.022$; $\rspearman=-0.4$; $\pspearman=0.036$; see \autoref{fig:dmexcess-axisratio-n}).

In addition, we re-run the fit to \DMExcessRest{} against \ba{} with the data points for the host galaxies of FRBs 20190608B, 20220725A and 20231226A removed (that is, on the $\dagger$ sample); in these, the bulge was fit instead of the disk and the apparent inclinations are exaggerated as a result. The correlation, shown in \autoref{fig:dmexcess-axisratio-bulge}, is not greatly impacted, remaining significant and moderately strong.
This demonstrates that our result is not dependent on those potentially spurious morphological fits.

All correlation scores and equations of best fit are provided in Tables \ref{tab:correlations-all}, \ref{tab:correlations-mag}, \ref{tab:correlations-n} and \ref{tab:correlations-no-bad-bulges}.
For comparison, we have performed the same tests for correlations between $q$ and several other DM component splits derived from \DMFRB{}. Of these, the \DMExcessRest{} correlation described above remains the strongest.

In addition, we have performed similar checks of \DMExcessRest{} against measurements of FRB separation from the host nucleus, including separation projected to the host redshift (\offsetproj{}), de-projected using our derived inclination (\offsetdeproj), and normalised to the host's effective radius (\offsetnorm{}; \offsetdeprojnorm{}). We also performed the checks on the colours, either $(g-I)$ or $(R-K)$, of the hosts.
Most of these do not yield significant correlations. There are two exceptions in the $\mgalfit>21$ subsample ($\star$): \DMExcessRest{} correlates moderately with $\offsetnorm{}$ by both measures ($\rpearson{}=0.53$, $\rspearman{}=0.54$), to a statistically significant level ($\ppearson{}=0.023$, $\pspearman{}=0.021$); while \DMExcessRest{} and $\offsetdeprojnorm{}$ yield scores of $\rpearson{}=0.53$ ($\ppearson{}=0.023$) and $\rspearman{}=0.77$ ($\ppearson{}=0.00044$).
These correlations are not found in the full sample or other subsamples. This may again reflect the more reliable axis ratios (from which inclinations and deprojected offsets are calculated) and effective radii afforded by the brighter galaxies. It is easy to imagine that the amount of gas traversed by an FRB might vary with the distance of its source from a galactic nucleus. It should be noted that \offsetdeproj{} is an unreliable value, as it depends on the inclination $i$. Nonetheless, 
this correlation is well worth further investigation with a larger sample.


\input{tables/_correlations}

\section{Discussion}
\label{discussion}

\subsection{Inclinations and ACS-GC}
\label{discussion:acsgc}

According to the cosmological principle, no particular orientation relative to Earth should be preferred. However, 
the $\cos(i)$ distribution we derive for the ACS-GC sample (using the same formula presented in \autoref{equ:inclination}) is not consistent with a uniform distribution, presenting a sigmoid empirical distribution function which appears to prefer edge-on to face-on, and moderately inclined to both, as also discussed by B+24.
The low inclination roll-off can be explained by the fact that real galaxies do not form perfect circles (or even, necessarily, ellipses; see e.g. \citealp{Lokas2022}); that is, a real face-on galaxy does not have precisely $\ba=1$, nor will all galaxies share a common face-on axis ratio. Therefore, when the inclination is calculated by \autoref{equ:inclination}, there will be only a few or no completely `face-on' galaxies, even for a large sample. 
At the high end, the roll-off is an artefact of the choice of \qint{}, and that some galaxies are flatter than others (for a catalogue of extreme cases, see \citealp{UltraFlatGalaxies}). For example, a galaxy that is truly viewed edge-on, but has a true $\qint=0.3$, will have a calculated inclination of only \SI{77}{\degree} using $\qint=0.2$. 
So, while it is unlikely that any orientation with respect to Earth is actually preferred, this is not reflected in the measured distribution of inclination values. 
The axis ratio is more directly observable and thus more useful for cross-sample comparison. It also avoids an artificial cut at $\ba<q_0$, which occurs when converting to inclination as discussed in \autoref{morph}.

\subsection{The correlation between inclination and host dispersion measure}
\label{discussion:correlation}

The results presented in \autoref{correlations}, consistent with the simulation-based studies of \cite{Kovacs2024} and \cite{Orr2024}, indicate that some significant fraction of the variation in \DMHost{} can be explained geometrically. This has significant ramifications for our understanding of where FRBs occur, and hence what causes them. 
Additionally, reducing the observed scatter in the Macquart relation by accounting for this effect will ultimately affect the inferred amount of feedback in galaxy evolution, which reduces scatter by redistributing baryons \citep{Baptista2023}.

A correlation with inclination places at least some significant fraction, perhaps the majority, of CRAFT FRB sources in the disks of their hosts. If they were preferentially hosted in the stellar halo, as the M81 repeater is (found in a globular
cluster; \citealp{GlobularCluster}), then the host galaxy's orientation would be of little 
importance except, perhaps, in that a face-on galaxy would block the lines-of-sight from a greater fraction of its stellar halo. However, this would cause a trend in the opposite direction to what is found here, or a weak selection bias against FRBs in face-on hosts.

Placement in the disk strengthens the connection to both magnetars, which have thus far been found almost exclusively in the plane of the disk of the Milky Way \citep{Magnetars}, and star formation, which also tends to occur in the disk.
In addition, these findings offer a new empirical method of estimating individual values for \DMHost{}. 
While there is clearly a great deal of scatter in this `relation', there is significant value in a \DMHost{} calculated from easily-measured observables taken directly from sufficiently deep imaging.

\begin{figure}
    \centering
    \includegraphics[width=\linewidth]{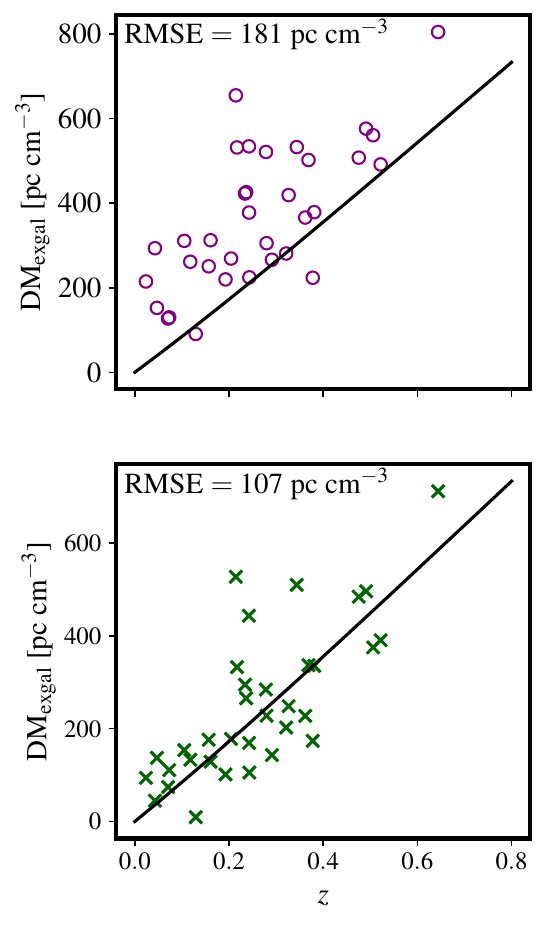}
    
    \caption[Axis ratio-corrected \DMExgal{} against redshift.]{The Macquart relation for CRAFT FRB host galaxies (\DMExgal{} versus redshift), before and after correction using the axis ratio-\DMExcessRest{} relation given in \autoref{fig:dmexcess-axisratio}. The green crosses show the corrected values, and the purple circles the original. The black line provides the theoretical Macquart relation \DMCosmicAvg{}.
    }
    \label{fig:dmexgalcorr-z}
\end{figure}

We use the equation of the line of best fit provided in \autoref{fig:dmexcess-axisratio} as a model of \DMHost{} as a function of \ba{}:
\begin{equation}
    \DMRestsup{model}{host} = [-309.3\times(\ba{}-0.5) + 155.9]\ \mathrm{pc\ cm^{-3}}
\end{equation} We calculate \DMRestsup{model}{host} from this equation, then for each host convert it to the observer frame as \DMsup{model}{host}. 
In \autoref{fig:dmexgalcorr-z}, we subtract \DMsup{model}{host} from the extragalactic dispersion measure contribution \DMExgal{} and plot this against redshift. This significantly reduces the scatter about the theoretical Macquart relation: before correction, the points have a root-mean-square error (RMSE) from $\DMCosmic{}$ of \DMSI{181}; after correction, the RMSE is \DMSI{107}, a $40\%$ reduction.

Although potentially useful, the fact that changing the bounds of the sample 
affects the slope of the relation to the degree shown in Figures \ref{fig:dmexcess-axisratio} to \ref{fig:dmexcess-axisratio-bulge} indicates that the sample size is not yet large enough for precise estimate of the mean correction.
However, the correlation appears real; in all of our sample cuts, the trend remains a negative correlation between axis ratio and \DMExcessRest{}, corresponding to a positive trend in inclination and \DMExcessRest{}. In fact, the correlation becomes stronger with a magnitude cut. Likewise, \pspearman{} and \ppearson{}, computed prior to fitting lines or clipping outliers, indicate that the correlation is statistically significant in each case. We will need to continue expanding our sample to better measure the shape of the relation, test its generality, and understand genuine outliers.




\subsection{A tension with B+24}
\label{discussion:bhardwaj}

\begin{figure}
    \centering
    \includegraphics[width=\linewidth]{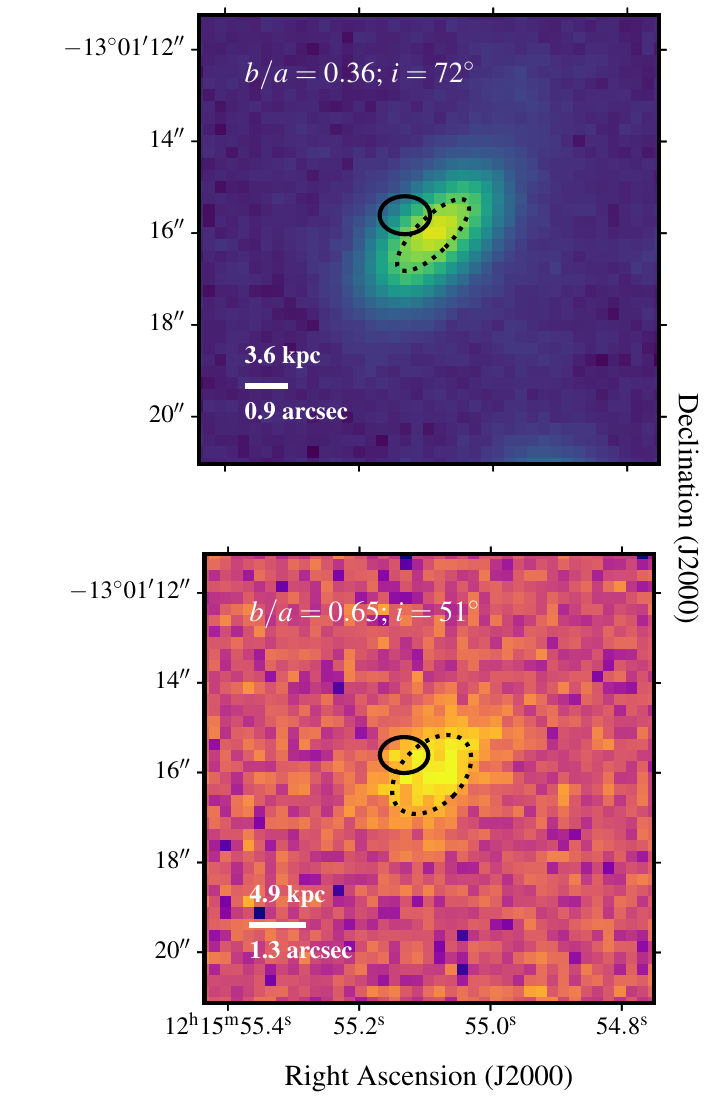}
    \caption{Comparison of imaging for \FRB{20190714A} from the Very Large Telescope as used by this work (top); and Pan-STARRS, as used by B+24 (bottom). The dotted ellipses illustrate the axis ratios fitted by us and B+24, respectively, and from which inclination is derived. The solid ellipses provide the \xsigma{1} FRB localisation region. The angular scale provides the PSF FWHM for the image, and the physical scale is the projected distance of the FWHM at the host's redshift.}
    \label{fig:frb20190714A}
\end{figure}

\begin{figure}
    \centering
    \includegraphics[width=\linewidth]{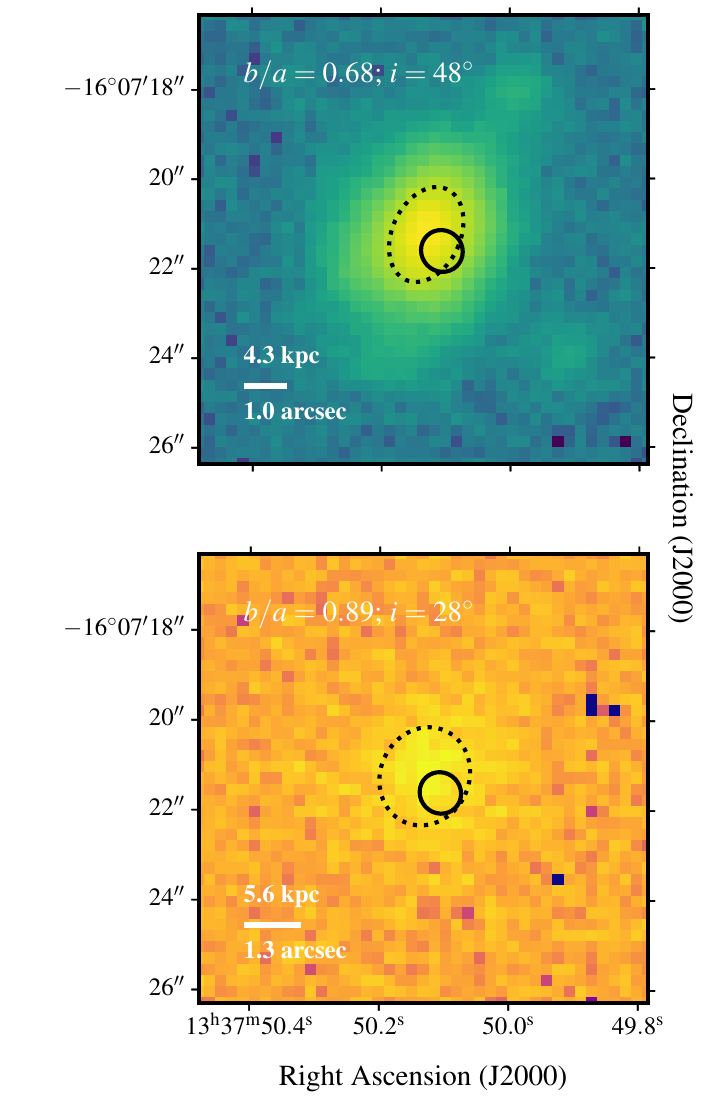}
    \caption{Comparison of imaging for \FRB{20210320C} from the Very Large Telescope as used by this work (top); and Pan-STARRS as used by B+24 (bottom). Ellipses and scales as described in \autoref{fig:frb20190714A}.}
    \label{fig:frb20210320C}
\end{figure}

\begin{figure}
    \centering
    \includegraphics[width=\linewidth]{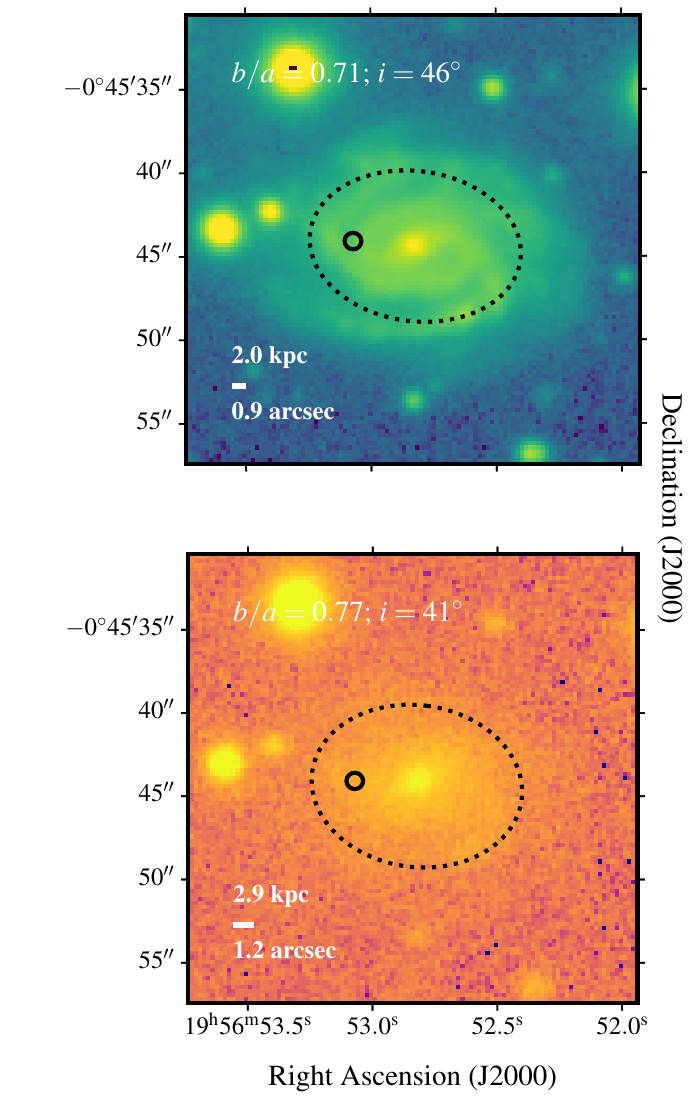}
    \caption{Comparison of imaging for \FRB{20210807D} from the Very Large Telescope as used by this work (top); and Pan-STARRS, as used by B+24 (bottom). Ellipses and scales as described in \autoref{fig:frb20190714A}.}
    \label{fig:frb20210807D}
\end{figure}

\begin{figure}
    \centering
    \includegraphics[width=\linewidth]{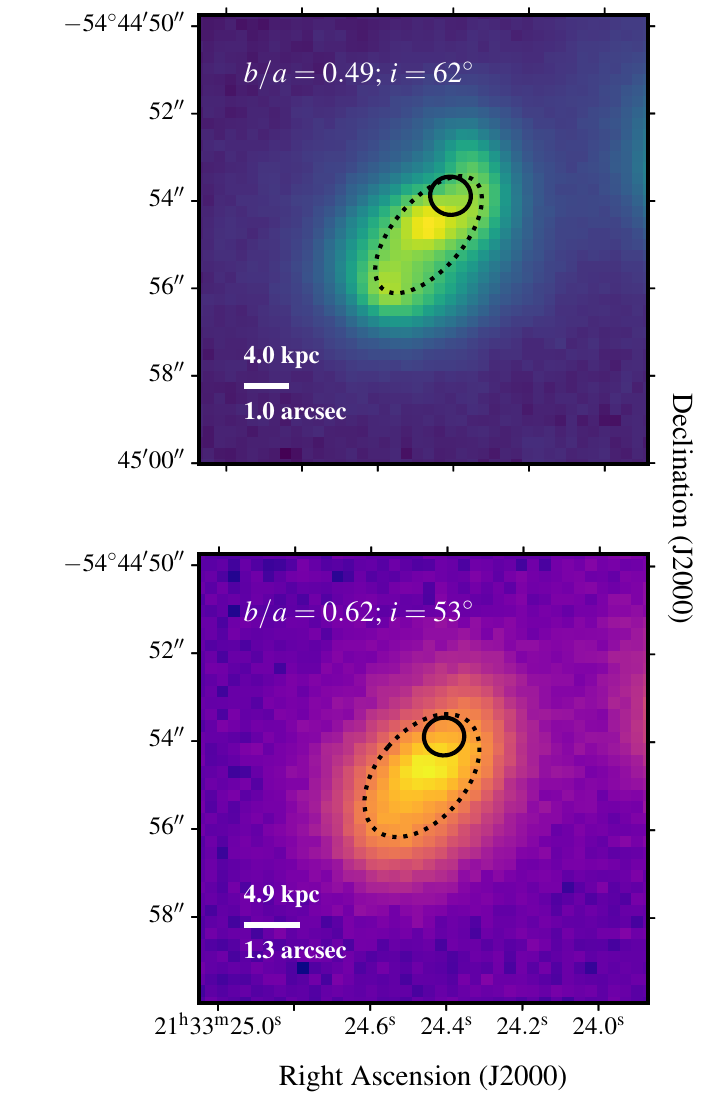}
    \caption{Comparison of imaging for \FRB{20191001A} from the Very Large Telescope as used by this work (top); and the DESI Legacy Survey as used by B+24 (bottom). Ellipses and scales as described in \autoref{fig:frb20190714A}.}
    \label{fig:frb20191001A}
\end{figure}

\begin{figure}
    \centering
    \includegraphics[width=\linewidth]{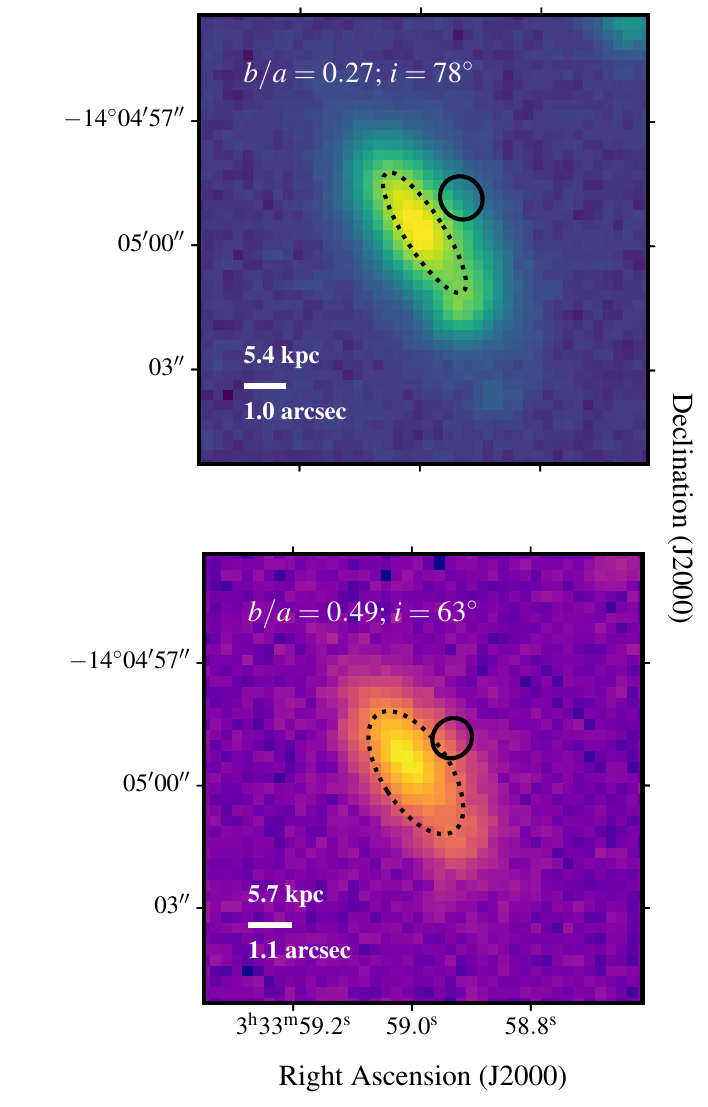}
    \caption{Comparison of imaging for \FRB{20200906A} from the Very Large Telescope as used by this work (top); and the DESI Legacy Survey as used by B+24 (bottom). Ellipses and scales as described in \autoref{fig:frb20190714A}.}
    \label{fig:frb20200906A}
\end{figure}

    

    

\begin{figure*}
    \centering
    \includegraphics[width=0.7\linewidth]{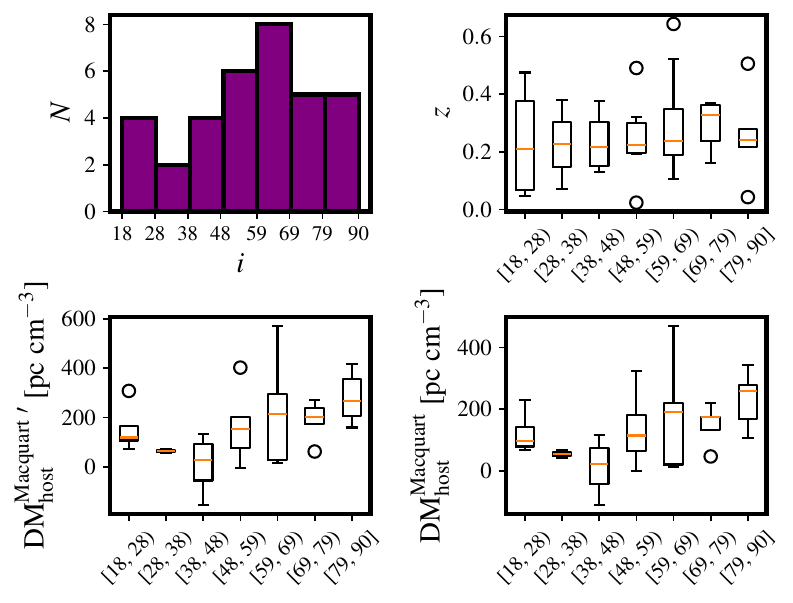}
    
    \caption[Reproduction of B+24 Figure 2, with our full \galfit{} sample and bins freely determined]{A partial reproduction of Figure 2 of B+24 for our CRAFT FRB host sample, using our full \galfit{} sample and with inclination bins freely determined. The top-left panel is a histogram, and the rest are box-and-whisker plots. For each inclination bin, the orange line gives the median, the upper and lower bounds of the box give the first and third quartiles, and the whiskers extend to the farthest data point that is within 1.5 inter-quartile ranges of the box. Circles show outliers past this range.}
    \label{fig:bhardwaj-like-any}
\end{figure*}



The results presented in \autoref{populations} are in severe tension with those presented by B+24, who have performed a similar study on a heterogeneous sample of 23 hosts; 10 from the Deep Synoptic Array (DSA; e.g., \citealp{LawDSA-110Cat}) and 13 from the CRAFT survey. They find that highly-inclined FRB hosts are under-represented in their host sample, to an A-D \pval{} of 0.0003. By contrast, \autoref{populations} finds no significant difference between the CRAFT FRB host sample and the ACS-GC background 
population.
This is surprising, given that we have adopted similar
methods.
Some potential explanations are explored below. 

B+24 perform their primary comparison on values of $\cos{(i)}$ derived with \autoref{equ:inclination}. Although we elected instead to make the axis ratio \ba{} the primary point of comparison, as it is more directly observable and not subject to the same biases and assumptions, we have also derived $\cos{(i)}$ in the same fashion and performed the same tests against it.
Neither figure \ref{fig:acsgc-inclinations} nor \ref{fig:sdss-inclinations}, the latter using the same SDSS DR16 sample as B+24, demonstrate significant differences in the $\cos(i)$ distributions
of our FRB host galaxy sample and our control sample.
It appears that the two samples, B+24 and our own, trace fundamentally different empirical cumulative distribution functions, as we can see when comparing these figures to Figure 1 of B+24.
The ECDF presented in B+24 Figure 1 has an obvious, visible deficit at the high inclination (small $\cos{i}$) end of the distribution when compared to our own, even without looking at the control samples.

\input{tables/overlap}

We test whether this tension remains when adopting
the sample cut of 
$m_\galfit{} < 21$. 
While this does not perfectly match the cut of $m_R < 21$ used in B+24, it is the closest approximation available to us given that approximately half of our sample was imaged in filters other than $R$-band.
No significant difference is observed from the ECDF of our full sample. Again, we do not observe the edge-on deficit.
It appears that the tension between our results cannot be resolved with sample cuts.

For comparison, 
we have partially reproduced the Figure 2 of B+24 
in \autoref{fig:bhardwaj-like-any} using our sample, with different binning set freely according to the sample size (using \numpy{'s} \href{https://numpy.org/doc/stable/reference/generated/numpy.histogram_bin_edges.html}{\texttt{bin\_edges(bins="auto")}}).

Not only do we not see the deficit of edge-on galaxies, the \DMExcessRest{} box plot demonstrates the same trend shown in \autoref{fig:dmexcess-inclination}: that is, increasing inclination yields increasing (in this case, median) \DMExcessRest{}; this trend is not visible in B+24's equivalent plot, and they explicitly report looking for and \textit{not} seeing any such trend. 

The fact that our results differ so significantly is especially surprising considering that our samples overlap; that is, we have used 11 of the same FRB hosts, those of ASKAP FRBs 20180924B, 20190102C, 20190608B, 20190714A, 20191001A, 20200906A, 20210320C, 20210807D, 20211127I, 20211203C and 20211212A, with the remaining 23 of B+24's sample being DSA-110 hosts. We have \textit{not}, however, used the same imaging, and derive different inclinations for several of them. 
The intersection of our two samples is tabulated in \autoref{tab:overlap}. B+24 inclinations are systematically smaller (that is, they trend towards being more face-on), with a mean difference of \SI{-9}{\degree} and a median difference of \SI{-13}{\degree}. B+24 have used a combination of Dark Energy Spectroscopic Instrument (DESI) Legacy Imaging Survey\footnote{\url{https://www.legacysurvey.org/}} DR10 (for southern FRBs, predominantly the CRAFT ones) and Pan-STARRS DR2 (northern, predominantly DSA-110; \citealp{PanSTARRS}) imaging to derive their results, whereas we have used only the (significantly deeper) VLT/FORS2 imaging; albeit, in three different bands and with slightly varied depths. As B+24 note, the Pan-STARRS imaging is shallower still than that from DESI. In Figures \ref{fig:frb20190714A} to \ref{fig:frb20210807D} (Pan-STARRS) and \ref{fig:frb20191001A} to \ref{fig:frb20200906A} (DESI), the VLT imaging of a selection of hosts used in both samples is compared to the corresponding survey imaging. Given the difference in depth, it is perhaps little surprise that we derive differing inclinations, even before considering the methodology difference highlighted in the following paragraph. This may not fully explain a systematic bias, but it might contribute---galaxies often have bulges or spheroidal inner components, which will be emphasised in shallower imaging due to their higher surface brightness, potentially leading to the inference of rounder axis ratios. In addition, the broader point-spread function (indicated in the scale bars of Figures \ref{fig:frb20190714A} to \ref{fig:frb20200906A})  of the survey imaging will tend to make galaxies appear rounder. While our respective software packages are supposed to account for this by including a PSF model, they may do so imperfectly. 

B+24 use a similar method to derive morphological models, and from them the axis ratios and thus inclinations. They do use different software, \autoprof{},
which is primarily intended for extracting non-parametric surface brightness profiles \citep{AutoProf}; it can then be set to fit parametric functions, such as the \sersic{}, to the  extracted profile, which is how B+24 use it. On the other hand, we have used \galfit{} to derive \sersic{} profiles directly from the imaging data. This may result in some systematic differences between the derived models.

Methodologies aside, the CRAFT sample contains several galaxies that are, by eye alone, clearly high-inclination: \FRB{20240201A},
\FRB{20231230},
\FRB{20220105A};
as well as possibly \FRB{20200906A} 
and \FRB{20190714A}. Each of these are shown in Figure 1 of \cite{ShannonICS}.
The three former hosts alone account for $\sim8\%$ of our sample, arguing against a significant deficit of edge-on galaxies. 

B+24 find no trend between \DMHost{} (derived using approximately the same method as our \DMExcessRest{}{}) and inclination, despite expecting to see one, and connect this to the deficit in edge-on hosts; that is, the high-\DMHost{} FRBs are missing for the same reason that the highly-inclined ones are, as they are one and the same. However, we \textit{do} see this trend in our data, presented in
\autoref{correlations}. 
When we reproduce the B+24 magnitude cut, the trend remains. 
Ultimately our imaging is deeper and taken under better seeing conditions, and our sample size is larger and more homogeneous, both in FRB detection and in optical imaging.

\section{Conclusions}

We have performed tests that continue to strengthen the connection between FRBs and recent star formation, consistent with findings by \cite{James2022B, Bhandari2022, GordonProspector}, and others. The presence of a correlation between the orientation of a galaxy and its contribution to \DMFRB{} is consistent with the analyses of \cite{Kovacs2024} and \cite{Orr2024}. It is also a strong indicator that FRBs are preferentially hosted in the disk, home to both star formation and magnetars. 
A disk preference also indicates that occurrences in globular clusters are probably infrequent, at least in the CRAFT FRB sample analysed here. These results are supported by the work of \cite{GordonSpatial}.

In addition, we have used this relationship to demonstrate a new technique for estimating \DMHost{}, and used this to reduce the scatter in the Macquart relation by 40\%.

Although \nGALFIT{} galaxies does not, in most fields of astronomy, constitute a large sample, it is an important step in the study of FRB hosts. With it, we are finally beginning to see how the host population relates to the galaxy population at large. Future, large-scale samples will only improve upon this.

On the topic of parametric morphology, further work could benefit from decomposition of components representing the bulge, disk and/or nucleus of some galaxies. The Bayesian Information Criterion and similar measurements may be helpful in deciding how many components are needed for each galaxy. For this work we have used a single-component fit for the sake of comparison to the literature, in which large samples of multi-component parametric morphology are scarce, but work with a greater focus on the \DMHost{}--\ba{} correlation will benefit from the multi-component approach.

An important step is a search for a correlation between inclination with the scattering timescale or with rotation measure. A trend in scattering with \DMHost{} is broadly expected. A trend with RM is predicted by simulation \citep{Kovacs2024}, and would be expected given that \DMHost{} and RM have been found to correlate \citep{Mannings2023}. \cite{GlowackiScattering} have explored these using the morphological dataset that we derived here, finding no correlation between inclination and scattering, but a positive trend between $|\RM{}|$ and \ba{}.

In exploring the \DMHost{}--inclination relationship, we might improve our cross-sample comparison by using advanced software such as \textsc{Galapagos-2/Galfitm} \citep{galapagos,GALFITM} or \textsc{ProFit/ProFuse} \citep{ProFit, ProFuse}, both designed for many-object and multi-band parametric morphology, to construct a larger catalogue of VLT field galaxies. We might also improve the control sample by better replicating the CRAFT sample in terms of magnitude and redshift, or by analysing the rest of the VLT images themselves.
While we could improve our understanding of the relationship incrementally by fine-tuning the methodology and adding a handful of other hosts to the sample, only a larger sample of deeply-imaged FRB host galaxies (preferably, in consistent bandpasses) will truly advance it. This will also be necessary for exploring any correlation with colour. Fortunately, by the time samples of thousands of FRB host galaxies become available, so will deep imaging from the Legacy Survey of Space and Time \citep{LSST} in up to 6 bandpasses for all southern hosts; this will enable such a study, among many others \citep{Clancy_LSST}.

\section*{Data Availability}

Code and data for reproducing the results reported here are available at \url{https://github.com/Lachimax/publications/tree/main/Marnoch\%2B2026_Inclinations}; further input data will be downloaded by the script \texttt{00-retrieve.py}.
The fully-reduced VLT/FORS2 images used for this work can be supplied upon request from the lead author. 

\section{Acknowledgments}

The authors thank Aaron Robotham, Jason Hessels and Clancy James, whose comments elevated this work.
Based on observations collected at the European Southern Observatory under ESO programmes 105.204W.003 and 108.21ZF.005.
L.M. acknowledges the receipt of an MQ-RES scholarship from Macquarie University and a Studentship from CSIRO.
R.M.S. acknowledges support through Australian Research Council Future Fellowship FT190100155 and Discovery Project  DP220102305. N.T. acknowledges support from FONDECYT grant 1252229. W.F. and A.C.G. acknowledge support by National Science Foundation under grant Nos. AST-2206494, AST-2308182, AST-2432037, and CAREER grant No. AST-2047919, the David and Lucile Packard Foundation and the Research Corporation for Science Advancement through Cottrell Scholar Award \#28284. A.C.G. and J.X.P. acknowledge support from NSF grants AST-1911140, AST-1910471 and AST-2206490 as a member of the Fast and Fortunate for FRB Follow-up team. S.S. acknowledges the support of the Brinson Foundation as the NU-UC Brinson Postodctoral Fellow. M.G. is supported through UK STFC Grant ST/Y001117/1. M.G. acknowledges support from the Inter-University Institute for Data Intensive Astronomy (IDIA). IDIA is a partnership of the University of Cape Town, the University of Pretoria and the University of the Western Cape.

\bibliographystyle{apalike}
\bibliography{references}

\appendix

\section{Supplementary tables and figures}
\label{app:residuals}

    

\input{residuals.tex}

\end{document}

%% file: commands/siunits.tex
\usepackage{siunitx}
\DeclareSIUnit{\adu}{ADU}
\DeclareSIUnit{\electron}{e^-}
\DeclareSIUnit{\hubbleunits}{\kilo\metre\per\second\per\mega\parsec}
\DeclareSIUnit{\parsec}{pc}
\DeclareSIUnit{\kpc}{kpc}
\DeclareSIUnit{\pixel}{pix}
\DeclareSIUnit{\rmunits}{\radian\ \metre^{-2}}
\DeclareSIUnit{\solarmass}{M_\odot}
\DeclareSIUnit{\solarlum}{L_\odot}
\DeclareSIUnit{\year}{yr}
\DeclareSIUnit{\jansky}{Jy}
\DeclareSIUnit{\erg}{erg}
\DeclareSIUnit{\gauss}{G}

\DeclareSIUnit{\dmunits}{pc\ cm^{-3}}

\DeclareSIUnit{\Funits}{\parsec^{-2/3}\ \kilo\metre^{-1/3}}

\newcommand{\DMSI}[1]{\SI{#1}{\dmunits}}

%% file: commands/commands.tex
\newcommand{\dn}[1]{\,\mathrm{d}#1}

\newcommand{\degrees}{^\circ}

\newcommand{\tavg}[1]{\langle #1 \rangle}

\newcommand{\pval}{$p$-value}
\newcommand{\pks}{\ensuremath{p_\textsc{ks}}}
\newcommand{\pad}{\ensuremath{p_\textsc{ad}}}
\newcommand{\pspearman}{\ensuremath{p_\textsc{s}}}
\newcommand{\ppearson}{\ensuremath{p_\textsc{p}}}

\newcommand{\rspearman}{\ensuremath{r_\textsc{s}}}
\newcommand{\rpearson}{\ensuremath{r_\textsc{p}}}

\newcommand{\HII}{\ion{H}{ii}}

\newcommand{\xsigma}[1]{#1\text{-}\ensuremath{\sigma}}

\newcommand{\m}[1]{\ensuremath{m_\mathrm{#1}}}

\newcommand{\M}[1]{\ensuremath{M_\mathrm{#1}}}

\newcommand{\en}{\ensuremath{n_\mathrm{e}}}
\newcommand{\electron}{\ensuremath{\mathrm{e}^-}}

\newcommand{\Reff}{\ensuremath{R_\mathrm{eff}}}

\newcommand{\ba}{\ensuremath{b/a}}
\newcommand{\mgalfit}{\m{galfit}}
\newcommand{\qint}{\ensuremath{q_0}}

\newcommand{\massstar}{\M{\star}} 

\newcommand{\z}[1]{\ensuremath{z_\mathrm{#1}}}

\newcommand{\zhost}{\z{host}}

\newcommand{\offsetnorm}{\ensuremath{\rho / \Reff}}
\newcommand{\offsetproj}{\ensuremath{\rho_\mathrm{proj}}}

\newcommand{\offsetdeproj}{\ensuremath{\rho_\mathrm{deproj}}}
\newcommand{\offsetdeprojnorm}{\ensuremath{\rho_\mathrm{deproj} / \Reff}}

\newcommand{\DMRest}[1]{\ensuremath{\mathrm{DM^\prime_{#1}}}}
\newcommand{\DM}[1]{\ensuremath{\mathrm{DM_{#1}}}}
\newcommand{\DMsup}[2]{\ensuremath{\mathrm{DM^{#1}_{#2}}}}
\newcommand{\DMRestsup}[2]{\ensuremath{\mathrm{DM^{#1\ \prime}_{#2}}}}
\newcommand{\DMFRB}{\DM{FRB}}

\newcommand{\DMMW}{\DM{MW}}
\newcommand{\DMMWHalo}{\ensuremath{\DMsup{halo}{MW}}}
\newcommand{\DMMWISM}{\DMsup{ISM}{MW}}
\newcommand{\DMNE}{\DM{NE2001}}
\newcommand{\DMYMW}{\DM{YMW16}}

\newcommand{\DMExgal}{\DM{exgal}}
\newcommand{\DMExcess}{\DMsup{Macquart}{host}}
\newcommand{\DMExcessRest}{\DMRestsup{Macquart}{host}}
\newcommand{\DMCosmic}{\DM{cosmic}}
\newcommand{\DMCosmicAvg}{\ensuremath{\tavg{\DMCosmic}}}

\newcommand{\DMHost}{\DM{host}}
\newcommand{\DMHostRest}{\DMRest{host}}
\newcommand{\DMHostHalo}{\DMsup{halo}{host}}
\newcommand{\DMHostHaloRest}{\DMRestsup{halo}{host}}
\newcommand{\DMHostISM}{\DMsup{ISM}{host}}
\newcommand{\DMHostISMRest}{\DMRestsup{ISM}{host}}

\newcommand{\DMSource}{\DM{source}}
\newcommand{\DMSourceRest}{\DMRest{source}}

\newcommand{\RM}[1]{\mathrm{RM_{#1}}}

\newcommand{\FRB}[1]{FRB\,#1}

\newcommand{\POx}{\ensuremath{P(O_i|x)}}

\newcommand{\dmunits}{\ensuremath{\ \mathrm{pc}\ \mathrm{cm}^{-3}}}
\newcommand{\rmunits}{\mathrm{rad\ m^{-2}}}

\newcommand{\sersic}{Sérsic}


%% file: commands/commands_other.tex
\newcommand{\autoprof}{\textsc{AutoProf}}
\newcommand{\ccdproc}{\textsc{ccdproc}}

\newcommand{\galapagos}{\textsc{galapagos}}
\newcommand{\galfit}{\textsc{Galfit}}
\newcommand{\numpy}{\textsc{NumPy}}
\newcommand{\sextractor}{\textsc{Source-Extractor}}
\newcommand{\scipy}{\textsc{SciPy}}
\newcommand{\ltsfit}{\textsc{LtsFit}}
\newcommand{\esoreflex}{\textsc{ESOReflex}}
\newcommand{\montage}{\textsc{Montage}}
\newcommand{\psfex}{\textsc{PSFEx}}
\newcommand{\astronet}{\textsc{Astrometry.net}}

\newcommand{\magmax}{24}

%% file: commands/commands_galfit_generated.tex
\newcommand{\nGALFIT}{34}
\newcommand{\nGALFITz}{34}
\newcommand{\nGALFITmag}{16}
\newcommand{\nGALFITn}{28}

%% file: commands/commands_acsgc_generated.tex
\newcommand{\NBAGALFITCRAFTvACSGC}{15667}
\newcommand{\KSgalfitaxisratioCRAFTvACSGC}{0.848}
\newcommand{\ADgalfitaxisratioCRAFTvACSGC}{0.25}
\newcommand{\NIGALFITCRAFTvACSGC}{15667}
\newcommand{\KSgalfitinclinationCRAFTvACSGC}{0.848}
\newcommand{\ADgalfitinclinationCRAFTvACSGC}{0.25}
\newcommand{\NCOSIGALFITCRAFTvACSGC}{15667}
\newcommand{\KSgalfitcosinclinationCRAFTvACSGC}{0.848}
\newcommand{\ADgalfitcosinclinationCRAFTvACSGC}{0.25}

\newcommand{\NBAGALFITCRAFTmagvACSGCmag}{1002}
\newcommand{\KSgalfitaxisratioCRAFTmagvACSGCmag}{0.495}
\newcommand{\ADgalfitaxisratioCRAFTmagvACSGCmag}{0.25}
\newcommand{\NIGALFITCRAFTmagvACSGCmag}{1002}
\newcommand{\KSgalfitinclinationCRAFTmagvACSGCmag}{0.495}
\newcommand{\ADgalfitinclinationCRAFTmagvACSGCmag}{0.25}
\newcommand{\NCOSIGALFITCRAFTmagvACSGCmag}{1002}
\newcommand{\KSgalfitcosinclinationCRAFTmagvACSGCmag}{0.495}
\newcommand{\ADgalfitcosinclinationCRAFTmagvACSGCmag}{0.25}

\newcommand{\NBAGALFITCRAFTnvACSGCn}{12052}
\newcommand{\KSgalfitaxisratioCRAFTnvACSGCn}{0.842}
\newcommand{\ADgalfitaxisratioCRAFTnvACSGCn}{0.25}
\newcommand{\NIGALFITCRAFTnvACSGCn}{12052}
\newcommand{\KSgalfitinclinationCRAFTnvACSGCn}{0.842}
\newcommand{\ADgalfitinclinationCRAFTnvACSGCn}{0.25}
\newcommand{\NCOSIGALFITCRAFTnvACSGCn}{12052}
\newcommand{\KSgalfitcosinclinationCRAFTnvACSGCn}{0.842}
\newcommand{\ADgalfitcosinclinationCRAFTnvACSGCn}{0.25}

%% file: commands/commands_lts_generated.tex
\newcommand{\Pearsongalfitaxisratiodmexcessrest}{-0.402}
\newcommand{\PPearsongalfitaxisratiodmexcessrest}{0.018}
\newcommand{\Spearmangalfitaxisratiodmexcessrest}{-0.437}
\newcommand{\PSpearmangalfitaxisratiodmexcessrest}{0.01}

\newcommand{\Pearsongalfitaxisratiodmexcessrestmagcut}{-0.538}

\newcommand{\Spearmangalfitaxisratiodmexcessrestmagcut}{-0.544}

\newcommand{\Pearsongalfitaxisratiodmexcesscraftgalfit}{-0.399}
\newcommand{\PPearsongalfitaxisratiodmexcesscraftgalfit}{0.02}
\newcommand{\Spearmangalfitaxisratiodmexcesscraftgalfit}{-0.425}
\newcommand{\PSpearmangalfitaxisratiodmexcesscraftgalfit}{0.012}

%% file: tables/craft_sample.tex
\begin{table*}
\centering
\caption{FRB host sample properties. Galaxies with an axis ratio less than 0.2 have been set to $i=90\degrees$; as our statistical results are founded on the \ba{} measurement, they are not affected by this choice. 
\POx\ is the PATH posterior for host association. Symbols in the second column indicate sample membership: $\star:$ sample cut by magnitude ($\mgalfit\leq21.0$); $\diamond:$ sample cut by \sersic{} index ($n<2.0$); $\dagger:$ main sample with the hosts of FRBs 20190608B, 20220725A and 20231226A removed (see \autoref{morph}). Although statistical uncertainties are estimated for each quantity (with the uncertainties in $\ba$, $n$ and $\mgalfit$ listed as reported by \galfit{}), in the case of inclination the systematics will overwhelm these, and so we report only the fiducial values for inclination to avoid false precision.}
\label{tab:craft_sample}
\begin{tabular}{llccccccccc}
\hline FRB & \phantom{S} & $z$ & \POx & Band & $\DMExcess$ & $b/a$ & $i$ & $n$ & $m_\mathrm{galfit}$ & Ref. \\
  &   &   &   &   & \dmunits &   & ($\degrees$) &   &   \\ \hline
20180924B & ${\dagger}$ & 0.3214 & 0.99 & $g$ & $0(30)$ & $0.669(8)$ & 49 & $4.4(2)$ & $21.07(2)$ & \cite{BannisterFRB20180924B} \\
20181112A & ${\dagger}$ & 0.4755 & 0.761 & $I$ & $80(50)$ & $0.89(2)$ & 27 & $2.2(2)$ & $21.19(3)$ & \cite{ProchaskaFRB20181112A} \\
20190102C & ${\diamond\dagger}$ & 0.2912 & 0.981 & $g$ & $10(30)$ & $0.49(1)$ & 62 & $0.67(5)$ & $22.554(9)$ & \cite{MacquartCosmicDM} \\
20190608B & ${\star}$ & 0.11778 & 0.992 & $g$ & $160(20)$ & $0.540(2)$ & 59 & $3.60(4)$ & $18.108(6)$ & \cite{MacquartCosmicDM} \\
20190611B & ${\diamond\dagger}$ & 0.3778 & 0.803 & $g$ & $-110(40)$ & $0.78(6)$ & 39 & $0.1(6)$ & $24.00(2)$ & \cite{MacquartCosmicDM} \\
20190711A & ${\diamond\dagger}$ & 0.522 & 0.935 & $I$ & $20(60)$ & $0.5(2)$ & 61 & $0(1)$ & $22.89(6)$ & \cite{MacquartCosmicDM} \\
20190714A & ${\diamond\dagger}$ & 0.2365 & 0.999 & $g$ & $220(20)$ & $0.363(2)$ & 72 & $0.434(7)$ & $21.065(2)$ & \cite{Heintz2020} \\
20191001A & ${\star\diamond\dagger}$ & 0.234 & 0.997 & $g$ & $220(30)$ & $0.493(2)$ & 62 & $0.264(6)$ & $19.292(2)$ & \cite{BhandariFRB20191001A}\\
20191228A & ${\diamond\dagger}$ & 0.2432 & 0.987 & $I$ & $10(30)$ & $0.53(2)$ & 60 & $0.9(1)$ & $21.64(2)$ & \cite{Bhandari2022}\\
20200430A & ${\diamond\dagger}$ & 0.1608 & 0.998 & $g$ & $170(20)$ & $0.316(6)$ & 75 & $1.11(4)$ & $21.854(3)$ & \cite{Heintz2020}\\
20200906A & ${\star\diamond\dagger}$ & 0.3688 & 0.998 & $g$ & $180(40)$ & $0.274(2)$ & 78 & $0.297(7)$ & $20.960(2)$ & \cite{Bhandari2022}\\
20210117A & ${\dagger}$ & 0.2145 & 0.993 & $I$ & $470(20)$ & $0.50(8)$ & 61 & $6(3)$ & $22.3(3)$ & \cite{BhandariFRB20210117A} \\
20210320C & ${\star\diamond\dagger}$ & 0.2797 & 0.999 & $g$ & $60(30)$ & $0.684(3)$ & 48 & $0.919(9)$ & $20.493(2)$ & \cite{JamesH0} \\
20210807D & ${\star\diamond\dagger}$ & 0.12927 & 0.972 & $g$ & $-20(30)$ & $0.707(4)$ & 46 & $1.37(2)$ & $17.92(1)$ & \cite{JamesH0} \\
20211127I & ${\star\diamond\dagger}$ & 0.0469 & 0.998 & $g$ & $110(10)$ & $0.953(2)$ & 18 & $1.777(7)$ & $15.548(4)$ & \cite{JamesH0} \\
20211203C & ${\star\diamond\dagger}$ & 0.3439 & 0.998 & $R$ & $230(40)$ & $0.905(3)$ & 25 & $0.543(8)$ & $20.291(1)$ & \cite{GordonProspector} \\
20211212A & ${\star\diamond\dagger}$ & 0.0707 & 0.997 & $g$ & $70(10)$ & $0.822(2)$ & 35 & $0.783(4)$ & $17.159(2)$ & \cite{JamesH0} \\
20220105A & ${\star\diamond\dagger}$ & 0.2785 & 0.957 & $R$ & $280(30)$ & $0.0259(9)$ & 90 & $1.05(3)$ & $20.963(7)$ & \cite{GordonProspector} \\
20220501C & ${\diamond\dagger}$ & 0.381 & 0.984 & $I$ & $40(40)$ & $0.81(5)$ & 36 & $0.4(2)$ & $21.20(2)$ & \cite{ShannonICS} \\
20220725A & ${\star\diamond}$ & 0.1926 & 0.999 & $R$ & $50(30)$ & $0.548(2)$ & 58 & $1.188(0)$ & $17.886(3)$ & \cite{ShannonICS} \\
20220918A & ${\diamond\dagger}$ & 0.491 & 0.973 & $R$ & $140(50)$ & $0.62(6)$ & 53 & $1.0(4)$ & $23.52(2)$ & \cite{ShannonICS} \\
20221106A & ${\star\dagger}$ & 0.2044 & 0.995 & $R$ & $90(20)$ & $0.650(2)$ & 50 & $2.65(2)$ & $18.262(5)$ & \cite{ShannonICS} \\
20230708A & ${\dagger}$ & 0.105 & 0.997 & $R$ & $220(20)$ & $0.44(6)$ & 66 & $8(4)$ & $22.4(1)$ & \cite{ShannonICS}\\
20230902A & ${\diamond\dagger}$ & 0.3619 & 0.999 & $R$ & $50(40)$ & $0.39(1)$ & 69 & $1.37(8)$ & $21.509(3)$ & \cite{ShannonICS} \\
20231226A & ${\star\diamond}$ & 0.1569 & 0.996 & $R$ & $120(20)$ & $0.725(3)$ & 44 & $1.20(1)$ & $18.939(6)$ & \cite{ShannonICS} \\
20231230D & ${\diamond\dagger}$ & 0.506 & 0.992 & $R$ & $110(60)$ & $0.101(2)$ & 90 & $1.90(4)$ & $21.01(1)$ & \cite{GordonSpatial} \\
20240117B & ${\diamond\dagger}$ & 0.644 & 0.984 & $R$ & $220(70)$ & $0.508(9)$ & 61 & $0.06(8)$ & $22.334(5)$ & \cite{GordonSpatial} \\
20240201A & ${\star\diamond\dagger}$ & 0.042729 & 0.989 & $R$ & $260(10)$ & $0.1666(2)$ & 90 & $0.825(2)$ & $17.0146(8)$ & \cite{ShannonICS}\\
20240203C & ${\diamond\dagger}$ & 0.2426 & 0.991 & $R$ & $170(20)$ & $0.166(8)$ & 90 & $1.40(7)$ & $22.12(2)$ & \cite{GordonSpatial} \\
20240210A & ${\star\diamond\dagger}$ & 0.023686 & 0.954 & $R$ & $200(10)$ & $0.6051(5)$ & 54 & $0.992(2)$ & $14.8546(9)$ & \cite{ShannonICS} \\
20240304A & ${\star\diamond\dagger}$ & 0.2423 & 0.994 & $R$ & $320(30)$ & $0.638(5)$ & 51 & $1.45(3)$ & $20.884(7)$ & \cite{ShannonICS} \\
20240310A & ${\star\diamond\dagger}$ & 0.217 & 0.998 & $R$ & $340(20)$ & $0.220(2)$ & 84 & $0.79(1)$ & $20.199(1)$  & \cite{ShannonICS}\\
20240525A & ${\star\diamond\dagger}$ & 0.3266 & 0.998 & $R$ & $130(30)$ & $0.273(1)$ & 79 & $0.455(8)$ & $19.956(2)$ & \cite{GordonSpatial} \\
20240615B & ${\star\diamond\dagger}$ & 0.073 & 0.999 & $R$ & $69(9)$ & $0.938(1)$ & 20 & $1.172(3)$ & $16.015(2)$ & \cite{GordonSpatial} \\\hline
\end{tabular}
\end{table*}

%% file: tables/acsgc.tex
\begin{table*}[t]
    \centering
        \caption[Results of K-S and A-D tests on \galfit{} parameters, between the CRAFT host galaxy samples and the ACS-GC]{Results of K-S and A-D tests on \galfit{} parameters, between the CRAFT host galaxy samples and the ACS-GC. $N$ gives the size of the ACS-GC sample used for comparison.}
    \label{tab:population-tests:acsgc}
    \begin{tabular}{c|ccc|ccc|ccc}
        Quantity
        & \multicolumn{3}{c|}{Full sample}  
        & \multicolumn{3}{c|}{$\star:\ \mgalfit < 21\ $}
        & \multicolumn{3}{c}{$\diamond:\ n<2$}
        \\ \hline
        $N$ CRAFT
        & \multicolumn{3}{c|}{\nGALFITz}
        & \multicolumn{3}{c|}{\nGALFITmag}
        & \multicolumn{3}{c}{\nGALFITn}
        \\ \hline
        & $N$ & \pks{} & \pad{}
        & $N$ & \pks{} & \pad{}
        & $N$ & \pks{} & \pad{}      
        \\ \hline
        \ba{}
        &\NBAGALFITCRAFTvACSGC
        &\KSgalfitaxisratioCRAFTvACSGC
        &$\geq\ADgalfitaxisratioCRAFTvACSGC$
        
        &\NBAGALFITCRAFTmagvACSGCmag
        &\KSgalfitaxisratioCRAFTmagvACSGCmag
        &$\geq\ADgalfitaxisratioCRAFTmagvACSGCmag$
        
        &\NBAGALFITCRAFTnvACSGCn
        &\KSgalfitaxisratioCRAFTnvACSGCn
        &$\geq\ADgalfitaxisratioCRAFTnvACSGCn$
        \\
        $i$
        &\NIGALFITCRAFTvACSGC
        &\KSgalfitinclinationCRAFTvACSGC
        &$\geq\ADgalfitinclinationCRAFTvACSGC$
        
        &\NIGALFITCRAFTmagvACSGCmag
        &\KSgalfitinclinationCRAFTmagvACSGCmag
        &$\geq\ADgalfitinclinationCRAFTmagvACSGCmag$
        
        &\NIGALFITCRAFTnvACSGCn
        &\KSgalfitinclinationCRAFTnvACSGCn
        &$\geq\ADgalfitinclinationCRAFTnvACSGCn$
        \\
        $\cos(i)$
        &\NCOSIGALFITCRAFTvACSGC
        &\KSgalfitcosinclinationCRAFTvACSGC
        &$\geq\ADgalfitcosinclinationCRAFTvACSGC$
        
        &\NCOSIGALFITCRAFTmagvACSGCmag
        &\KSgalfitcosinclinationCRAFTmagvACSGCmag
        &$\geq\ADgalfitcosinclinationCRAFTmagvACSGCmag$
        
        &\NCOSIGALFITCRAFTnvACSGCn
        &\KSgalfitcosinclinationCRAFTnvACSGCn
        &$\geq\ADgalfitcosinclinationCRAFTnvACSGCn$
        
        
        
        
        
        
        
        
        
        

    \end{tabular}
\end{table*}

%% file: tables/_correlations.tex
\begin{table*}
\centering
\resizebox{\textwidth}{!}{%
\begin{tabular}{l|l|c|c|c|c|c|l}
$\mathbf{y}$               & $\mathbf{x}$   &$\mathbf{N}$ &$\mathbf{\rpearson}$&$\mathbf{\ppearson}$ &$\mathbf{\rspearman}$&$\mathbf{\pspearman}$& \textbf{Line of best fit}\\\hline

$\mathrm{DM^{Macquart}_{host}}$ & $q$ & 34 & \textbf{-0.4} & \textbf{0.02} & \textbf{-0.42} & \textbf{0.012} & $y = [(-193.8\pm65.6)\times(x-0.5) + (137.8\pm16.6))]\ \mathrm{pc\ cm^{-3}}$ \\
$\mathrm{DM^{Macquart\ \prime}_{host}}$ & $q$ & 34 & \textbf{-0.4} & \textbf{0.018} & \textbf{-0.44} & \textbf{0.01} & $y = [(-309.3\pm69.5)\times(x-0.5) + (155.9\pm17.4))]\ \mathrm{pc\ cm^{-3}}$  \\
$\mathrm{DM^{YMW16\prime}_{excess}}$ & $q$ & 34 & \textbf{-0.38} & \textbf{0.029} & \textbf{-0.41} & \textbf{0.016} & $y = [(-216.6\pm59.6)\times(x-0.5) + (159.7\pm15.0))]\ \mathrm{pc\ cm^{-3}}$ \\
$\mathrm{DM_{exgal}}$ & $q$ & 34 & \textbf{-0.37} & \textbf{0.029} & \textbf{-0.42} & \textbf{0.012} & $y = [(-470.1\pm54.8)\times(x-0.5) + (318.4\pm13.5))]\ \mathrm{pc\ cm^{-3}}$ \\
$\mathrm{DM_{FRB}}$ & $q$ & 34 & \textbf{-0.36} & \textbf{0.038} & \textbf{-0.42} & \textbf{0.012} & $y = [(-398.2\pm62.8)\times(x-0.5) + (401.5\pm15.7))]\ \mathrm{pc\ cm^{-3}}$ \\
$\mathrm{DM_{FRB}} - \mathrm{DM_{cosmic}}$ & $q$ & 34 & \textbf{-0.38} & \textbf{0.026} & \textbf{-0.44} & \textbf{0.01} & $y = [(-230.2\pm55.3)\times(x-0.5) + (206.7\pm13.8))]\ \mathrm{pc\ cm^{-3}}$ \\
$\mathrm{DM^{Macquart\ \prime}_{host}}$ & $i$ & 34 & \textbf{0.38} & \textbf{0.025} & \textbf{0.44} & \textbf{0.01} &  $y = (3.6\pm0.9)\ \mathrm{\frac{pc}{{}^{\circ}\,cm^{3}}}\times(x-59.7) + (150.3\pm17.9)\ \mathrm{pc\ cm^{-3}}$  \\
$\mathrm{DM^{Macquart\ \prime}_{host}}$ & $\cos(i)$ & 34 & \textbf{-0.4} & \textbf{0.019} & \textbf{-0.44} & \textbf{0.01} & $y = [(-270.2\pm60.5)\times(x-0.5) + (143.1\pm17.3))]\ \mathrm{pc\ cm^{-3}}$ \\
$\mathrm{DM^{Macquart\ \prime}_{host}}$ & $\rho$ & 34 & -0.03 & 0.887 & -0.04 & 0.84 & $y = (5790\pm18387)\ \mathrm{\frac{pc}{{}^{\prime\prime}\,cm^{3}}}\times(x-1.1) + (2078\pm6085)\ \mathrm{pc\ cm^{-3}}$ \\
$\mathrm{DM^{Macquart\ \prime}_{host}}$ & $\frac{\rho}{R_\mathrm{eff}}$ & 32 & -0.06 & 0.751 & 0.2 & 0.277 & $y = [(167.3\pm42.6)\times(x-0.7) + (145.4\pm19.2))]\ \mathrm{pc\ cm^{-3}}$ \\
$\mathrm{DM^{Macquart\ \prime}_{host}}$ & $\rho_\mathrm{proj}$ & 34 & -0.05 & 0.796 & -0.0 & 0.979 & $y = (33.1\pm9.7)\ \mathrm{\frac{pc}{kpc\,cm^{3}}}\times(x-3.3) + (153.3\pm21.3)\ \mathrm{pc\ cm^{-3}}$ \\
$\mathrm{DM^{Macquart\ \prime}_{host}}$ & $\rho_\mathrm{deproj}$ & 29 & 0.11 & 0.571 & 0.13 & 0.488 & $y = (115.6\pm26.4)\ \mathrm{\frac{pc}{{}^{\prime\prime}\,cm^{3}}}\times(x-1.4) + (69.3\pm20.1)\ \mathrm{pc\ cm^{-3}}$ \\
$\mathrm{DM^{Macquart\ \prime}_{host}}$ & $\frac{\rho_\mathrm{deproj}}{R_\mathrm{eff}}$ & 29 & 0.11 & 0.58 & \textbf{0.36} & 0.057 & $y = [(274.5\pm51.1)\times(x-1.2) + (183.4\pm19.3))]\ \mathrm{pc\ cm^{-3}}$ \\

$\mathrm{DM^{Macquart\ \prime}_{host}}$ & $g-I$ & 17 & -0.03 & 0.912 & -0.0 & 0.993 & $y = (-193.6\pm53.9)\ \mathrm{\frac{pc}{mag\,cm^{3}}}\times(x-1.2) + (32.9\pm15.3)\ \mathrm{pc\ cm^{-3}}$ \\
$\mathrm{DM^{Macquart\ \prime}_{host}}$ & $R-K$ & 14 & 0.06 & 0.842 & 0.1 & 0.737 & $y = (8.5\pm46.7)\ \mathrm{\frac{pc}{mag\,cm^{3}}}\times(x-1.6) + (215.4\pm35.3)\ \mathrm{pc\ cm^{-3}}$ \\
\end{tabular}
}
\caption{Correlations and linear fits for the maximal sample. 
We have \textbf{bolded} correlation scores greater than 0.3 and $p$-values less than 0.05 (our adopted threshold for statistical significance). The fit for \DMExcessRest{} versus $q$ is visualised in \autoref{fig:dmexcess-axisratio}, and for \DMExcessRest{} versus $i$ in \autoref{fig:dmexcess-inclination}.
}
\label{tab:correlations-all}
\end{table*}

\begin{table*}
\centering
\resizebox{\textwidth}{!}{%
\begin{tabular}{l|l|c|c|c|c|c|l}
$\mathbf{y}$               & $\mathbf{x}$   &$\mathbf{N}$ &$\mathbf{\rpearson}$&$\mathbf{\ppearson}$ &$\mathbf{\rspearman}$&$\mathbf{\pspearman}$& \textbf{Line of best fit}\\\hline
$\mathrm{DM^{Macquart}_{host}}$ & $q$ & 18 & \textbf{-0.54} & \textbf{0.022} & \textbf{-0.56} & \textbf{0.017} & $y = [(-202.7\pm76.0)\times(x-0.6) + (152.7\pm20.5))]\ \mathrm{pc\ cm^{-3}}$ \\
$\mathrm{DM^{Macquart\ \prime}_{host}}$ & $q$ & 18 & \textbf{-0.54} & \textbf{0.021} & \textbf{-0.54} & \textbf{0.02} & $y = [(-257.4\pm50.9)\times(x-0.6) + (157.9\pm13.8))]\ \mathrm{pc\ cm^{-3}}$ \\
$\mathrm{DM^{YMW16\prime}_{excess}}$ & $q$ & 18 & \textbf{-0.52} & \textbf{0.026} & \textbf{-0.58} & \textbf{0.011} & $y = [(-271.4\pm56.9)\times(x-0.6) + (158.4\pm15.4))]\ \mathrm{pc\ cm^{-3}}$ \\
$\mathrm{DM_{exgal}}$ & $q$ & 18 & \textbf{-0.56} & \textbf{0.016} & \textbf{-0.51} & \textbf{0.029} & $y = [(-499.7\pm66.0)\times(x-0.6) + (277.9\pm17.4))]\ \mathrm{pc\ cm^{-3}}$ \\
$\mathrm{DM_{FRB}}$ & $q$ & 18 & \textbf{-0.51} & \textbf{0.029} & \textbf{-0.54} & \textbf{0.02} & $y = [(-478.5\pm61.3)\times(x-0.6) + (357.7\pm16.0))]\ \mathrm{pc\ cm^{-3}}$ \\
$\mathrm{DM_{FRB}} - \mathrm{DM_{cosmic}}$ & $q$ & 18 & \textbf{-0.47} & \textbf{0.047} & \textbf{-0.54} & \textbf{0.02} & $y = [(-196.6\pm50.3)\times(x-0.6) + (204.8\pm13.4))]\ \mathrm{pc\ cm^{-3}}$ \\
$\mathrm{DM^{Macquart\ \prime}_{host}}$ & $i$ & 18 & \textbf{0.51} & \textbf{0.031} & \textbf{0.54} & \textbf{0.02} & $y = (3.0\pm0.7)\ \mathrm{\frac{pc}{{}^{\circ}\,cm^{3}}}\times(x-53.1) + (159.7\pm15.4)\ \mathrm{pc\ cm^{-3}}$ \\
$\mathrm{DM^{Macquart\ \prime}_{host}}$ & $\cos(i)$ & 18 & \textbf{-0.54} & \textbf{0.02} & \textbf{-0.54} & \textbf{0.02} & $y = [(-225.2\pm46.8)\times(x-0.6) + (150.2\pm14.5))]\ \mathrm{pc\ cm^{-3}}$ \\
$\mathrm{DM^{Macquart\ \prime}_{host}}$ & $\rho$ & 18 & -0.1 & 0.693 & -0.08 & 0.766 & $y = (-325.3\pm129.6)\ \mathrm{\frac{pc}{{}^{\prime\prime}\,cm^{3}}}\times(x-1.4) + (163.5\pm52.9)\ \mathrm{pc\ cm^{-3}}$ \\
$\mathrm{DM^{Macquart\ \prime}_{host}}$ & $\frac{\rho}{R_\mathrm{eff}}$ & 18 & \textbf{0.53} & \textbf{0.023} & \textbf{0.54} & \textbf{0.021} & Not well-fit by linear function\\
$\mathrm{DM^{Macquart\ \prime}_{host}}$ & $\rho_\mathrm{proj}$ & 18 & 0.09 & 0.715 & 0.16 & 0.521 & $y = (48.4\pm10.0)\ \mathrm{\frac{pc}{kpc\,cm^{3}}}\times(x-3.9) + (163.1\pm13.6)\ \mathrm{pc\ cm^{-3}}$ \\
$\mathrm{DM^{Macquart\ \prime}_{host}}$ & $\rho_\mathrm{deproj}$ & 16 & 0.04 & 0.896 & 0.05 & 0.854 & $y = (61.3\pm17.5)\ \mathrm{\frac{pc}{{}^{\prime\prime}\,cm^{3}}}\times(x-1.8) + (106.0\pm14.1)\ \mathrm{pc\ cm^{-3}}$ \\
$\mathrm{DM^{Macquart\ \prime}_{host}}$ & $\frac{\rho_\mathrm{deproj}}{R_\mathrm{eff}}$ & 16 & \textbf{0.57} & \textbf{0.021} & \textbf{0.77} & \textbf{0.0} & $y = [(201.3\pm54.0)\times(x-0.9) + (137.2\pm17.2))]\ \mathrm{pc\ cm^{-3}}$ \\

\end{tabular}}
\caption{Correlations and linear fits for the sample ($\star$) cut to $\mgalfit < 21$. See the caption of \autoref{tab:correlations-all} for further details. The fit for \DMExcessRest{} versus \ba{} is visualised in \autoref{fig:dmexcess-axisratio-mag}.
}
\label{tab:correlations-mag}
\end{table*}

\begin{table*}
\centering
\resizebox{\textwidth}{!}{%
\begin{tabular}{l|l|c|c|c|c|c|l}
$\mathbf{y}$               & $\mathbf{x}$   &$\mathbf{N}$ &$\mathbf{\rpearson}$&$\mathbf{\ppearson}$ &$\mathbf{\rspearman}$&$\mathbf{\pspearman}$& \textbf{Line of best fit}\\\hline
$\mathrm{DM^{Macquart}_{host}}$ & $q$ & 28 & \textbf{-0.42} & \textbf{0.024} & \textbf{-0.37} & \textbf{0.049} &  $y = [(-183.9\pm72.1)\times(x-0.5) + (137.0\pm18.9))]\ \mathrm{pc\ cm^{-3}}$ \\
$\mathrm{DM^{Macquart\ \prime}_{host}}$ & $q$ & 28 & \textbf{-0.43} & \textbf{0.022} & \textbf{-0.4} & \textbf{0.036} & $y = [(-310.7\pm76.4)\times(x-0.5) + (152.3\pm20.0))]\ \mathrm{pc\ cm^{-3}}$ \\
$\mathrm{DM^{YMW16\prime}_{excess}}$ & $q$ & 28 & \textbf{-0.4} & \textbf{0.035} & \textbf{-0.37} & \textbf{0.049} & $y = [(-217.0\pm64.4)\times(x-0.5) + (156.2\pm16.7))]\ \mathrm{pc\ cm^{-3}}$ \\
$\mathrm{DM_{exgal}}$ & $q$ & 28 & \textbf{-0.44} & \textbf{0.018} & \textbf{-0.46} & \textbf{0.015} & $y = [(-405.6\pm69.5)\times(x-0.5) + (320.9\pm18.5))]\ \mathrm{pc\ cm^{-3}}$ \\
$\mathrm{DM_{FRB}}$ & $q$ & 28 & \textbf{-0.42} & \textbf{0.026} & \textbf{-0.46} & \textbf{0.013} & $y = [(-398.9\pm69.2)\times(x-0.5) + (403.1\pm18.4))]\ \mathrm{pc\ cm^{-3}}$ \\
$\mathrm{DM_{FRB}} - \mathrm{DM_{cosmic}}$ & $q$ & 28 & \textbf{-0.4} & \textbf{0.034} & \textbf{-0.4} & \textbf{0.037} & $y = [(-228.0\pm59.8)\times(x-0.5) + (203.0\pm15.3))]\ \mathrm{pc\ cm^{-3}}$ \\
$\mathrm{DM^{Macquart\ \prime}_{host}}$ & $i$ & 28 & \textbf{0.41} & \textbf{0.032} & \textbf{0.4} & \textbf{0.037} & $y = (3.7\pm1.0)\ \mathrm{\frac{pc}{{}^{\circ}\,cm^{3}}}\times(x-60.9) + (151.8\pm20.7)\ \mathrm{pc\ cm^{-3}}$ \\
$\mathrm{DM^{Macquart\ \prime}_{host}}$ & $\cos(i)$ & 28 & \textbf{-0.43} & \textbf{0.022} & \textbf{-0.4} & \textbf{0.037} & $y = [(-271.4\pm66.4)\times(x-0.5) + (139.1\pm20.1))]\ \mathrm{pc\ cm^{-3}}$ \\
$\mathrm{DM^{Macquart\ \prime}_{host}}$ & $\rho$ & 28 & -0.02 & 0.938 & -0.02 & 0.934 & $y = (1520\pm1938)\ \mathrm{\frac{pc}{{}^{\prime\prime}\,cm^{3}}}\times(x-1.2) + (791\pm795)\ \mathrm{pc\ cm^{-3}}$ \\
$\mathrm{DM^{Macquart\ \prime}_{host}}$ & $\frac{\rho}{R_\mathrm{eff}}$ & 27 & -0.1 & 0.636 & 0.17 & 0.41 & $y = [(186.6\pm29.3)\times(x-0.7) + (124.1\pm11.9))]\ \mathrm{pc\ cm^{-3}}$ \\
$\mathrm{DM^{Macquart\ \prime}_{host}}$ & $\rho_\mathrm{proj}$ & 28 & -0.02 & 0.917 & 0.03 & 0.897 & $y = (63.7\pm15.7)\ \mathrm{\frac{pc}{kpc\,cm^{3}}}\times(x-3.0) + (135.2\pm21.2)\ \mathrm{pc\ cm^{-3}}$ \\
$\mathrm{DM^{Macquart\ \prime}_{host}}$ & $\rho_\mathrm{deproj}$ & 23 & 0.14 & 0.536 & 0.16 & 0.466 & $y = (147.2\pm55.5)\ \mathrm{\frac{pc}{{}^{\prime\prime}\,cm^{3}}}\times(x-1.4) + (142.4\pm40.6)\ \mathrm{pc\ cm^{-3}}$ \\
$\mathrm{DM^{Macquart\ \prime}_{host}}$ & $\frac{\rho_\mathrm{deproj}}{R_\mathrm{eff}}$ & 23 & 0.04 & 0.858 & 0.27 & 0.205 & $y = [(210.0\pm56.7)\times(x-1.2) + (196.1\pm17.7))]\ \mathrm{pc\ cm^{-3}}$ \\

\end{tabular}}
\caption{Correlations and linear fits for the sample ($\diamond$) cut to $n<2$. See the caption of \autoref{tab:correlations-all} for further details. The fit for \DMExcessRest{} versus \ba{} is visualised in \autoref{fig:dmexcess-axisratio-n}.
}
\label{tab:correlations-n}
\end{table*}

\begin{table*}
\centering
\resizebox{\textwidth}{!}{%
\begin{tabular}{l|l|c|c|c|c|c|l}
$\mathbf{y}$               & $\mathbf{x}$   &$\mathbf{N}$ &$\mathbf{\rpearson}$&$\mathbf{\ppearson}$ &$\mathbf{\rspearman}$&$\mathbf{\pspearman}$& \textbf{Line of best fit}\\\hline
$\mathrm{DM^{Macquart}_{host}}$ & $q$ & 31 & \textbf{-0.4} & \textbf{0.026} & \textbf{-0.43} & \textbf{0.016} & $y = [(-195.2\pm68.9)\times(x-0.5) + (138.3\pm18.1))]\ \mathrm{pc\ cm^{-3}}$ \\
$\mathrm{DM^{Macquart\ \prime}_{host}}$ & $q$ & 31 & \textbf{-0.4} & \textbf{0.025} & \textbf{-0.43} & \textbf{0.015} & $y = [(-314.4\pm73.1)\times(x-0.5) + (155.2\pm19.1))]\ \mathrm{pc\ cm^{-3}}$ \\
$\mathrm{DM^{YMW16\prime}_{excess}}$ & $q$ & 31 & \textbf{-0.37} & \textbf{0.038} & \textbf{-0.4} & \textbf{0.024} & $y = [(-288.2\pm76.8)\times(x-0.5) + (166.5\pm19.8))]\ \mathrm{pc\ cm^{-3}}$ \\
$\mathrm{DM_{exgal}}$ & $q$ & 31 & \textbf{-0.37} & \textbf{0.042} & \textbf{-0.38} & \textbf{0.033} & $y = [(-471.9\pm57.4)\times(x-0.5) + (321.8\pm14.8))]\ \mathrm{pc\ cm^{-3}}$ \\
$\mathrm{DM_{FRB}}$ & $q$ & 31 & \textbf{-0.35} & 0.053 & \textbf{-0.38} & \textbf{0.033} & $y = [(-395.0\pm65.8)\times(x-0.5) + (406.7\pm17.2))]\ \mathrm{pc\ cm^{-3}}$ \\
$\mathrm{DM_{FRB}} - \mathrm{DM_{cosmic}}$ & $q$ & 31 & \textbf{-0.38} & \textbf{0.033} & \textbf{-0.44} & \textbf{0.013} & $y = [(-233.8\pm57.6)\times(x-0.5) + (206.4\pm14.9))]\ \mathrm{pc\ cm^{-3}}$ \\
$\mathrm{DM^{Macquart\ \prime}_{host}}$ & $i$ & 31 & \textbf{0.39} & \textbf{0.032} & \textbf{0.43} & \textbf{0.015} & $y = (3.7\pm0.9)\ \mathrm{\frac{pc}{{}^{\circ}\,cm^{3}}}\times(x-61.5) + (156.8\pm19.8)\ \mathrm{pc\ cm^{-3}}$ \\
$\mathrm{DM^{Macquart\ \prime}_{host}}$ & $\cos(i)$ & 31 & \textbf{-0.4} & \textbf{0.026} & \textbf{-0.43} & \textbf{0.015} & $y = [(-274.7\pm63.8)\times(x-0.5) + (141.9\pm19.1))]\ \mathrm{pc\ cm^{-3}}$ \\
$\mathrm{DM^{Macquart\ \prime}_{host}}$ & $\rho$ & 31 & -0.04 & 0.839 & -0.06 & 0.729 & $y = (9439\pm49618)\ \mathrm{\frac{pc}{{}^{\prime\prime}\,cm^{3}}}\times(x-1.1) + (3113\pm15472)\ \mathrm{pc\ cm^{-3}}$ \\
$\mathrm{DM^{Macquart\ \prime}_{host}}$ & $\frac{\rho}{R_\mathrm{eff}}$ & 29 & -0.08 & 0.667 & 0.17 & 0.385 & $y = [(1780.7\pm2069.4)\times(x-0.6) + (166.1\pm139.7))]\ \mathrm{pc\ cm^{-3}}$ \\
$\mathrm{DM^{Macquart\ \prime}_{host}}$ & $\rho_\mathrm{proj}$ & 31 & -0.07 & 0.725 & -0.04 & 0.835 & $y = (68.9\pm20.8)\ \mathrm{\frac{pc}{kpc\,cm^{3}}}\times(x-3.0) + (124.9\pm27.2)\ \mathrm{pc\ cm^{-3}}$ \\
$\mathrm{DM^{Macquart\ \prime}_{host}}$ & $\rho_\mathrm{deproj}$ & 26 & 0.09 & 0.653 & 0.09 & 0.679 & $y = (165.6\pm61.6)\ \mathrm{\frac{pc}{{}^{\prime\prime}\,cm^{3}}}\times(x-1.4) + (117.8\pm38.6)\ \mathrm{pc\ cm^{-3}}$ \\
$\mathrm{DM^{Macquart\ \prime}_{host}}$ & $\frac{\rho_\mathrm{deproj}}{R_\mathrm{eff}}$ & 26 & 0.08 & 0.688 & \textbf{0.32} & 0.116 & $y = [(214.1\pm66.3)\times(x-1.2) + (198.0\pm18.4))]\ \mathrm{pc\ cm^{-3}}$ \\

\end{tabular}}
\caption{Correlations and linear fits for the sample with the poorly-fit (by \galfit{}) hosts of FRBs 20190608B, 20220725A and 20231226A removed from the sample ($\dagger$). See the caption of \autoref{tab:correlations-all} for further details. The fit for \DMExcessRest{} versus \ba{} is visualised in \autoref{fig:dmexcess-axisratio-bulge}}
\label{tab:correlations-no-bad-bulges}
\end{table*}

%% file: tables/overlap.tex
\begin{table*}[t]
\centering
\caption{Properties of the FRB host galaxies used by both \cite{Bhardwaj2024} and in this work.}
\label{tab:overlap}
\begin{tabular}{cc|ccc|ccc}
FRB         & B+24 imaging  & \multicolumn{3}{c|}{$i$ ($^\circ$)} & \multicolumn{3}{c}{$\DMHostRest{}$ (\dmunits{})} \\
            & source        & This work & B+24  & $\Delta$ & This work (\DMExcessRest{}) & B+24 & $\Delta$\\ \hline
20180924B   & DESI DR10     & 49        & 34    & -15   & 0     & 80    & 80\\
20190102C   & DESI DR10     & 63        & 50    & -13   & 10    & 94    & 84\\
20190608B   & DESI DR10     & 59        & 37    & -22   & 180   & 238   & 58\\
20190714A   & Pan-STARRS DR2& 72        & 51    & -21   & 273   & 335   & 62\\
20191001A   & DESI DR10     & 63        & 53    & -10   & 270   & 340   & 70\\
20200906A   & DESI DR10     & 79        & 63    & -16   & 241   & 309   & 68\\
20210320C   & Pan-STARRS DR2& 28        & 48    & 20    & 80    & 148   & 68 \\
20210807D   & Pan-STARRS DR2& 46        & 41    & -5    & -20   & 66    & 86\\
20211127I   & Pan-STARRS DR2& 18        & 21    & 3     & 120   & 171   & 51\\
20211203C   & DESI DR10     & 26        & 7     & -19   & 310   &395    &85\\
20211212A   & DESI DR10     & 36        & 33    & -3    & 70    & 127   &57 \\
\hline 
Mean        && 49        & 40    & -9    & 140   & 209   & 70\\
Median      && 49        & 41    & -13   & 120   & 171   & 68\\
$\sigma$    && 20        & 16    & 12    & 121   & 118   & 12\\

\end{tabular}
\end{table*}

%% file: residuals.tex

\begin{figure}
    \centering
    \includegraphics[trim={0.3cm 0.31cm 0.25cm 0.26cm}, width=0.49\textwidth, clip]{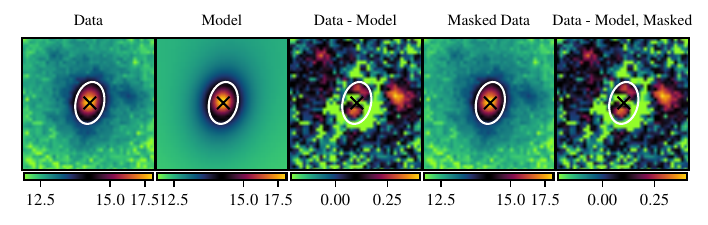}
    
    \caption[\galfit{} subtraction of HG\,20180924B]{\galfit{} model subtraction for FRB\,20180924B. In this and the figures below, the white ellipse demonstrates the fitted \Reff{} and axis ratio, with the long axis equal to $2\times\Reff{}$. The black cross marks the model centroid. Units of pixel brightness are arbitrary.
}
    \label{fig:residuals:FRB20180924B}

\end{figure}
                    
\begin{figure}
    \centering
    \includegraphics[trim={0.3cm 0.31cm 0.25cm 0.26cm}, width=0.49\textwidth, clip]{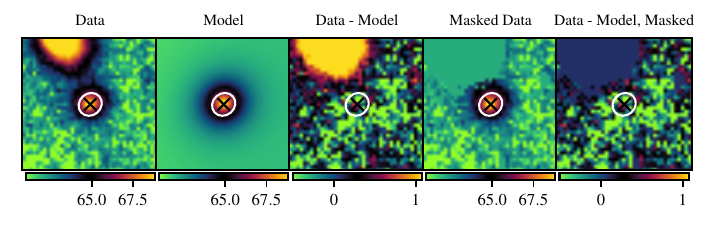}
    
    \caption[\galfit{} subtraction of HG\,20181112A]{\galfit{} model subtraction for FRB\,20181112A. 
}
    \label{fig:residuals:FRB20181112A}
    
\end{figure}
                    
\begin{figure}
    \centering
    \includegraphics[trim={0.3cm 0.31cm 0.25cm 0.26cm}, width=0.49\textwidth, clip]{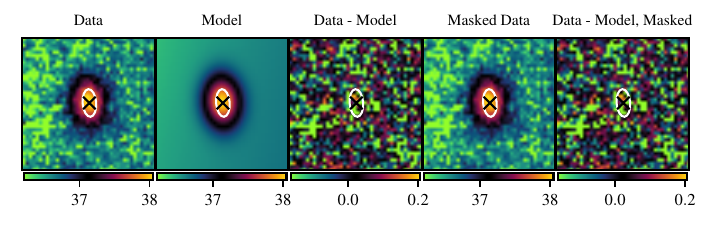}
    
    \caption[\galfit{} subtraction of HG\,20190102C]{\galfit{} model subtraction for FRB\,20190102C. 
}
    \label{fig:residuals:FRB20190102C}
    
\end{figure}
                    
\begin{figure}
    \centering
    \includegraphics[trim={0.3cm 0.31cm 0.25cm 0.26cm}, width=0.49\textwidth, clip]{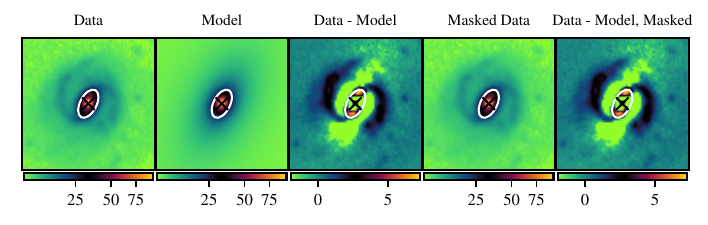}
    
    \caption[\galfit{} subtraction of HG\,20190608B]{\galfit{} model subtraction for FRB\,20190608B. 
}
    \label{fig:residuals:FRB20190608B}
    
\end{figure}
                    
\begin{figure}
    \centering
    \includegraphics[trim={0.3cm 0.31cm 0.25cm 0.26cm}, width=0.49\textwidth, clip]{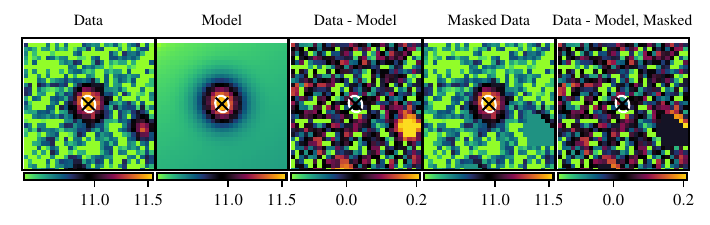}
    
    \caption[\galfit{} subtraction of HG\,20190611B]{\galfit{} model subtraction for FRB\,20190611B. 
}
    \label{fig:residuals:FRB20190611B}
    
\end{figure}
                    
\begin{figure}
    \centering
    \includegraphics[trim={0.3cm 0.31cm 0.25cm 0.26cm}, width=0.49\textwidth, clip]{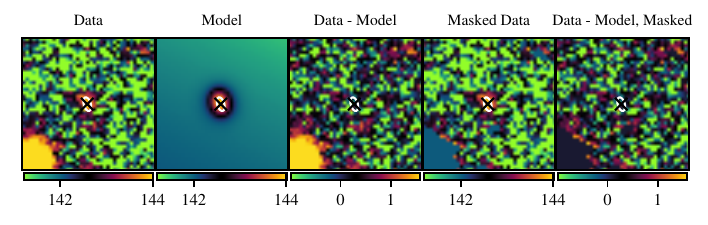}
    
    \caption[\galfit{} subtraction of HG\,20190711A]{\galfit{} model subtraction for FRB\,20190711A. 
}
    \label{fig:residuals:FRB20190711A}
    
\end{figure}
                    
\begin{figure}
    \centering
    \includegraphics[trim={0.3cm 0.31cm 0.25cm 0.26cm}, width=0.49\textwidth, clip]{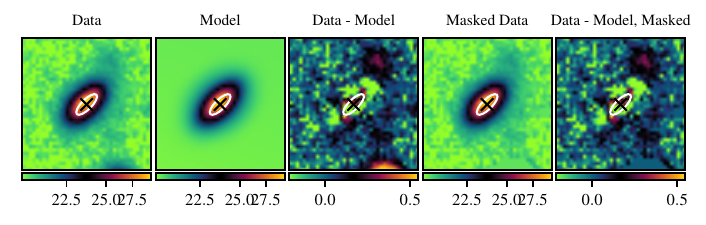}
    
    \caption[\galfit{} subtraction of HG\,20190714A]{\galfit{} model subtraction for FRB\,20190714A.
}
    \label{fig:residuals:FRB20190714A}
    
\end{figure}
                    
\begin{figure}
    \centering
    \includegraphics[trim={0.3cm 0.31cm 0.25cm 0.26cm}, width=0.49\textwidth, clip]{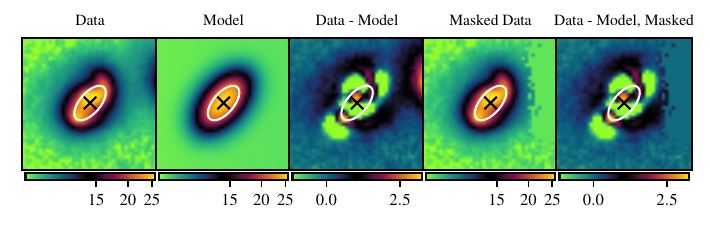}
    
    \caption[\galfit{} subtraction of HG\,20191001A]{\galfit{} model subtraction for FRB\,20191001A. 
}
    \label{fig:residuals:FRB20191001A}
    
\end{figure}
                    
\begin{figure}
    \centering
    \includegraphics[trim={0.3cm 0.31cm 0.25cm 0.26cm}, width=0.49\textwidth, clip]{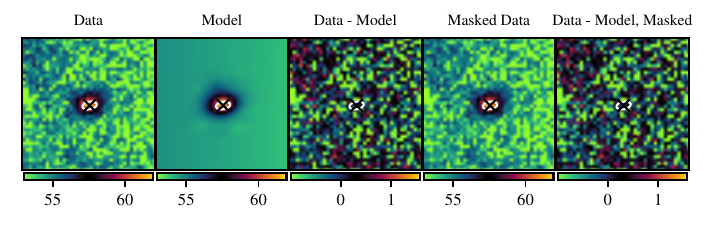}
    
    \caption[\galfit{} subtraction of HG\,20191228A]{\galfit{} model subtraction for FRB\,20191228A. 
}
    \label{fig:residuals:FRB20191228A}
    
\end{figure}
                    
\begin{figure}
    \centering
    \includegraphics[trim={0.3cm 0.31cm 0.25cm 0.26cm}, width=0.49\textwidth, clip]{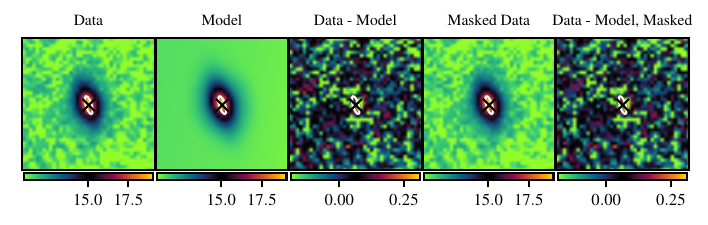}
    
    \caption[\galfit{} subtraction of HG\,20200430A]{\galfit{} model subtraction for FRB\,20200430A. 
}
    \label{fig:residuals:FRB20200430A}
\end{figure}
                    
    
    
                    
\begin{figure}
    \centering
    \includegraphics[trim={0.3cm 0.31cm 0.25cm 0.26cm}, width=0.49\textwidth, clip]{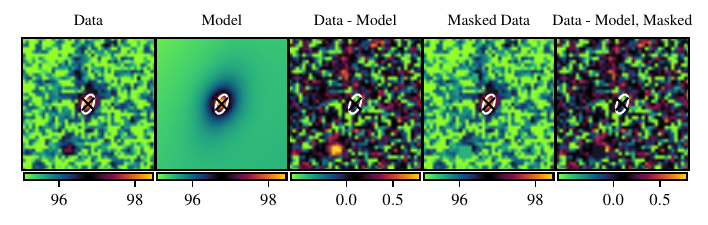}
    
    \caption[\galfit{} subtraction of HG\,20210117A]{\galfit{} model subtraction for FRB\,20210117A. 
}
    \label{fig:residuals:FRB20210117A}
    
\end{figure}
                    
\begin{figure}
    \centering
    \includegraphics[trim={0.3cm 0.31cm 0.25cm 0.26cm}, width=0.49\textwidth, clip]{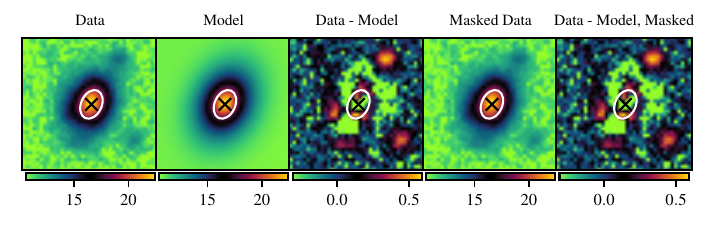}
    
    \caption[\galfit{} subtraction of HG\,20210320C]{\galfit{} model subtraction for FRB\,20210320C. 
}
    \label{fig:residuals:FRB20210320C}
    
\end{figure}
                    
    
    
                    
    
    
                    
\begin{figure}
    \centering
    \includegraphics[trim={0.3cm 0.31cm 0.25cm 0.26cm}, width=0.49\textwidth, clip]{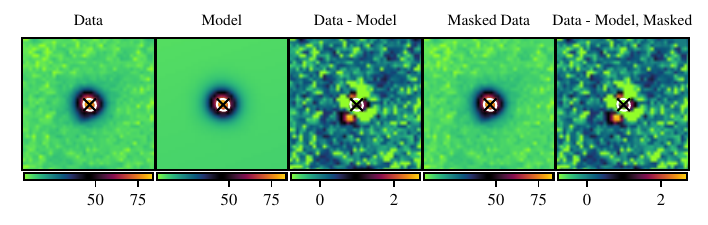}
    
    \caption[\galfit{} subtraction of HG\,20211203C]{\galfit{} model subtraction for FRB\,20211203C. 
}
    \label{fig:residuals:FRB20211203C}
    
\end{figure}
                    
\begin{figure}
    \centering
    \includegraphics[trim={0.3cm 0.31cm 0.25cm 0.26cm}, width=0.49\textwidth, clip]{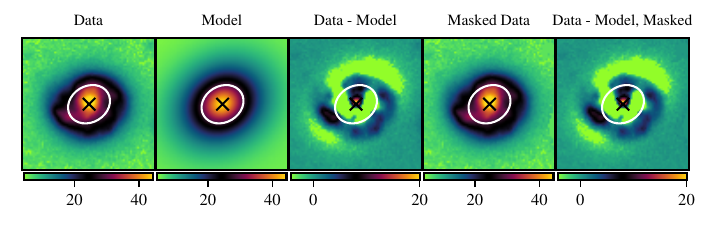}
    
    \caption[\galfit{} subtraction of HG\,20211212A]{\galfit{} model subtraction for FRB\,20211212A. 
}
    \label{fig:residuals:FRB20211212A}
    
\end{figure}
                    
\begin{figure}
    \centering
    \includegraphics[trim={0.3cm 0.31cm 0.25cm 0.26cm}, width=0.49\textwidth, clip]{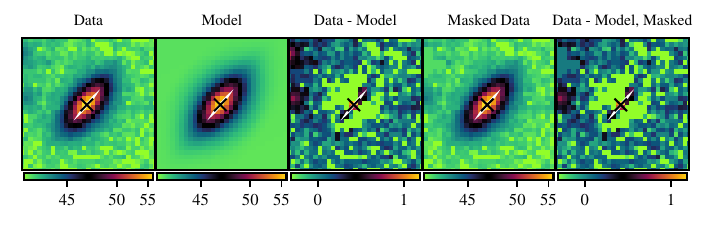}
    
    \caption[\galfit{} subtraction of HG\,20220105A]{\galfit{} model subtraction for FRB\,20220105A. 
}
    \label{fig:residuals:FRB20220105A}
    
\end{figure}
                    
\begin{figure}
    \centering
    \includegraphics[trim={0.3cm 0.31cm 0.25cm 0.26cm}, width=0.49\textwidth, clip]{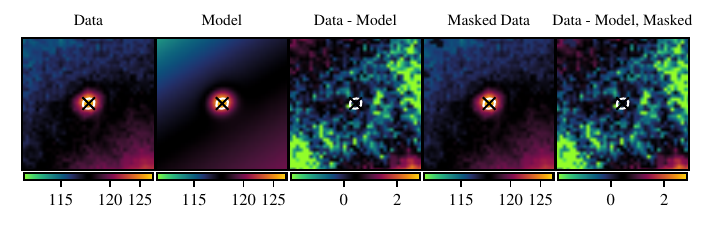}
    
    \caption[\galfit{} subtraction of HG\,20220501C]{\galfit{} model subtraction for FRB\,20220501C. 
}
    \label{fig:residuals:FRB20220501C}
    
\end{figure}
                    
    
                    
\begin{figure}
    \centering
    \includegraphics[trim={0.3cm 0.31cm 0.25cm 0.26cm}, width=0.49\textwidth, clip]{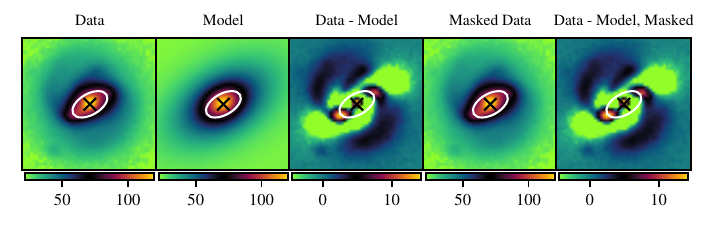}
    
    \caption[\galfit{} subtraction of HG\,20220725A]{\galfit{} model subtraction for FRB\,20220725A. 
}
    \label{fig:residuals:FRB20220725A}
\end{figure}
                    
\begin{figure}
    \centering
    \includegraphics[trim={0.3cm 0.31cm 0.25cm 0.26cm}, width=0.49\textwidth, clip]{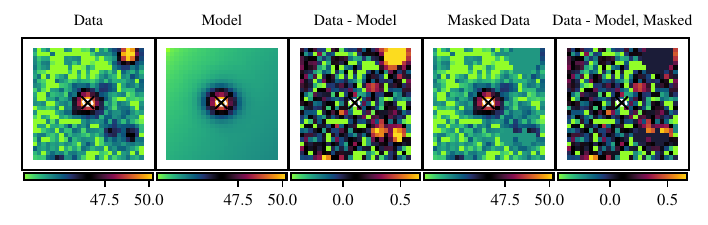}
    
    \caption[\galfit{} subtraction of HG\,20220918A]{\galfit{} model subtraction for FRB\,20220918A. 
}
    \label{fig:residuals:FRB20220918A}
    
\end{figure}
                    
\begin{figure}
    \centering
    \includegraphics[trim={0.3cm 0.31cm 0.25cm 0.26cm}, width=0.49\textwidth, clip]{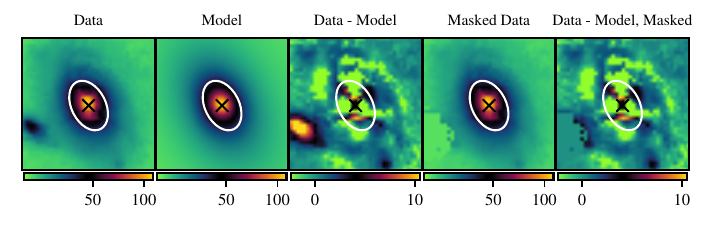}
    
    \caption[\galfit{} subtraction of HG\,20221106A]{\galfit{} model subtraction for FRB\,20221106A. 
}
    \label{fig:residuals:FRB20221106A}
    
\end{figure}

\begin{figure}
    \centering
    \includegraphics[trim={0.3cm 0.31cm 0.25cm 0.26cm}, width=0.49\textwidth, clip]{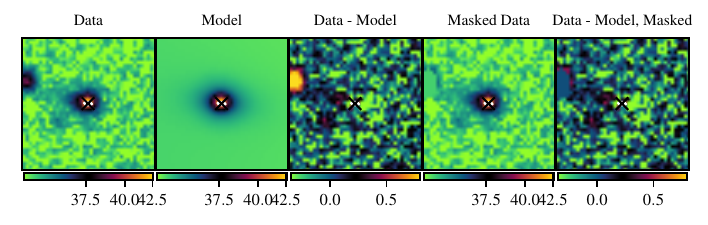}
    
    \caption[\galfit{} subtraction of HG\,20230708A]{\galfit{} model subtraction for FRB\,20230708A. 
}
    \label{fig:residuals:FRB20230708A}
    
\end{figure}
                    
\begin{figure}
    \centering
    \includegraphics[trim={0.3cm 0.31cm 0.25cm 0.26cm}, width=0.49\textwidth, clip]{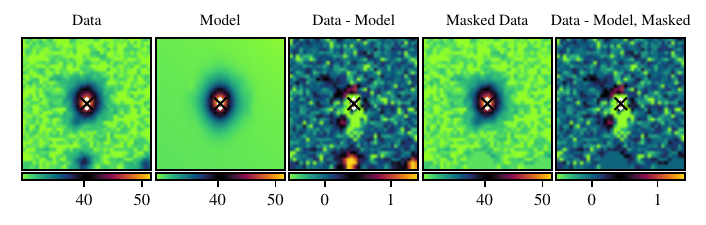}
    
    \caption[\galfit{} subtraction of HG\,20230902A]{\galfit{} model subtraction for FRB\,20230902A. 
}
    \label{fig:residuals:FRB20230902A}
    
\end{figure}
                    
    
    
                    
\begin{figure}
    \centering
    \includegraphics[trim={0.3cm 0.31cm 0.25cm 0.26cm}, width=0.49\textwidth, clip]{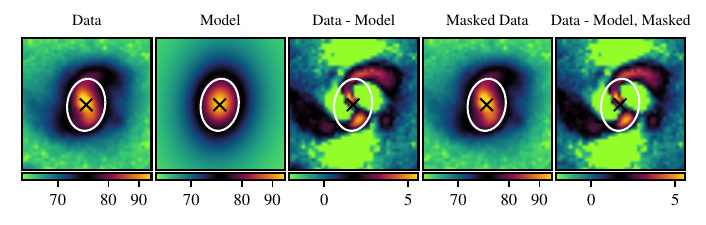}
    
    \caption[\galfit{} subtraction of HG\,20231226A]{\galfit{} model subtraction for FRB\,20231226A. 
}
    \label{fig:residuals:FRB20231226A}
    
\end{figure}
                    
\begin{figure}
    \centering
    \includegraphics[trim={0.3cm 0.31cm 0.25cm 0.26cm}, width=0.49\textwidth, clip]{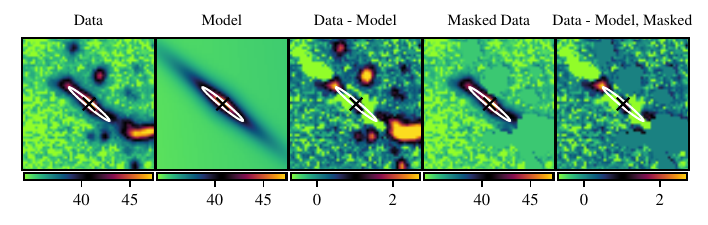}
    
    \caption[\galfit{} subtraction of HG\,20231230]{\galfit{} model subtraction for FRB\,20231230. 
}
    \label{fig:residuals:FRB20231230}
    
\end{figure}
                    
\begin{figure}
    \centering
    \includegraphics[trim={0.3cm 0.31cm 0.25cm 0.26cm}, width=0.49\textwidth, clip]{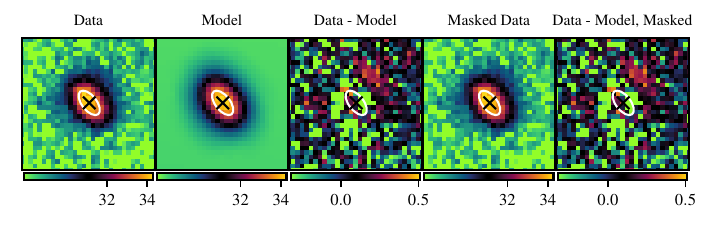}
    
    \caption[\galfit{} subtraction of HG\,20240117]{\galfit{} model subtraction for FRB\,20240117. 
}
    \label{fig:residuals:FRB20240117}
    
\end{figure}
                    
    
    
                    
\begin{figure}
    \centering
    \includegraphics[trim={0.3cm 0.31cm 0.25cm 0.26cm}, width=0.49\textwidth, clip]{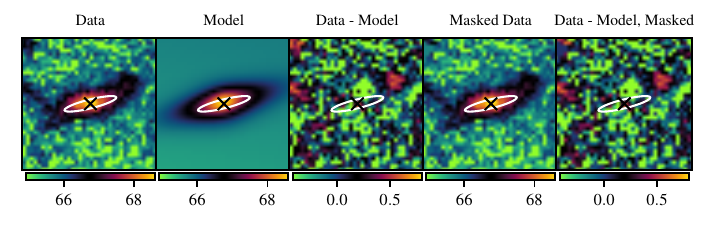}
    
    \caption[\galfit{} subtraction of HG\,20240203]{\galfit{} model subtraction for FRB\,20240203. 
}
    \label{fig:residuals:FRB20240203}
    
\end{figure}
                    
\begin{figure}
    \centering
    \includegraphics[trim={0.3cm 0.31cm 0.25cm 0.26cm}, width=0.49\textwidth, clip]{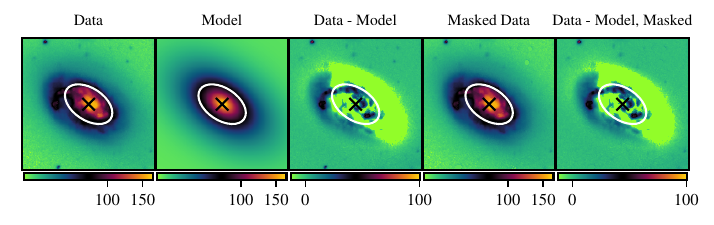}
    
    \caption[\galfit{} subtraction of HG\,20240210A]{\galfit{} model subtraction for FRB\,20240210A. 
}
    \label{fig:residuals:FRB20240210A}
    
\end{figure}
                    
\begin{figure}
    \centering
    \includegraphics[trim={0.3cm 0.31cm 0.25cm 0.26cm}, width=0.49\textwidth, clip]{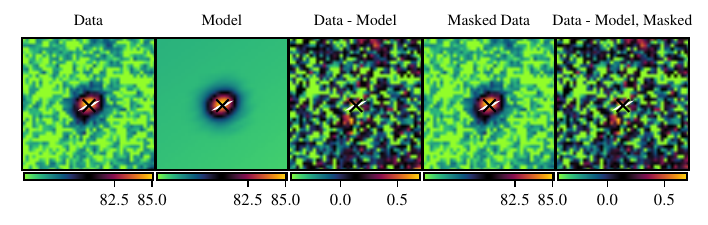}
    
    \caption[\galfit{} subtraction of HG\,20240210b]{\galfit{} model subtraction for FRB\,20240210b. 
}
    \label{fig:residuals:FRB20240210b}
    
\end{figure}
                    
\begin{figure}
    \centering
    \includegraphics[trim={0.3cm 0.31cm 0.25cm 0.26cm}, width=0.49\textwidth, clip]{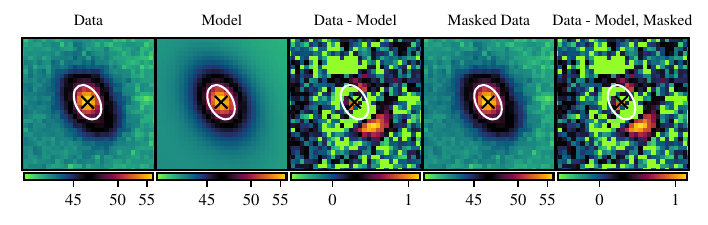}
    
    \caption[\galfit{} subtraction of HG\,20240304A]{\galfit{} model subtraction for FRB\,20240304A. 
}
    \label{fig:residuals:FRB20240304A}
    
\end{figure}
                    
\begin{figure}
    \centering
    \includegraphics[trim={0.3cm 0.31cm 0.25cm 0.26cm}, width=0.49\textwidth, clip]{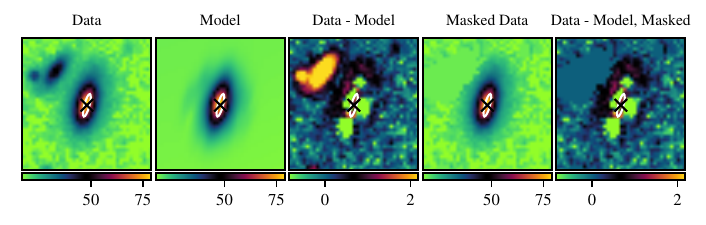}
    
    \caption[\galfit{} subtraction of HG\,20240310A]{\galfit{} model subtraction for FRB\,20240310A. 
}
    \label{fig:residuals:FRB20240310A}
    
\end{figure}
                    
\begin{figure}
    \centering
    \includegraphics[trim={0.3cm 0.31cm 0.25cm 0.26cm}, width=0.49\textwidth, clip]{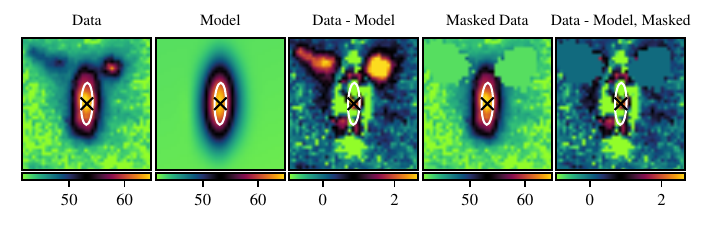}
    
    \caption[\galfit{} subtraction of HG\,20240525A]{\galfit{} model subtraction for FRB\,20240525A. 
}
    \label{fig:residuals:FRB20240525A}
    
\end{figure}
                    
\begin{figure}
    \centering
    \includegraphics[trim={0.3cm 0.31cm 0.25cm 0.26cm}, width=0.49\textwidth, clip]{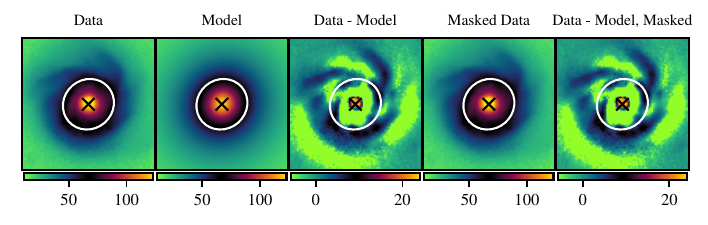}
    
    \caption[\galfit{} subtraction of HG\,20240615]{\galfit{} model subtraction for FRB\,20240615. 
}
    \label{fig:residuals:FRB20240615}
    
\end{figure}

%% file: references.bib
@ARTICLE{Pardo-Araujo2026,
       author = {{Pardo-Araujo}, Celsa and {Rea}, Nanda and {Ronchi}, Michele and {Graber}, Vanessa},
        title = "{Magnetar fraction in Core-Collapse Supernovae}",
      journal = {preprint (arXiv:2601.16159)},
         year = 2026,
        month = jan,
          eid = {arXiv:2601.16159},
          doi = {10.48550/arXiv.2601.16159},
archivePrefix = {arXiv},
       eprint = {2601.16159},
 primaryClass = {astro-ph.HE},
       adsurl = {https://ui.adsabs.harvard.edu/abs/2026arXiv260116159P}
}

@ARTICLE{Clancy_LSST,
       author = {{James}, C.~W. and {Smith}, B. and {Dage}, K. and {Chies Santos}, A.~L. and {Bannister}, K.~W. and {Caleb}, M. and {Crenshaw}, J.~F. and {Deller}, A.~T. and {Lee}, K.~G. and {Marnoch}, L. and {Rajwade}, K.~M. and {Ryder}, S.~D. and {Shannon}, R.~M. and {Stappers}, B. and {Zhang}, T.},
        title = "{Fast radio bursts in the era of the Vera C. Rubin Observatory's Legacy Survey of Space and Time}",
      journal = {Publications of the Astronomical Society of Australia},
         year = 2026,
        month = jun,
       volume = {43},
          eid = {e081},
        pages = {e081},
          doi = {10.1017/pasa.2026.10212},
archivePrefix = {arXiv},
       eprint = {2603.00371},
 primaryClass = {astro-ph.HE},
       adsurl = {https://ui.adsabs.harvard.edu/abs/2026PASA...43...81J}
}

@article{CHIMESGR1935,
    title = {{A bright millisecond-duration radio burst from a Galactic magnetar}},
    year = {2020},
    journal = {Nature},
    author = {{The CHIME/FRB Collaboration} and Andersen, B. C. and Bandura, K. M. and Bhardwaj, M. and Bij, A. and Boyce, M. M. and Boyle, P. J. and Brar, C. and Cassanelli, T. and Chawla, P. and Chen, T. and Cliche, J. F. and Cook, A. and Cubranic, D. and Curtin, A. P. and Denman, N. T. and Dobbs, M. and Dong, F. Q. and Fandino, M. and Fonseca, E. and Gaensler, B. M. and Giri, U. and Good, D. C. and Halpern, M. and Hill, A. S. and Hinshaw, G. F. and H{\"{o}}fer, C. and Josephy, A. and Kania, J. W. and Kaspi, V. M. and Landecker, T. L. and Leung, C. and Li, D. Z. and Lin, H. H. and Masui, K. W. and Mckinven, R. and Mena-Parra, J. and Merryfield, M. and Meyers, B. W. and Michilli, D. and Milutinovic, N. and Mirhosseini, A. and M{\"{u}}nchmeyer, M. and Naidu, A. and Newburgh, L. B. and Ng, C. and Patel, C. and Pen, U. L. and Pinsonneault-Marotte, T. and Pleunis, Z. and Quine, B. M. and Rafiei-Ravandi, M. and Rahman, M. and Ransom, S. M. and Renard, A. and Sanghavi, P. and Scholz, P. and Shaw, J. R. and Shin, K. and Siegel, S. R. and Singh, S. and Smegal, R. J. and Smith, K. M. and Stairs, I. H. and Tan, C. M. and Tendulkar, S. P. and Tretyakov, I. and Vanderlinde, K. and Wang, H. and Wulf, D. and Zwaniga, A. V.},
    number = {7832},
    month = {11},
    pages = {54--58},
    volume = {587},
    publisher = {Nature Research},
    url = {https://doi.org/10.1038/s41586-020-2863-y},
    doi = {10.1038/s41586-020-2863-y},
    issn = {14764687},
    pmid = {33149292},
    arxivId = {2005.10324}
}

@article{MacquartCosmicDM,
    title = {{A census of baryons in the Universe from localized fast radio bursts}},
    year = {2020},
    journal = {Nature},
    author = {Macquart, J. P. and Prochaska, J. X. and McQuinn, M. and Bannister, K. W. and Bhandari, S. and Day, C. K. and Deller, A. T. and Ekers, R. D. and James, C. W. and Marnoch, L. and Os{\l}owski, S. and Phillips, C. and Ryder, S. D. and Scott, D. R. and Shannon, R. M. and Tejos, N.},
    number = {7809},
    month = {5},
    pages = {391--395},
    volume = {581},
    publisher = {Nature Research},
    url = {https://doi.org/10.1038/s41586-020-2300-2},
    doi = {10.1038/s41586-020-2300-2},
    issn = {14764687},
    pmid = {32461651},
    arxivId = {2005.13161}
}

@article{MargalitMetzger2018,
    title = {{A Concordance Picture of FRB 121102 as a Flaring Magnetar Embedded in a Magnetized Ion–Electron Wind Nebula}},
    year = {2018},
    journal = {The Astrophysical Journal Letters},
    author = {Margalit, Ben and Metzger, Brian D.},
    number = {1},
    month = {11},
    pages = {L4},
    volume = {868},
    publisher = {IOP Publishing},
    url = {https://doi.org/10.3847/2041-8213/aaedad},
    doi = {10.3847/2041-8213/AAEDAD},
    issn = {2041-8205},
    arxivId = {1808.09969}
}

@article{BochenekSGR1935,
    title = {{A fast radio burst associated with a Galactic magnetar}},
    year = {2020},
    journal = {Nature},
    author = {Bochenek, C. D. and Ravi, V. and Belov, K. V. and Hallinan, G. and Kocz, J. and Kulkarni, S. R. and McKenna, D. L.},
    number = {7832},
    month = {11},
    pages = {59--62},
    volume = {587},
    publisher = {Nature Research},
    url = {https://doi.org/10.1038/s41586-020-2872-x},
    doi = {10.1038/s41586-020-2872-x},
    issn = {14764687},
    pmid = {33149288},
    arxivId = {2005.10828}
}

@article{CalebFRB20240304B,
    title = {{A fast radio burst from the first 3 billion years of the Universe}},
    year = {2025},
    journal = {preprint (arXiv:2508.01648)},
    author = {Caleb, Manisha and Nanayakkara, Themiya and Stappers, Benjamin and Pastor-Marazuela, Inés and Khrykin, Ilya S. and Glazebrook, Karl and Tejos, Nicolas and Prochaska, J. Xavier and Rajwade, Kaustubh and Mas-Ribas, Lluis and Driessen, Laura N. and Fong, Wen-fai and Gordon, Alexa C. and Hoffmann, Jordan and James, Clancy W. and Jankowski, Fabian and Kahinga, Lordrick and Kramer, Michael and Simha, Sunil and Barr, Ewan D. and Bezuidenhout, Mechiel Christiaan and Deng, Xihan and Lin, Zeren and Marnoch, Lachlan and Martin, Christopher D. and Nugent, Anya and Shaji, Kavya and Tian, Jun},
    month = {8},
    url = {https://arxiv.org/abs/2508.01648},
    arxivId = {2508.01648}
}

@article{GordonFRB20220610A,
    title = {{A Fast Radio Burst in a Compact Galaxy Group at z ~ 1}},
    year = {2024},
    journal = {The Astrophysical Journal Letters},
    author = {Gordon, Alexa C. and Fong, Wen-fai and Simha, Sunil and Dong, Yuxin and Kilpatrick, Charles D. and Deller, Adam T. and Ryder, Stuart D. and Eftekhari, Tarraneh and Glowacki, Marcin and Marnoch, Lachlan and Muller, August R. and Nugent, Anya E. and Palmese, Antonella and Prochaska, J. Xavier and Rafelski, Marc and Shannon, Ryan M. and Tejos, Nicolas},
    number = {2},
    month = {3},
    pages = {L34},
    volume = {963},
    publisher = {IOP Publishing},
    url = {https://doi.org/10.3847/2041-8213/ad2773},
    doi = {10.3847/2041-8213/AD2773},
    issn = {2041-8205},
    arxivId = {2311.10815}
}

@article{RyderFRB20220610A,
    title = {{A luminous fast radio burst that probes the Universe at redshift 1}},
    year = {2023},
    journal = {Science},
    author = {Ryder, S. D. and Bannister, K. W. and Bhandari, S. and Deller, A. T. and Ekers, R. D. and Glowacki, M. and Gordon, A. C. and Gourdji, K. and James, C. W. and Kilpatrick, C. D. and Lu, W. and Marnoch, L. and Moss, V. A. and Prochaska, J. X. and Qiu, H. and Sadler, E. M. and Simha, S. and Sammons, M. W. and Scott, D. R. and Tejos, N. and Shannon, R. M.},
    number = {6668},
    month = {10},
    pages = {294--299},
    volume = {382},
    publisher = {American Association for the Advancement of Science},
    url = {https://doi.org/10.1126/science.adf2678},
    doi = {10.1126/SCIENCE.ADF2678},
    issn = {0036-8075}
}

@article{JamesH0,
    title = {{A measurement of Hubble’s Constant using Fast Radio Bursts}},
    year = {2022},
    journal = {Monthly Notices of the Royal Astronomical Society},
    author = {James, C W and Ghosh, E M and Prochaska, J X and Bannister, K W and Bhandari, S and Day, C K and Deller, A T and Glowacki, M and Gordon, A C and Heintz, K E and Marnoch, L and Ryder, S D and Scott, D R and Shannon, R M and Tejos, N},
    number = {4},
    month = {10},
    pages = {4862--4881},
    volume = {516},
    publisher = {Oxford Academic},
    url = {https://doi.org/10.1093/mnras/stac2524},
    doi = {10.1093/MNRAS/STAC2524},
    issn = {0035-8711},
    arxivId = {2208.00819}
}

@article{YMW16,
    title = {{A New Electron-density Model for Estimation of Pulsar and FRB Distances}},
    year = {2017},
    journal = {The Astrophysical Journal},
    author = {Yao, J. M. and Manchester, R. N. and Wang, N.},
    number = {1},
    month = {1},
    pages = {29},
    volume = {835},
    publisher = {IOP Publishing},
    url = {https://doi.org/10.3847/1538-4357/835/1/29},
    doi = {10.3847/1538-4357/835/1/29},
    issn = {1538-4357}
}

@article{BhandariFRB20210117A,
    title = {{A Nonrepeating Fast Radio Burst in a Dwarf Host Galaxy}},
    year = {2023},
    journal = {The Astrophysical Journal},
    author = {Bhandari, Shivani and Gordon, Alexa C. and Scott, Danica R. and Marnoch, Lachlan and Sridhar, Navin and Kumar, Pravir and James, Clancy W. and Qiu, Hao and Bannister, Keith W. and Deller, Adam T. and Eftekhari, Tarraneh and Fong, Wen-fai and Glowacki, Marcin and Prochaska, J. Xavier and Ryder, Stuart D. and Shannon, Ryan M. and Simha, Sunil},
    number = {1},
    month = {5},
    pages = {67},
    volume = {948},
    publisher = {IOP Publishing},
    url = {https://doi.org/10.3847/1538-4357/acc178},
    doi = {10.3847/1538-4357/ACC178},
    issn = {0004-637X}
}

@article{GlobularCluster,
    title = {{A repeating fast radio burst source in a globular cluster}},
    year = {2022},
    journal = {Nature},
    author = {Kirsten, F. and Marcote, B. and Nimmo, K. and Hessels, J. W.T. and Bhardwaj, M. and Tendulkar, S. P. and Keimpema, A. and Yang, J. and Snelders, M. P. and Scholz, P. and Pearlman, A. B. and Law, C. J. and Peters, W. M. and Giroletti, M. and Paragi, Z. and Bassa, C. and Hewitt, D. M. and Bach, U. and Bezrukovs, V. and Burgay, M. and Buttaccio, S. T. and Conway, J. E. and Corongiu, A. and Feiler, R. and Forss{\'{e}}n, O. and Gawro{\'{n}}ski, M. P. and Karuppusamy, R. and Kharinov, M. A. and Lindqvist, M. and Maccaferri, G. and Melnikov, A. and Ould-Boukattine, O. S. and Possenti, A. and Surcis, G. and Wang, N. and Yuan, J. and Aggarwal, K. and Anna-Thomas, R. and Bower, G. C. and Blaauw, R. and Burke-Spolaor, S. and Cassanelli, T. and Clarke, T. E. and Fonseca, E. and Gaensler, B. M. and Gopinath, A. and Kaspi, V. M. and Kassim, N. and Lazio, T. J.W. and Leung, C. and Li, D. Z. and Lin, H. H. and Masui, K. W. and Mckinven, R. and Michilli, D. and Mikhailov, A. G. and Ng, C. and Orbidans, A. and Pen, U. L. and Petroff, E. and Rahman, M. and Ransom, S. M. and Shin, K. and Smith, K. M. and Stairs, I. H. and Vlemmings, W.},
    number = {7898},
    month = {2},
    pages = {585--589},
    volume = {602},
    publisher = {Nature Publishing Group},
    url = {https://doi.org/10.1038/s41586-021-04354-w},
    doi = {10.1038/s41586-021-04354-w},
    issn = {1476-4687},
    pmid = {35197615},
    arxivId = {2105.11445}
}

@article{BannisterFRB20180924B,
    title = {{A single fast radio burst localized to a massive galaxy at cosmological distance}},
    year = {2019},
    journal = {Science},
    author = {Bannister, K W and Deller, A T and Phillips, C and Macquart, J-P and Prochaska, J X and Tejos, N and Ryder, S D and Sadler, E M and Shannon, R M and Simha, S and Day, C K and McQuinn, M and North-Hickey, F O and Bhandari, S and Arcus, W R and Bennert, V N and Burchett, J and Bouwhuis, M and Dodson, R and Ekers, R D and Farah, W and Flynn, C and James, C W and Kerr, M and Lenc, E and Mahony, E K and O'Meara, J and Os{\l}owski, S and Qiu, H and Treu, T and U, V and Bateman, T J and Bock, D C-J and Bolton, R J and Brown, A and Bunton, J D and Chippendale, A P and Cooray, F R and Cornwell, T and Gupta, N and Hayman, D B and Kesteven, M and Koribalski, B S and MacLeod, A and McClure-Griffiths, N M and Neuhold, S and Norris, R P and Pilawa, M A and Qiao, R-Y and Reynolds, J and Roxby, D N and Shimwell, T W and Voronkov, M A and Wilson, C D},
    number = {6453},
    month = {8},
    pages = {565--570},
    volume = {365},
    publisher = {American Association for the Advancement of Science},
    url = {https://doi.org/10.1126/science.aaw5903},
    doi = {10.1126/science.aaw5903},
    issn = {1095-9203},
    pmid = {31249136}
}

@article{Mas-Ribas2026,
    title = {{A {$\tau$}–DM Relation for Fast Radio Burst Hosts?}},
    year = {2026},
    journal = {The Astrophysical Journal},
    author = {Mas-Ribas, Lluis and James, Clancy W.},
    number = {1},
    month = {2},
    pages = {1},
    volume = {998},
    publisher = {IOP Publishing},
    url = {https://iopscience.iop.org/article/10.3847/1538-4357/ae36a9},
    doi = {10.3847/1538-4357/AE36A9},
    issn = {0004-637X}
}

@article{Cook2023,
    title = {{An FRB Sent Me a DM: Constraining the Electron Column of the Milky Way Halo with Fast Radio Burst Dispersion Measures from CHIME/FRB}},
    year = {2023},
    journal = {The Astrophysical Journal},
    author = {Cook, Amanda M. and Bhardwaj, Mohit and Gaensler, B. M. and Scholz, Paul and Eadie, Gwendolyn M. and Hill, Alex S. and Kaspi, Victoria M. and Masui, Kiyoshi W. and Curtin, Alice P. and Dong, Fengqiu Adam and Fonseca, Emmanuel and Herrera-Martin, Antonio and Kaczmarek, Jane and Lanman, Adam E. and Lazda, Mattias and Leung, Calvin and Meyers, Bradley W. and Michilli, Daniele and Pandhi, Ayush and Pearlman, Aaron B. and Pleunis, Ziggy and Ransom, Scott and Rahman, Mubdi and Sand, Ketan R. and Shin, Kaitlyn and Smith, Kendrick and Stairs, Ingrid and Stenning, David C.},
    number = {2},
    month = {3},
    pages = {58},
    volume = {946},
    publisher = {IOP Publishing},
    url = {https://doi.org/10.3847/1538-4357/ACBBD0},
    doi = {10.3847/1538-4357/ACBBD0},
    issn = {0004-637X},
    arxivId = {2301.03502}
}

@article{GlowackiScattering,
    title = {{An investigation into correlations between FRB and host galaxy properties}},
    year = {2025},
    journal = {Publications of the Astronomical Society of Australia},
    author = {Glowacki, Marcin and Bera, Apurba and James, Clancy and Patterson, Jasper and Deller, Adam T. and Gordon, Alexa and Marnoch, Lachlan and Muller, August and Prochaska, Xavier and Ryder, Stuart and Shannon, Ryan M. and Tejos, Nicolas and Mannings, Alexandra G.},
    month = {11},
    pages = {e157},
    volume = {42},
    publisher = {Cambridge University Press},
    url = {https://doi.org/10.1017/PASA.2025.10122},
    doi = {10.1017/PASA.2025.10122},
    issn = {1323-3580},
    arxivId = {2506.23403}
}

@article{ccdproc,
    title = {{astropy/ccdproc: v1.3.0.post1}},
    year = {2017},
    journal = {Zenodo},
    author = {Craig, Matt and Crawford, Steve and Seifert, Michael and Robitaille, Thomas and Sipocz, Brigitta and Walawender, Josh and Vin{\'{i}}cius, Zé and Ninan, Joe Philip and Droettboom, Michael and Youn, Jiyong and Tollerud, Erik and Bray, Erik and {walkerna22} and Janga, VSN Reddy and {stottsco} and G{\"{u}}nther, Hans Moritz and Rol, Evert and Bach, Yoonsoo P. and Bradley, Larry and Deil, Christoph and Price-Whelan, Adrian and Barbary, Kyle and Horton, Anthony and Schoenell, William and {Nathan} and Gasdia, Forrest and Nelson, Stefan and Streicher, Ole},
    month = {12},
    url = {https://zenodo.org/record/1043589},
    doi = {10.5281/ZENODO.1069648}
}

@article{ADtest,
    title = {{Asymptotic Theory of Certain "Goodness of Fit" Criteria Based on Stochastic Processes}},
    year = {1952},
    journal = {The Annals of Mathematical Statistics},
    author = {Anderson, T. W. and Darling, D. A.},
    number = {2},
    month = {6},
    pages = {193--212},
    volume = {23},
    publisher = {Institute of Mathematical Statistics},
    url = {https://doi.org/10.1214/aoms/1177729437},
    doi = {10.1214/AOMS/1177729437},
    issn = {0003-4851}
}

@book{SersicProfile2,
    title = {{Atlas de Galaxias Australes}},
    year = {1968},
    author = {S{\'{e}}rsic, Jose Luis},
    url = {https://ui.adsabs.harvard.edu/abs/1968adga.book.....S}
}

@article{ASKAP,
    title = {{Australian square kilometre array pathfinder: I. system description}},
    year = {2021},
    journal = {Publications of the Astronomical Society of Australia},
    author = {Hotan, A. W. and Bunton, J. D. and Chippendale, A. P. and Whiting, M. and Tuthill, J. and Moss, V. A. and McConnell, D. and Amy, S. W. and Huynh, M. T. and Allison, J. R. and Anderson, C. S. and Bannister, K. W. and Bastholm, E. and Beresford, R. and Bock, D. C.-J. and Bolton, R. and Chapman, J. M. and Chow, K. and Collier, J. D. and Cooray, F. R. and Cornwell, T. J. and Diamond, P. J. and Edwards, P. G. and Feain, I. J. and Franzen, T. M. O. and George, D. and Gupta, N. and Hampson, G. A. and Harvey-Smith, L. and Hayman, D. B. and Heywood, I. and Jacka, C. and Jackson, C. A. and Jackson, S. and Jeganathan, K. and Johnston, S. and Kesteven, M. and Kleiner, D. and Koribalski, B. S. and Lee-Waddell, K. and Lenc, E. and Lensson, E. S. and Mackay, S. and Mahony, E. K. and McClure-Griffiths, N. M. and McConigley, R. and Mirtschin, P. and Ng, A. K. and Norris, R. P. and Pearce, S. E. and Phillips, C. and Pilawa, M. A. and Raja, W. and Reynolds, J. E. and Roberts, P. and Roxby, D. N. and Sadler, E. M. and Shields, M. and Schinckel, A. E. T. and Serra, P. and Shaw, R. D. and Sweetnam, T. and Troup, E. R. and Tzioumis, A. and Voronkov, M. A. and Westmeier, T.},
    pages = {e009},
    volume = {38},
    publisher = {Cambridge University Press},
    url = {https://doi.org/10.1017/PASA.2021.1},
    doi = {10.1017/PASA.2021.1},
    issn = {1323-3580},
    arxivId = {2102.01870}
}

@article{ESOReflex,
    title = {{Automated data reduction workflows for astronomy - The ESO Reflex environment}},
    year = {2013},
    journal = {Astronomy {\&} Astrophysics},
    author = {Freudling, W. and Romaniello, M. and Bramich, D. M. and Ballester, P. and Forchi, V. and Garc{\'{i}}a-Dabl{\'{o}}, C. E. and Moehler, S. and Neeser, M. J.},
    month = {11},
    pages = {A96},
    volume = {559},
    publisher = {EDP Sciences},
    url = {https://doi.org/10.1051/0004-6361/201322494},
    doi = {10.1051/0004-6361/201322494},
    issn = {0004-6361}
}

@inproceedings{PSFEx,
    title = {{Automated Morphometry with SExtractor and PSFEx}},
    year = {2011},
    booktitle = {Astronomical Data Analysis Software and Systems XX},
    author = {Bertin, E.},
    month = {7},
    pages = {435},
    volume = {442},
    url = {http://aspbooks.org/custom/publications/paper/442-0435.html}
}

@article{AutoProf,
    title = {{AutoProf -- I. An automated non-parametric light profile pipeline for modern galaxy surveys}},
    year = {2021},
    journal = {Monthly Notices of the Royal Astronomical Society},
    author = {Stone, Connor and Arora, Nikhil and Courteau, Stéphane and Cuillandre, Jean-Charles},
    number = {2},
    month = {6},
    pages = {1870--1887},
    volume = {508},
    publisher = {Oxford University Press},
    url = {http://doi.org/10.1093/mnras/stab2709},
    doi = {10.1093/mnras/stab2709},
    arxivId = {2106.13809v2}
}

@article{Bhandari2022,
    title = {{Characterizing the Fast Radio Burst Host Galaxy Population and its Connection to Transients in the Local and Extragalactic Universe}},
    year = {2022},
    journal = {The Astronomical Journal},
    author = {Bhandari, Shivani and Heintz, Kasper E and Aggarwal, Kshitij and Marnoch, Lachlan and Day, Cherie K and Sydnor, Jessica and Burke-Spolaor, Sarah and Law, Casey J and Prochaska, J Xavier and Tejos, Nicolas and Bannister, Keith W and Butler, Bryan J and Deller, Adam T and Ekers, R D and Flynn, Chris and Fong, Wen-Fai and James, Clancy W and Joseph, T and Lazio, W and Luo, Rui and Mahony, Elizabeth K and Ryder, Stuart D and Sadler, Elaine M and Shannon, Ryan M and Han, Jinlin and Lee, Kejia and Zhang, Bing},
    number = {2},
    month = {1},
    pages = {69},
    volume = {163},
    publisher = {IOP Publishing},
    url = {https://doi.org/10.3847/1538-3881/ac3aec},
    doi = {10.3847/1538-3881/AC3AEC},
    issn = {1538-3881},
    arxivId = {2108.01282}
}

@article{MathewsProchaska2017,
    title = {{Circumgalactic Oxygen Absorption and Feedback}},
    year = {2017},
    journal = {The Astrophysical Journal Letters},
    author = {Mathews, William G. and Prochaska, J. Xavier},
    number = {2},
    month = {9},
    pages = {L24},
    volume = {846},
    publisher = {IOP Publishing},
    url = {https://doi.org/10.3847/2041-8213/aa8861},
    doi = {10.3847/2041-8213/AA8861},
    issn = {2041-8205},
    arxivId = {1708.07140}
}

@article{Kremer2023,
    title = {{Connecting the young pulsars in Milky Way globular clusters with white dwarf mergers and the M81 fast radio burst}},
    year = {2023},
    journal = {Monthly Notices of the Royal Astronomical Society: Letters},
    author = {Kremer, Kyle and Fuller, Jim and Piro, Anthony L. and Ransom, Scott M.},
    number = {1},
    month = {7},
    pages = {L22-L27},
    volume = {525},
    publisher = {Oxford Academic},
    url = {https://doi.org/10.1093/mnrasl/slad088},
    doi = {10.1093/MNRASL/SLAD088},
    issn = {1745-3925},
    arxivId = {2305.11933}
}

@article{FLIMFLAM,
    title = {{Constraining the Cosmic Baryon Distribution with Fast Radio Burst Foreground Mapping}},
    year = {2022},
    journal = {The Astrophysical Journal},
    author = {Lee, Khee-Gan and Ata, Metin and Khrykin, Ilya S. and Huang, Yuxin and Prochaska, J. Xavier and Cooke, Jeff and Zhang, Jielai and Batten, Adam},
    number = {1},
    month = {3},
    pages = {9},
    volume = {928},
    publisher = {IOP Publishing},
    url = {https://doi.org/10.3847/1538-4357/ac4f62},
    doi = {10.3847/1538-4357/AC4F62},
    issn = {0004-637X},
    arxivId = {2109.00386}
}

@article{Tully2013,
    title = {{Cosmicflows-2: The Data}},
    year = {2013},
    journal = {The Astronomical Journal},
    author = {Tully, R. Brent and Courtois, Hélène M. and Dolphin, Andrew E. and Fisher, J. Richard and H{\'{e}}raudeau, Philippe and Jacobs, Bradley A. and Karachentsev, Igor D. and Makarov, Dmitry and Makarova, Lidia and Mitronova, Sofia and Rizzi, Luca and Shaya, Edward J. and Sorce, Jenny G. and Wu, Po Feng},
    number = {4},
    month = {9},
    pages = {86},
    volume = {146},
    publisher = {IOP Publishing},
    url = {https://doi.org/10.1088/0004-6256/146/4/86},
    doi = {10.1088/0004-6256/146/4/86},
    issn = {1538-3881},
    arxivId = {1307.7213}
}

@article{COSMOS,
    title = {{COSMOS: Hubble Space Telescope Observations}},
    year = {2007},
    journal = {The Astrophysical Journal Supplement Series},
    author = {Scoville, N. and Abraham, R. G. and Aussel, H. and Barnes, J. E. and Benson, A. and Blain, A. W. and Calzetti, D. and Comastri, A. and Capak, P. and Carilli, C. and Carlstrom, J. E. and Carollo, C. M. and Colbert, J. and Daddi, E. and Ellis, R. S. and Elvis, M. and Ewald, S. P. and Fall, M. and Franceschini, A. and Giavalisco, M. and Green, W. and Griffiths, R. E. and Guzzo, L. and Hasinger, G. and Impey, C. and Kneib, J.‐P. and Koda, J. and Koekemoer, A. and Lefevre, O. and Lilly, S. and Liu, C. T. and McCracken, H. J. and Massey, R. and Mellier, Y. and Miyazaki, S. and Mobasher, B. and Mould, J. and Norman, C. and Refregier, A. and Renzini, A. and Rhodes, J. and Rich, M. and Sanders, D. B. and Schiminovich, D. and Schinnerer, E. and Scodeggio, M. and Sheth, K. and Shopbell, P. L. and Taniguchi, Y. and Tyson, N. D. and Urry, C. M. and Van Waerbeke, L. and Vettolani, P. and White, S. D. M. and Yan, L.},
    number = {1},
    month = {9},
    pages = {38--45},
    volume = {172},
    publisher = {American Astronomical Society},
    url = {https://doi.org/10.1086/516580},
    doi = {10.1086/516580},
    issn = {0067-0049},
    arxivId = {astro-ph/0612306}
}

@article{TNSCRAFT2026,
    title = {{CRAFT Transient FRB Discovery Report for 2026-01-01}},
    year = {2026},
    journal = {Transient Name Server Fast Radio Bursts},
    author = {Shannon, R. M. and {Shannon} and M., R.},
    month = {1},
    pages = {1},
    volume = {4},
    url = {https://ui.adsabs.harvard.edu/abs/2026TNSFR...4....1S/abstract}
}

@article{Ravi2025,
    title = {{Deep Synoptic Array Science: A 50 Mpc Fast Radio Burst Constrains the Mass of the Milky Way Circumgalactic Medium}},
    year = {2025},
    journal = {The Astronomical Journal},
    author = {Ravi, Vikram and Catha, Morgan and Chen, Ge and Connor, Liam and Cordes, James M. and Faber, Jakob T. and Lamb, James W. and Hallinan, Gregg and Harnach, Charlie and Hellbourg, Greg and Hobbs, Rick and Hodge, David and Hodges, Mark and Law, Casey and Rasmussen, Paul and Sharma, Kritti and Sherman, Myles B. and Shi, Jun and Simard, Dana and Somalwar, Jean J. and Squillace, Reynier and Weinreb, Sander and Woody, David P. and Yadlapalli, Nitika and Team), (The Deep Synoptic Array},
    number = {6},
    month = {5},
    pages = {330},
    volume = {169},
    publisher = {IOP Publishing},
    url = {https://doi.org/10.3847/1538-3881/adc725},
    doi = {10.3847/1538-3881/ADC725},
    issn = {1538-3881},
    arxivId = {2301.01000}
}

@article{LawDSA-110Cat,
    title = {{Deep Synoptic Array Science: First FRB and Host Galaxy Catalog}},
    year = {2024},
    journal = {The Astrophysical Journal},
    author = {Law, Casey J. and Sharma, Kritti and Ravi, Vikram and Chen, Ge and Catha, Morgan and Connor, Liam and Faber, Jakob T. and Hallinan, Gregg and Harnach, Charlie and Hellbourg, Greg and Hobbs, Rick and Hodge, David and Hodges, Mark and Lamb, James W. and Rasmussen, Paul and Sherman, Myles B. and Shi, Jun and Simard, Dana and Squillace, Reynier and Weinreb, Sander and Woody, David P. and Yurk, Nitika Yadlapalli},
    number = {1},
    month = {5},
    pages = {29},
    volume = {967},
    publisher = {IOP Publishing},
    url = {https://doi.org/10.3847/1538-4357/ad3736},
    doi = {10.3847/1538-4357/AD3736},
    issn = {0004-637X},
    arxivId = {2307.03344}
}

@article{Galfit2,
    title = {{Detailed decomposition of galaxy images. II. beyond axisymmetric models}},
    year = {2010},
    journal = {Astronomical Journal},
    author = {Peng, Chien Y. and Ho, Luis C. and Impey, Chris D. and Rix, Hans Walter},
    number = {6},
    month = {4},
    pages = {2097--2129},
    volume = {139},
    publisher = {IOP Publishing},
    url = {https://doi.org/10.1088/0004-6256/139/6/2097},
    doi = {10.1088/0004-6256/139/6/2097},
    issn = {00046256},
    arxivId = {0912.0731}
}

@article{Galfit1,
    title = {{Detailed Structural Decomposition of Galaxy Images}},
    year = {2002},
    journal = {The Astronomical Journal},
    author = {Peng, Chien Y. and Ho, Luis C. and Impey, Chris D. and Rix, Hans-Walter},
    number = {1},
    month = {7},
    pages = {266--293},
    volume = {124},
    publisher = {American Astronomical Society},
    url = {https://doi.org/10.1086/340952},
    doi = {10.1086/340952},
    issn = {00046256},
    arxivId = {astro-ph/0204182}
}

@article{SimhaFRB20190608B,
    title = {{Disentangling the Cosmic Web toward FRB 190608}},
    year = {2020},
    journal = {The Astrophysical Journal},
    author = {Simha, Sunil and Burchett, Joseph N. and Prochaska, J. Xavier and Chittidi, Jay S. and Elek, Oskar and Tejos, Nicolas and Jorgenson, Regina and Bannister, Keith W. and Bhandari, Shivani and Day, Cherie K. and Deller, Adam T. and Forbes, Angus G. and Macquart, Jean-Pierre and Ryder, Stuart D. and Shannon, Ryan M.},
    number = {2},
    month = {9},
    pages = {134},
    volume = {901},
    publisher = {IOP Publishing},
    url = {https://doi.org/10.3847/1538-4357/ABAFC3},
    doi = {10.3847/1538-4357/ABAFC3},
    issn = {0004-637X}
}

@article{Kovacs2024,
    title = {{Dispersion and rotation measures from fast radio burst (FRB) host galaxies based on the TNG50 simulation}},
    year = {2024},
    journal = {Astronomy {\&} Astrophysics},
    author = {Kovacs, Timea Orsolya and Mao, Sui Ann and Basu, Aritra and Ma, Yik Ki and Pakmor, Ruediger and Spitler, Laura G. and Walker, Charles R.H.},
    month = {10},
    pages = {A47},
    volume = {690},
    publisher = {EDP Sciences},
    url = {https://doi.org/10.1051/0004-6361/202347459},
    doi = {10.1051/0004-6361/202347459},
    issn = {0004-6361}
}

@article{ChittidiFRB20190608B,
    title = {{Dissecting the Local Environment of FRB 190608 in the Spiral Arm of its Host Galaxy}},
    year = {2021},
    journal = {The Astrophysical Journal},
    author = {Chittidi, Jay S. and Simha, Sunil and Mannings, Alexandra and Prochaska, J. Xavier and Ryder, Stuart D. and Rafelski, Marc and Neeleman, Marcel and Macquart, Jean-Pierre and Tejos, Nicolas and Jorgenson, Regina A. and Day, Cherie K. and Marnoch, Lachlan and Bhandari, Shivani and Deller, Adam T. and Qiu, Hao and Bannister, Keith W. and Shannon, Ryan M. and Heintz, Kasper E.},
    number = {2},
    month = {12},
    pages = {173},
    volume = {922},
    url = {https://doi.org/10.3847/1538-4357/ac2818},
    doi = {10.3847/1538-4357/AC2818},
    issn = {0004-637X}
}

@article{Bernales-Cortes2025,
    title = {{Empirical estimation of host galaxy dispersion measure toward well-localized fast radio bursts}},
    year = {2025},
    journal = {Astronomy {\&} Astrophysics},
    author = {Bernales-Cortes, Lucas and Tejos, Nicolas and Prochaska, J. Xavier and Khrykin, Ilya S. and Marnoch, Lachlan and Ryder, Stuart D. and Shannon, Ryan M.},
    month = {4},
    pages = {A81},
    volume = {696},
    publisher = {EDP Sciences},
    url = {https://doi.org/10.1051/0004-6361/202452026},
    doi = {10.1051/0004-6361/202452026},
    issn = {0004-6361}
}

@article{Hubble1926,
    title = {{Extragalactic nebulae.}},
    year = {1926},
    journal = {Astrophysical Journal},
    author = {Hubble, E. P.},
    month = {12},
    pages = {321},
    volume = {64},
    publisher = {American Astronomical Society},
    url = {https://doi.org/10.1086/143018},
    doi = {10.1086/143018},
    issn = {0004-637X}
}

@article{Mannings2023,
    title = {{Fast Radio Bursts as Probes of Magnetic Fields in Galaxies at z < 0.5}},
    year = {2023},
    journal = {The Astrophysical Journal},
    author = {Mannings, Alexandra G. and Pakmor, Rüdiger and Prochaska, J. Xavier and Voort, Freeke van de and Simha, Sunil and Shannon, R. M. and Tejos, Nicolas and Deller, Adam and Rafelski, Marc},
    number = {2},
    month = {9},
    pages = {179},
    volume = {954},
    publisher = {IOP Publishing},
    url = {https://iopscience.iop.org/article/10.3847/1538-4357/ace7bb https://iopscience.iop.org/article/10.3847/1538-4357/ace7bb/meta},
    doi = {10.3847/1538-4357/ACE7BB},
    issn = {0004-637X},
    arxivId = {2209.15113}
}

@article{COBRADR1,
    title = {{FLIMFLAM DR1: The First Constraints on the Cosmic Baryon Distribution from Eight Fast Radio Burst Sight Lines}},
    year = {2024},
    journal = {The Astrophysical Journal},
    author = {Khrykin, Ilya S. and Ata, Metin and Lee, Khee-Gan and Simha, Sunil and Huang, Yuxin and Prochaska, J. Xavier and Tejos, Nicolas and Bannister, Keith W. and Cooke, Jeff and Day, Cherie K. and Deller, Adam and Glowacki, Marcin and Gordon, Alexa C. and James, Clancy W. and Marnoch, Lachlan and Shannon, Ryan. M. and Zhang, Jielai and Bernales-Cortes, Lucas},
    number = {2},
    month = {9},
    pages = {151},
    volume = {973},
    publisher = {IOP Publishing},
    url = {https://doi.org/10.3847/1538-4357/AD6567},
    doi = {10.3847/1538-4357/AD6567},
    issn = {0004-637X},
    arxivId = {2402.00505}
}

@article{Moster2013,
    title = {{Galactic star formation and accretion histories from matching galaxies to dark matter haloes}},
    year = {2013},
    journal = {Monthly Notices of the Royal Astronomical Society},
    author = {Moster, Benjamin P. and Naab, Thorsten and White, Simon D.M.},
    number = {4},
    month = {2},
    pages = {3121--3138},
    volume = {428},
    publisher = {Oxford Academic},
    url = {https://doi.org/10.1093/MNRAS/STS261},
    doi = {10.1093/MNRAS/STS261},
    issn = {0035-8711},
    arxivId = {1205.5807}
}

@article{GALFITM,
    title = {{Galapagos-2/Galfitm/Gama – Multi-wavelength measurement of galaxy structure: Separating the properties of spheroid and disk components in modern surveys}},
    year = {2022},
    journal = {Astronomy {\&} Astrophysics},
    author = {H{\"{a}}u{\ss}ler, Boris and Vika, Marina and Bamford, Steven P. and Johnston, Evelyn J. and Brough, Sarah and Casura, Sarah and Holwerda, Benne W. and Kelvin, Lee S. and Popescu, Cristina},
    month = {8},
    pages = {A92},
    volume = {664},
    publisher = {EDP Sciences},
    url = {https://doi.org/10.1051/0004-6361/202142935},
    doi = {10.1051/0004-6361/202142935},
    issn = {0004-6361},
    arxivId = {2204.05907}
}

@article{galapagos,
    title = {{galapagos: from pixels to parameters}},
    year = {2012},
    journal = {Monthly Notices of the Royal Astronomical Society},
    author = {Barden, Marco and H{\"{a}}u{\ss}ler, Boris and Peng, Chien Y. and McIntosh, Daniel H. and Guo, Yicheng},
    number = {1},
    month = {5},
    pages = {449--468},
    volume = {422},
    publisher = {Oxford Academic},
    url = {https://doi.org/10.1111/j.1365-2966.2012.20619.x},
    doi = {10.1111/J.1365-2966.2012.20619.X},
    issn = {0035-8711},
    arxivId = {1203.1831}
}

@article{Dewsnap2023,
    title = {{GALFIT-ing AGN Host Galaxies in COSMOS: HST versus Subaru}},
    year = {2023},
    journal = {The Astrophysical Journal},
    author = {Dewsnap, Callum and Barmby, Pauline and Gallagher, Sarah C. and Urry, C. Megan and Ghosh, Aritra and Powell, Meredith C.},
    number = {2},
    month = {2},
    pages = {137},
    volume = {944},
    publisher = {IOP Publishing},
    url = {https://doi.org/10.3847/1538-4357/ac9400},
    doi = {10.3847/1538-4357/AC9400},
    issn = {0004-637X},
    arxivId = {2209.11325}
}

@article{GEMS,
    title = {{GEMS Survey Data and Catalog}},
    year = {2008},
    journal = {The Astrophysical Journal Supplement Series},
    author = {Caldwell, John A. R. and McIntosh, Daniel H. and Rix, Hans‐Walter and Barden, Marco and Beckwith, Steven V. W. and Bell, Eric F. and Borch, Andrea and Heymans, Catherine and H{\"{a}}u{\ss}ler, Boris and Jahnke, Knud and Jogee, Shardha and Meisenheimer, Klaus and Peng, Chien Y. and S{\'{a}}nchez, Sebastian F. and Somerville, Rachel S. and Wisotzki, Lutz and Wolf, Christian},
    number = {1},
    month = {1},
    pages = {136--144},
    volume = {174},
    publisher = {American Astronomical Society},
    url = {https://doi.org/10.1086/521080},
    doi = {10.1086/521080},
    issn = {0067-0049}
}

@article{Day2020,
    title = {{High time resolution and polarization properties of ASKAP-localized fast radio bursts}},
    year = {2020},
    journal = {Monthly Notices of the Royal Astronomical Society},
    author = {Day, Cherie K and Deller, Adam T and Shannon, Ryan M and Qiu, Hao and Bannister, Keith W and Bhandari, Shivani and Ekers, Ron and Flynn, Chris and James, Clancy W and Macquart, Jean-Pierre and Mahony, Elizabeth K and Phillips, Chris J and Xavier Prochaska, J},
    number = {3},
    month = {9},
    pages = {3335--3350},
    volume = {497},
    publisher = {Oxford University Press},
    url = {https://doi.org/10.1093/mnras/staa2138},
    doi = {10.1093/mnras/staa2138},
    issn = {0035-8711}
}

@article{Heintz2020,
    title = {{Host galaxy properties and offset distributions of fast radio bursts: Implications for their progenitors}},
    year = {2020},
    journal = {The Astrophysical Journal},
    author = {Heintz, Kasper Elm and Xavier Prochaska, J. and Simha, Sunil and Platts, Emma and Fong, Wen Fai and Tejos, Nicolas and Ryder, Stuart D. and Aggerwal, Kshitij and Bhandari, Shivani and Day, Cherie K. and Deller, Adam T. and Kilpatrick, Charles D. and Law, Casey J. and Macquart, Jean Pierre and Mannings, Alexandra and Marnoch, Lachlan J. and Sadler, Elaine M. and Shannon, Ryan M.},
    number = {2},
    month = {11},
    pages = {152},
    volume = {903},
    url = {https://doi.org/10.3847/1538-4357/abb6fb},
    doi = {10.3847/1538-4357/abb6fb},
    issn = {23318422},
    arxivId = {2009.10747}
}

@article{Hoffman2026,
    title = {{I can see your halo: Constraining the MilkyWay halo DM with FRB population studies}},
    year = {2026},
    journal = {Publications of the Astronomical Society of Australia},
    author = {Hoffmann, Jordan and James, Clancy and Prochaska, Jason Xavier and Glowacki, Marcin},
    month = {1},
    pages = {e017},
    volume = {43},
    publisher = {Cambridge University Press},
    url = {https://doi.org/10.1017/PASA.2026.10143},
    doi = {10.1017/PASA.2026.10143},
    issn = {1323-3580},
    arxivId = {2601.05496}
}

@article{Ocker2024,
    title = {{Implications for Galactic Electron Density Structure from Pulsar Sightlines Intersecting H ii Regions}},
    year = {2024},
    journal = {The Astrophysical Journal},
    author = {Ocker, Stella Koch and Anderson, Loren D. and Lazio, T. Joseph W. and Cordes, James M. and Ravi, Vikram},
    number = {1},
    month = {10},
    pages = {10},
    volume = {974},
    publisher = {IOP Publishing},
    url = {https://doi.org/10.3847/1538-4357/ad6a51},
    doi = {10.3847/1538-4357/AD6A51},
    issn = {0004-637X},
    arxivId = {2406.07664}
}

@article{SersicProfile,
    title = {{Influence of the atmospheric and instrumental dispersion on the brightness distribution in a galaxy}},
    year = {1963},
    journal = {Bolet{\'{i}}n de la Asociaci{\'{o}}n Argentina de Astronom{\'{i}}a, Vol. 6, pp. 41-43},
    author = {S{\'{e}}rsic, J. L.},
    month = {2},
    pages = {41},
    volume = {6},
    url = {https://ui.adsabs.harvard.edu/abs/1963BAAA....6...41S/abstract},
    issn = {0571-3285}
}

@article{KsampleAD,
    title = {{K-Sample Anderson–Darling Tests}},
    year = {1987},
    journal = {Journal of the American Statistical Association},
    author = {Scholz, F. W. and Stephens, M. A.},
    number = {399},
    pages = {918--924},
    volume = {82},
    publisher = {Taylor {\&} Francis Group},
    url = {https://doi.org/10.1080/01621459.1987.10478517},
    doi = {10.1080/01621459.1987.10478517},
    issn = {1537274X}
}

@article{LTS,
    title = {{Least median of squares regression}},
    year = {1984},
    journal = {Journal of the American Statistical Association},
    author = {Rousseeuw, Peter J.},
    number = {388},
    pages = {871--880},
    volume = {79},
    url = {https://doi.org/10.1080/01621459.1984.10477105},
    doi = {10.1080/01621459.1984.10477105},
    issn = {1537274X}
}

@article{BhandariFRB20191001A,
    title = {{Limits on Precursor and Afterglow Radio Emission from a Fast Radio Burst in a Star-forming Galaxy}},
    year = {2020},
    journal = {The Astrophysical Journal Letters},
    author = {Bhandari, Shivani and Bannister, Keith W. and Lenc, Emil and Cho, Hyerin and Ekers, Ron and Day, Cherie K. and Deller, Adam T. and Flynn, Chris and James, Clancy W. and Macquart, Jean-Pierre and Mahony, Elizabeth K. and Marnoch, Lachlan and Moss, Vanessa A. and Phillips, Chris and Prochaska, J. Xavier and Qiu, Hao and Ryder, Stuart D. and Shannon, Ryan M. and Tejos, Nicolas and Wong, O. Ivy},
    number = {2},
    month = {9},
    pages = {L20},
    volume = {901},
    publisher = {American Astronomical Society},
    url = {https://doi.org/10.3847/2041-8213/abb462},
    doi = {10.3847/2041-8213/abb462},
    issn = {2041-8205},
    arxivId = {2008.12488}
}

@article{Lokas2022,
    title = {{Lopsided galactic disks in IllustrisTNG}},
    year = {2022},
    journal = {Astronomy {\&} Astrophysics},
    author = {{\L}okas, Ewa L.},
    month = {6},
    pages = {A53},
    volume = {662},
    publisher = {EDP Sciences},
    url = {https://doi.org/10.1051/0004-6361/202142845},
    doi = {10.1051/0004-6361/202142845},
    issn = {0004-6361},
    arxivId = {2204.01456}
}

@article{LSST,
    title = {{LSST: From Science Drivers to Reference Design and Anticipated Data Products}},
    year = {2019},
    journal = {The Astrophysical Journal},
    author = {Ivezi{\'{c}}, \v{Z}eljko and Kahn, Steven M. and Tyson, J. Anthony and Abel, Bob and Acosta, Emily and Allsman, Robyn and Alonso, David and AlSayyad, Yusra and Anderson, Scott F. and Andrew, John and Angel, James Roger P. and Angeli, George Z. and Ansari, Reza and Antilogus, Pierre and Araujo, Constanza and Armstrong, Robert and Arndt, Kirk T. and Astier, Pierre and Aubourg, Eric and Auza, Nicole and Axelrod, Tim S. and Bard, Deborah J. and Barr, Jeff D. and Barrau, Aurelian and Bartlett, James G. and Bauer, Amanda E. and Bauman, Brian J. and Baumont, Sylvain and Bechtol, Ellen and Bechtol, Keith and Becker, Andrew C. and Becla, Jacek and Beldica, Cristina and Bellavia, Steve and Bianco, Federica B. and Biswas, Rahul and Blanc, Guillaume and Blazek, Jonathan and Blandford, Roger D. and Bloom, Josh S. and Bogart, Joanne and Bond, Tim W. and Booth, Michael T. and Borgland, Anders W. and Borne, Kirk and Bosch, James F. and Boutigny, Dominique and Brackett, Craig A. and Bradshaw, Andrew and Brandt, William Nielsen and Brown, Michael E. and Bullock, James S. and Burchat, Patricia and Burke, David L. and Cagnoli, Gianpietro and Calabrese, Daniel and Callahan, Shawn and Callen, Alice L. and Carlin, Jeffrey L. and Carlson, Erin L. and Chandrasekharan, Srinivasan and Charles-Emerson, Glenaver and Chesley, Steve and Cheu, Elliott C. and Chiang, Hsin-Fang and Chiang, James and Chirino, Carol and Chow, Derek and Ciardi, David R. and Claver, Charles F. and Cohen-Tanugi, Johann and Cockrum, Joseph J. and Coles, Rebecca and Connolly, Andrew J. and Cook, Kem H. and Cooray, Asantha and Covey, Kevin R. and Cribbs, Chris and Cui, Wei and Cutri, Roc and Daly, Philip N. and Daniel, Scott F. and Daruich, Felipe and Daubard, Guillaume and Daues, Greg and Dawson, William and Delgado, Francisco and Dellapenna, Alfred and Peyster, Robert de and Val-Borro, Miguel de and Digel, Seth W. and Doherty, Peter and Dubois, Richard and Dubois-Felsmann, Gregory P. and Durech, Josef and Economou, Frossie and Eifler, Tim and Eracleous, Michael and Emmons, Benjamin L. and Neto, Angelo Fausti and Ferguson, Henry and Figueroa, Enrique and Fisher-Levine, Merlin and Focke, Warren and Foss, Michael D. and Frank, James and Freemon, Michael D. and Gangler, Emmanuel and Gawiser, Eric and Geary, John C. and Gee, Perry and Geha, Marla and Gessner, Charles J. B. and Gibson, Robert R. and Gilmore, D. Kirk and Glanzman, Thomas and Glick, William and Goldina, Tatiana and Goldstein, Daniel A. and Goodenow, Iain and Graham, Melissa L. and Gressler, William J. and Gris, Philippe and Guy, Leanne P. and Guyonnet, Augustin and Haller, Gunther and Harris, Ron and Hascall, Patrick A. and Haupt, Justine and Hernandez, Fabio and Herrmann, Sven and Hileman, Edward and Hoblitt, Joshua and Hodgson, John A. and Hogan, Craig and Howard, James D. and Huang, Dajun and Huffer, Michael E. and Ingraham, Patrick and Innes, Walter R. and Jacoby, Suzanne H. and Jain, Bhuvnesh and Jammes, Fabrice and Jee, M. James and Jenness, Tim and Jernigan, Garrett and Jevremovi{\'{c}}, Darko and Johns, Kenneth and Johnson, Anthony S. and Johnson, Margaret W. G. and Jones, R. Lynne and Juramy-Gilles, Claire and Juri{\'{c}}, Mario and Kalirai, Jason S. and Kallivayalil, Nitya J. and Kalmbach, Bryce and Kantor, Jeffrey P. and Karst, Pierre and Kasliwal, Mansi M. and Kelly, Heather and Kessler, Richard and Kinnison, Veronica and Kirkby, David and Knox, Lloyd and Kotov, Ivan V. and Krabbendam, Victor L. and Krughoff, K. Simon and Kub{\'{a}}nek, Petr and Kuczewski, John and Kulkarni, Shri and Ku, John and Kurita, Nadine R. and Lage, Craig S. and Lambert, Ron and Lange, Travis and Langton, J. Brian and Guillou, Laurent Le and Levine, Deborah and Liang, Ming and Lim, Kian-Tat and Lintott, Chris J. and Long, Kevin E. and Lopez, Margaux and Lotz, Paul J. and Lupton, Robert H. and Lust, Nate B. and MacArthur, Lauren A. and Mahabal, Ashish and Mandelbaum, Rachel and Markiewicz, Thomas W. and Marsh, Darren S. and Marshall, Philip J. and Marshall, Stuart and May, Morgan and McKercher, Robert and McQueen, Michelle and Meyers, Joshua and Migliore, Myriam and Miller, Michelle and Mills, David J. and Miraval, Connor and Moeyens, Joachim and Moolekamp, Fred E. and Monet, David G. and Moniez, Marc and Monkewitz, Serge and Montgomery, Christopher and Morrison, Christopher B. and Mueller, Fritz and Muller, Gary P. and Arancibia, Freddy Muñoz and Neill, Douglas R. and Newbry, Scott P. and Nief, Jean-Yves and Nomerotski, Andrei and Nordby, Martin and O’Connor, Paul and Oliver, John and Olivier, Scot S. and Olsen, Knut and O’Mullane, William and Ortiz, Sandra and Osier, Shawn and Owen, Russell E. and Pain, Reynald and Palecek, Paul E. and Parejko, John K. and Parsons, James B. and Pease, Nathan M. and Peterson, J. Matt and Peterson, John R. and Petravick, Donald L. and Petrick, M. E. Libby and Petry, Cathy E. and Pierfederici, Francesco and Pietrowicz, Stephen and Pike, Rob and Pinto, Philip A. and Plante, Raymond and Plate, Stephen and Plutchak, Joel P. and Price, Paul A. and Prouza, Michael and Radeka, Veljko and Rajagopal, Jayadev and Rasmussen, Andrew P. and Regnault, Nicolas and Reil, Kevin A. and Reiss, David J. and Reuter, Michael A. and Ridgway, Stephen T. and Riot, Vincent J. and Ritz, Steve and Robinson, Sean and Roby, William and Roodman, Aaron and Rosing, Wayne and Roucelle, Cecille and Rumore, Matthew R. and Russo, Stefano and Saha, Abhijit and Sassolas, Benoit and Schalk, Terry L. and Schellart, Pim and Schindler, Rafe H. and Schmidt, Samuel and Schneider, Donald P. and Schneider, Michael D. and Schoening, William and Schumacher, German and Schwamb, Megan E. and Sebag, Jacques and Selvy, Brian and Sembroski, Glenn H. and Seppala, Lynn G. and Serio, Andrew and Serrano, Eduardo and Shaw, Richard A. and Shipsey, Ian and Sick, Jonathan and Silvestri, Nicole and Slater, Colin T. and Smith, J. Allyn and Smith, R. Chris and Sobhani, Shahram and Soldahl, Christine and Storrie-Lombardi, Lisa and Stover, Edward and Strauss, Michael A. and Street, Rachel A. and Stubbs, Christopher W. and Sullivan, Ian S. and Sweeney, Donald and Swinbank, John D. and Szalay, Alexander and Takacs, Peter and Tether, Stephen A. and Thaler, Jon J. and Thayer, John Gregg and Thomas, Sandrine and Thornton, Adam J. and Thukral, Vaikunth and Tice, Jeffrey and Trilling, David E. and Turri, Max and Berg, Richard Van and Berk, Daniel Vanden and Vetter, Kurt and Virieux, Francoise and Vucina, Tomislav and Wahl, William and Walkowicz, Lucianne and Walsh, Brian and Walter, Christopher W. and Wang, Daniel L. and Wang, Shin-Yawn and Warner, Michael and Wiecha, Oliver and Willman, Beth and Winters, Scott E. and Wittman, David and Wolff, Sidney C. and Wood-Vasey, W. Michael and Wu, Xiuqin and Xin, Bo and Yoachim, Peter and Zhan, Hu},
    number = {2},
    month = {3},
    pages = {111},
    volume = {873},
    publisher = {IOP Publishing},
    url = {https://doi.org/10.3847/1538-4357/ab042c},
    doi = {10.3847/1538-4357/AB042C},
    issn = {0004-637X},
    arxivId = {0805.2366}
}

@article{Magnetars,
    title = {{Magnetars}},
    year = {2017},
    journal = {Annual Review of Astronomy and Astrophysics},
    author = {Kaspi, Victoria M. and Beloborodov, Andrei M.},
    month = {8},
    pages = {261--301},
    volume = {55},
    publisher = {Annual Reviews},
    url = {https://doi.org/10.1146/annurev-astro-081915-023329},
    doi = {10.1146/ANNUREV-ASTRO-081915-023329},
    issn = {00664146},
    arxivId = {1703.00068}
}

@article{GordonSpatial,
    title = {{Mapping the Spatial Distribution of Fast Radio Bursts within their Host Galaxies}},
    year = {2025},
    journal = {The Astrophysical Journal},
    author = {Gordon, Alexa C. and Fong, Wen-fai and Deller, Adam T. and Marnoch, Lachlan and Lim, Sungsoon and Peng, Eric W. and Bannister, Keith W. and Bera, Apurba and Bhat, N. D. R. and Dial, Tyson and Dong, Yuxin and Eftekhari, Tarraneh and Glowacki, Marcin and Gourdji, Kelly and Gupta, Vivek and Jahns-Schindler, Joscha N. and Jaini, Akhil and Kilpatrick, Charles D. and Liu, Chang and Prochaska, J. Xavier and Ryder, Stuart D. and Shannon, Ryan M. and Simha, Sunil and Tejos, Nicolas and Wang, Yuanming and Wang, Ziteng},
    number = {1},
    month = {10},
    pages = {119},
    volume = {993},
    publisher = {IOP Publishing},
    url = {https://doi.org/10.3847/1538-4357/ae0298},
    doi = {10.3847/1538-4357/AE0298},
    issn = {0004-637X},
    arxivId = {2506.06453}
}

@article{Baptista2023,
    title = {{Measuring the Variance of the Macquart Relation in Redshift–Extragalactic Dispersion Measure Modeling}},
    year = {2024},
    journal = {The Astrophysical Journal},
    author = {Baptista, Jay and Prochaska, J. Xavier and Mannings, Alexandra G. and James, C. W. and Shannon, R. M. and Ryder, Stuart D. and Deller, A. T. and Scott, Danica R. and Glowacki, Marcin and Tejos, Nicolas},
    number = {1},
    month = {4},
    pages = {57},
    volume = {965},
    publisher = {IOP Publishing},
    url = {https://doi.org/10.3847/1538-4357/ad2705},
    doi = {10.3847/1538-4357/AD2705},
    issn = {0004-637X}
}

@article{NE2001-II,
    title = {{NE2001. II. Using Radio Propagation Data to Construct a Model for the Galactic Distribution of Free Electrons}},
    year = {2003},
    journal = {preprint (arXiv:astro-ph/0301598)},
    author = {Cordes, J M and Joseph, T and Lazio, W},
    month = {1},
    pages = {},
    url = {https://arxiv.org/abs/astro-ph/0301598v1},
    doi = {10.48550/arxiv.astro-ph/0301598},
    arxivId = {astro-ph/0301598}
}

@article{NE2001,
    title = {{NE2001.I. A New Model for the Galactic Distribution of Free Electrons and its Fluctuations}},
    year = {2002},
    journal = {preprint (arXiv:astro-ph/0207156)},
    author = {Cordes, J. M. and Lazio, T. J. W.},
    month = {7},
    url = {http://arxiv.org/abs/astro-ph/0207156},
    arxivId = {astro-ph/0207156}
}

@article{Pearson,
    title = {{Note on Regression and Inheritance in the Case of Two Parents}},
    year = {1895},
    journal = {Proceedings of the Royal Society of London},
    author = {Pearson, Karl},
    pages = {240--242},
    volume = {58},
    url = {https://ui.adsabs.harvard.edu/abs/1895RSPS...58..240P/abstract}
}

@article{Orr2024,
    title = {{Objects May Be Closer than They Appear: Significant Host Galaxy Dispersion Measures of Fast Radio Bursts in Zoom-in Simulations}},
    year = {2024},
    journal = {The Astrophysical Journal Letters},
    author = {Orr, Matthew E. and Burkhart, Blakesley and Lu, Wenbin and Ponnada, Sam B. and Hummels, Cameron B.},
    number = {2},
    month = {9},
    pages = {L26},
    volume = {972},
    publisher = {IOP Publishing},
    url = {https://doi.org/10.3847/2041-8213/ad725b},
    doi = {10.3847/2041-8213/AD725B},
    issn = {2041-8205},
    arxivId = {2406.03523}
}

@article{Sharma2025,
    title = {{Preferential occurrence of fast radio bursts in massive star-forming galaxies}},
    year = {2024},
    journal = {Nature},
    author = {Sharma, Kritti and Ravi, Vikram and Connor, Liam and Law, Casey and Ocker, Stella Koch and Sherman, Myles and Kosogorov, Nikita and Faber, Jakob and Hallinan, Gregg and Harnach, Charlie and Hellbourg, Greg and Hobbs, Rick and Hodge, David and Hodges, Mark and Lamb, James and Rasmussen, Paul and Somalwar, Jean and Weinreb, Sander and Woody, David and Leja, Joel and Anand, Shreya and Das, Kaustav Kashyap and Qin, Yu-Jing and Rose, Sam and Dong, Dillon Z. and Miller, Jessie and Yao, Yuhan},
    number = {8037},
    month = {11},
    pages = {61--66},
    volume = {635},
    publisher = {Nature Publishing Group},
    url = {https://doi.org/10.1038/s41586-024-08074-9},
    doi = {10.1038/s41586-024-08074-9},
    issn = {1476-4687},
    pmid = {39506155},
    arxivId = {2409.16964}
}

@article{AggarwalPATH,
    title = {{Probabilistic Association of Transients to their Hosts (PATH)}},
    year = {2021},
    journal = {The Astrophysical Journal},
    author = {Aggarwal, Kshitij and Budav{\'{a}}ri, Tamás and Deller, Adam T. and Eftekhari, Tarraneh and James, Clancy W. and Prochaska, J. Xavier and Tendulkar, Shriharsh P.},
    number = {2},
    month = {4},
    pages = {95},
    volume = {911},
    publisher = {American Astronomical Society},
    url = {https://doi.org/10.3847/1538-4357/abe8d2},
    doi = {10.3847/1538-4357/abe8d2},
    issn = {0004-637X},
    arxivId = {2102.10627}
}

@article{ProchaskaZheng2019,
    title = {{Probing Galactic haloes with fast radio bursts}},
    year = {2019},
    journal = {Monthly Notices of the Royal Astronomical Society},
    author = {Prochaska, J Xavier and Zheng, Yong},
    number = {1},
    month = {5},
    pages = {648--665},
    volume = {485},
    publisher = {Oxford Academic},
    url = {https://doi.org/10.1093/mnras/stz261},
    doi = {10.1093/MNRAS/STZ261},
    issn = {0035-8711}
}

@article{Inoue2004,
    title = {{Probing the cosmic reionization history and local environment of gamma-ray bursts through radio dispersion}},
    year = {2004},
    journal = {Monthly Notices of the Royal Astronomical Society},
    author = {Inoue, Susumu},
    number = {3},
    month = {3},
    pages = {999--1008},
    volume = {348},
    publisher = {Blackwell Publishing Ltd},
    url = {https://doi.org/10.1111/j.1365-2966.2004.07359.x},
    doi = {10.1111/j.1365-2966.2004.07359.x},
    issn = {00358711}
}

@article{ProFit,
    title = {{ProFit: Bayesian profile fitting of galaxy images}},
    year = {2017},
    journal = {Monthly Notices of the Royal Astronomical Society},
    author = {Robotham, A. S.G. and Taranu, D. S. and Tobar, R. and Moffett, A. and Driver, S. P.},
    number = {2},
    month = {4},
    pages = {1513--1541},
    volume = {466},
    publisher = {Oxford Academic},
    url = {https://doi.org/10.1093/mnras/stw3039},
    doi = {10.1093/MNRAS/STW3039},
    issn = {0035-8711},
    arxivId = {1611.08586}
}

@article{ProFuse,
    title = {{ProFuse: physical multiband structural decomposition of galaxies and the mass-size-age plane}},
    year = {2022},
    journal = {Monthly Notices of the Royal Astronomical Society},
    author = {Robotham, A. S.G. and Bellstedt, S. and Driver, S. P.},
    number = {2},
    month = {5},
    pages = {2985--3012},
    volume = {513},
    publisher = {Oxford Academic},
    url = {https://doi.org/10.1093/mnras/stac1032},
    doi = {10.1093/MNRAS/STAC1032},
    issn = {0035-8711},
    arxivId = {2204.04897}
}

@article{Cordes2022,
    title = {{Redshift Estimation and Constraints on Intergalactic and Interstellar Media from Dispersion and Scattering of Fast Radio Bursts}},
    year = {2022},
    journal = {The Astrophysical Journal},
    author = {Cordes, J. M. and Ocker, Stella Koch and Chatterjee, Shami},
    number = {2},
    month = {5},
    pages = {88},
    volume = {931},
    publisher = {IOP Publishing},
    url = {https://doi.org/10.3847/1538-4357/AC6873},
    doi = {10.3847/1538-4357/AC6873},
    issn = {0004-637X},
    arxivId = {2108.01172}
}

@article{scipy,
    title = {{SciPy 1.0: fundamental algorithms for scientific computing in Python}},
    year = {2020},
    journal = {Nature Methods},
    author = {Virtanen, Pauli and Gommers, Ralf and Oliphant, Travis E. and Haberland, Matt and Reddy, Tyler and Cournapeau, David and Burovski, Evgeni and Peterson, Pearu and Weckesser, Warren and Bright, Jonathan and van der Walt, Stéfan J. and Brett, Matthew and Wilson, Joshua and Millman, K. Jarrod and Mayorov, Nikolay and Nelson, Andrew R.J. and Jones, Eric and Kern, Robert and Larson, Eric and Carey, C. J. and Polat, Ilhan and Feng, Yu and Moore, Eric W. and VanderPlas, Jake and Laxalde, Denis and Perktold, Josef and Cimrman, Robert and Henriksen, Ian and Quintero, E. A. and Harris, Charles R. and Archibald, Anne M. and Ribeiro, Antônio H. and Pedregosa, Fabian and van Mulbregt, Paul and Vijaykumar, Aditya and Bardelli, Alessandro Pietro and Rothberg, Alex and Hilboll, Andreas and Kloeckner, Andreas and Scopatz, Anthony and Lee, Antony and Rokem, Ariel and Woods, C. Nathan and Fulton, Chad and Masson, Charles and H{\"{a}}ggstr{\"{o}}m, Christian and Fitzgerald, Clark and Nicholson, David A. and Hagen, David R. and Pasechnik, Dmitrii V. and Olivetti, Emanuele and Martin, Eric and Wieser, Eric and Silva, Fabrice and Lenders, Felix and Wilhelm, Florian and Young, G. and Price, Gavin A. and Ingold, Gert Ludwig and Allen, Gregory E. and Lee, Gregory R. and Audren, Hervé and Probst, Irvin and Dietrich, Jörg P. and Silterra, Jacob and Webber, James T. and Slavi{\v{c}}, Janko and Nothman, Joel and Buchner, Johannes and Kulick, Johannes and Sch{\"{o}}nberger, Johannes L. and de Miranda Cardoso, José Vinícius and Reimer, Joscha and Harrington, Joseph and Rodr{\'{i}}guez, Juan Luis Cano and Nunez-Iglesias, Juan and Kuczynski, Justin and Tritz, Kevin and Thoma, Martin and Newville, Matthew and K{\"{u}}mmerer, Matthias and Bolingbroke, Maximilian and Tartre, Michael and Pak, Mikhail and Smith, Nathaniel J. and Nowaczyk, Nikolai and Shebanov, Nikolay and Pavlyk, Oleksandr and Brodtkorb, Per A. and Lee, Perry and McGibbon, Robert T. and Feldbauer, Roman and Lewis, Sam and Tygier, Sam and Sievert, Scott and Vigna, Sebastiano and Peterson, Stefan and More, Surhud and Pudlik, Tadeusz and Oshima, Takuya and Pingel, Thomas J. and Robitaille, Thomas P. and Spura, Thomas and Jones, Thouis R. and Cera, Tim and Leslie, Tim and Zito, Tiziano and Krauss, Tom and Upadhyay, Utkarsh and Halchenko, Yaroslav O. and V{\'{a}}zquez-Baeza, Yoshiki},
    number = {3},
    month = {3},
    pages = {261--272},
    volume = {17},
    publisher = {Nature Research},
    url = {https://doi.org/10.1038/s41592-019-0686-2},
    doi = {10.1038/s41592-019-0686-2},
    issn = {15487105},
    pmid = {32015543},
    arxivId = {1907.10121}
}

@article{SimhaExcess,
    title = {{Searching for the Sources of Excess Extragalactic Dispersion of FRBs}},
    year = {2023},
    journal = {The Astrophysical Journal},
    author = {Simha, Sunil and Lee, Khee-Gan and Prochaska, J. Xavier and Khrykin, Ilya S. and Huang, Yuxin and Tejos, Nicolas and Marnoch, Lachlan and Ata, Metin and Bernales, Lucas and Bhandari, Shivani and Cooke, Jeff and Deller, Adam T. and Ryder, Stuart D. and Zhang, Jielai},
    number = {1},
    month = {8},
    pages = {71},
    volume = {954},
    publisher = {IOP Publishing},
    url = {https://doi.org/10.3847/1538-4357/ace324},
    doi = {10.3847/1538-4357/ACE324},
    issn = {0004-637X}
}

@article{Bhardwaj2024,
    title = {{Selection bias obfuscates the discovery of fast radio burst sources}},
    year = {2024},
    journal = {Nature},
    author = {Bhardwaj, Mohit and Lee, Jimin and Ji, Kevin},
    number = {8036},
    month = {10},
    pages = {1065--1069},
    volume = {634},
    publisher = {Nature Publishing Group},
    url = {https://doi.org/10.1038/s41586-024-08065-w},
    doi = {10.1038/s41586-024-08065-w},
    issn = {1476-4687},
    arxivId = {2408.01876}
}

@article{Leung2025,
    title = {{Stellar Mass–Dispersion Measure Correlations Constrain Baryonic Feedback in Fast Radio Burst Host Galaxies}},
    year = {2025},
    journal = {The Astrophysical Journal Letters},
    author = {Leung, Calvin and Simha, Sunil and Medlock, Isabel and Nagai, Daisuke and Masui, Kiyoshi W. and Kahinga, Lordrick A. and Lanman, Adam E. and Andrew, Shion and Bandura, Kevin and Curtin, Alice P. and Gaensler, B. M. and Gusinskaia, Nina and Joseph, Ronniy C. and Lazda, Mattias and Mas-Ribas, Lluis and Meyers, Bradley W. and Nimmo, Kenzie and Pearlman, Aaron B. and Prochaska, J. Xavier and Sammons, Mawson W. and Shin, Kaitlyn and Smith, Kendrick and Wang, Haochen and Collaboration), (CHIME/FRB},
    number = {1},
    month = {9},
    pages = {L25},
    volume = {991},
    publisher = {IOP Publishing},
    url = {https://iopscience.iop.org/article/10.3847/2041-8213/ae044d https://iopscience.iop.org/article/10.3847/2041-8213/ae044d/meta},
    doi = {10.3847/2041-8213/AE044D},
    issn = {2041-8205}
}

@article{Kravtsov2018,
    title = {{Stellar Mass—Halo Mass Relation and Star Formation Efficiency in High-Mass Halos}},
    year = {2018},
    journal = {Astronomy Letters},
    author = {Kravtsov, A. V. and Vikhlinin, A. A. and Meshcheryakov, A. V.},
    number = {1},
    month = {2},
    pages = {8--34},
    volume = {44},
    publisher = {Springer},
    url = {https://doi.org/10.1134/S1063773717120015},
    doi = {10.1134/S1063773717120015},
    issn = {1562-6873},
    arxivId = {1401.7329}
}

@article{FORS,
    title = {{Successful commissioning of FORS1 - the first optical instrument on the VLT.}},
    year = {1998},
    journal = {The Messenger},
    author = {Appenzeller, I. and Fricke, K. and F{\"{u}}rtig, W. and G{\"{a}}ssler, W. and H{\"{a}}fner, R. and Harke, R. and Hess, H.-J. and Hummel, W. and J{\"{u}}rgens, P. and Kudritzki, R.-P. and Mantel, K.-H. and Meisl, W. and Muschielok, B. and Nicklas, H. and Rupprecht, G. and Seifert, W. and Stahl, O. and Szeifert, T. and Tarantik, K.},
    pages = {1--6},
    volume = {94},
    publisher = {European Southern Observatory},
    url = {https://ui.adsabs.harvard.edu/abs/1998Msngr..94....1A}
}

@article{SDSS16,
    title = {{The 16th Data Release of the Sloan Digital Sky Surveys: First Release from the APOGEE-2 Southern Survey and Full Release of eBOSS Spectra}},
    year = {2020},
    journal = {The Astrophysical Journal Supplement Series},
    author = {Ahumada, Romina and Prieto, Carlos Allende and Almeida, Andrés and Anders, Friedrich and Anderson, Scott F. and Andrews, Brett H. and Anguiano, Borja and Arcodia, Riccardo and Armengaud, Eric and Aubert, Marie and Avila, Santiago and Avila-Reese, Vladimir and Badenes, Carles and Balland, Christophe and Barger, Kat and Barrera-Ballesteros, Jorge K. and Basu, Sarbani and Bautista, Julian and Beaton, Rachael L. and Beers, Timothy C. and Benavides, B. Izamar T. and Bender, Chad F. and Bernardi, Mariangela and Bershady, Matthew and Beutler, Florian and Bidin, Christian Moni and Bird, Jonathan and Bizyaev, Dmitry and Blanc, Guillermo A. and Blanton, Michael R. and Boquien, Médéric and Borissova, Jura and Bovy, Jo and Brandt, W. N. and Brinkmann, Jonathan and Brownstein, Joel R. and Bundy, Kevin and Bureau, Martin and Burgasser, Adam and Burtin, Etienne and Cano-D{\'{i}}az, Mariana and Capasso, Raffaella and Cappellari, Michele and Carrera, Ricardo and Chabanier, Solène and Chaplin, William and Chapman, Michael and Cherinka, Brian and Chiappini, Cristina and Choi, Peter Doohyun and Chojnowski, S. Drew and Chung, Haeun and Clerc, Nicolas and Coffey, Damien and Comerford, Julia M. and Comparat, Johan and Costa, Luiz da and Cousinou, Marie-Claude and Covey, Kevin and Crane, Jeffrey D. and Cunha, Katia and Ilha, Gabriele da Silva and Dai, Yu, Sophia and Damsted, Sanna B. and Darling, Jeremy and Davidson, James W. and Davies, Roger and Dawson, Kyle and De, Nikhil and Macorra, Axel de la and Lee, Nathan De and Queiroz, Anna Bárbara de Andrade and Machado, Alice Deconto and Torre, Sylvain de la and Dell’Agli, Flavia and Bourboux, Hélion du Mas des and Diamond-Stanic, Aleksandar M. and Dillon, Sean and Donor, John and Drory, Niv and Duckworth, Chris and Dwelly, Tom and Ebelke, Garrett and Eftekharzadeh, Sarah and Eigenbrot, Arthur Davis and Elsworth, Yvonne P. and Eracleous, Mike and Erfanianfar, Ghazaleh and Escoffier, Stephanie and Fan, Xiaohui and Farr, Emily and Fern{\'{a}}ndez-Trincado, José G. and Feuillet, Diane and Finoguenov, Alexis and Fofie, Patricia and Fraser-McKelvie, Amelia and Frinchaboy, Peter M. and Fromenteau, Sebastien and Fu, Hai and Galbany, Lluís and Garcia, Rafael A. and Garc{\'{i}}a-Hern{\'{a}}ndez, D. A. and Oehmichen, Luis Alberto Garma and Ge, Junqiang and Maia, Marcio Antonio Geimba and Geisler, Doug and Gelfand, Joseph and Goddy, Julian and Gonzalez-Perez, Violeta and Grabowski, Kathleen and Green, Paul and Grier, Catherine J. and Guo, Hong and Guy, Julien and Harding, Paul and Hasselquist, Sten and Hawken, Adam James and Hayes, Christian R. and Hearty, Fred and Hekker, S. and Hogg, David W. and Holtzman, Jon A. and Horta, Danny and Hou, Jiamin and Hsieh, Bau-Ching and Huber, Daniel and Hunt, Jason A. S. and Chitham, J. Ider and Imig, Julie and Jaber, Mariana and Angel, Camilo Eduardo Jimenez and Johnson, Jennifer A. and Jones, Amy M. and J{\"{o}}nsson, Henrik and Jullo, Eric and Kim, Yerim and Kinemuchi, Karen and IV, Charles C. Kirkpatrick and Kite, George W. and Klaene, Mark and Kneib, Jean-Paul and Kollmeier, Juna A. and Kong, Hui and Kounkel, Marina and Krishnarao, Dhanesh and Lacerna, Ivan and Lan, Ting-Wen and Lane, Richard R. and Law, David R. and Goff, Jean-Marc Le and Leung, Henry W. and Lewis, Hannah and Li, Cheng and Lian, Jianhui and Lin, Lihwai and Long, Dan and Longa-Pe{\~{n}}a, Penélope and Lundgren, Britt and Lyke, Brad W. and Mackereth, J. Ted and MacLeod, Chelsea L. and Majewski, Steven R. and Manchado, Arturo and Maraston, Claudia and Martini, Paul and Masseron, Thomas and Masters, Karen L. and Mathur, Savita and McDermid, Richard M. and Merloni, Andrea and Merrifield, Michael and M{\'{e}}sz{\'{a}}ros, Szabolcs and Miglio, Andrea and Minniti, Dante and Minsley, Rebecca and Miyaji, Takamitsu and Mohammad, Faizan Gohar and Mosser, Benoit and Mueller, Eva-Maria and Muna, Demitri and Mu{\~{n}}oz-Guti{\'{e}}rrez, Andrea and Myers, Adam D. and Nadathur, Seshadri and Nair, Preethi and Nandra, Kirpal and Nascimento, Janaina Correa do and Nevin, Rebecca Jean and Newman, Jeffrey A. and Nidever, David L. and Nitschelm, Christian and Noterdaeme, Pasquier and O’Connell, Julia E. and Olmstead, Matthew D. and Oravetz, Daniel and Oravetz, Audrey and Osorio, Yeisson and Pace, Zachary J. and Padilla, Nelson and Palanque-Delabrouille, Nathalie and Palicio, Pedro A. and Pan, Hsi-An and Pan, Kaike and Parker, James and Paviot, Romain and Peirani, Sebastien and Ramŕez, Karla Peña and Penny, Samantha and Percival, Will J. and Perez-Fournon, Ismael and P{\'{e}}rez-R{\`{a}}fols, Ignasi and Petitjean, Patrick and Pieri, Matthew M. and Pinsonneault, Marc and Poovelil, Vijith Jacob and Povick, Joshua Tyler and Prakash, Abhishek and Price-Whelan, Adrian M. and Raddick, M. Jordan and Raichoor, Anand and Ray, Amy and Rembold, Sandro Barboza and Rezaie, Mehdi and Riffel, Rogemar A. and Riffel, Rogério and Rix, Hans-Walter and Robin, Annie C. and Roman-Lopes, A. and Rom{\'{a}}n-Z{\'{u}}{\~{n}}iga, Carlos and Rose, Benjamin and Ross, Ashley J. and Rossi, Graziano and Rowlands, Kate and Rubin, Kate H. R. and Salvato, Mara and S{\'{a}}nchez, Ariel G. and S{\'{a}}nchez-Menguiano, Laura and S{\'{a}}nchez-Gallego, José R. and Sayres, Conor and Schaefer, Adam and Schiavon, Ricardo P. and Schimoia, Jaderson S. and Schlafly, Edward and Schlegel, David and Schneider, Donald P. and Schultheis, Mathias and Schwope, Axel and Seo, Hee-Jong and Serenelli, Aldo and Shafieloo, Arman and Shamsi, Shoaib Jamal and Shao, Zhengyi and Shen, Shiyin and Shetrone, Matthew and Shirley, Raphael and Aguirre, Víctor Silva and Simon, Joshua D. and Skrutskie, M. F. and Slosar, Anže and Smethurst, Rebecca and Sobeck, Jennifer and Sodi, Bernardo Cervantes and Souto, Diogo and Stark, David V. and Stassun, Keivan G. and Steinmetz, Matthias and Stello, Dennis and Stermer, Julianna and Storchi-Bergmann, Thaisa and Streblyanska, Alina and Stringfellow, Guy S. and Stutz, Amelia and Su{\'{a}}rez, Genaro and Sun, Jing and Taghizadeh-Popp, Manuchehr and Talbot, Michael S. and Tayar, Jamie and Thakar, Aniruddha R. and Theriault, Riley and Thomas, Daniel and Thomas, Zak C. and Tinker, Jeremy and Tojeiro, Rita and Toledo, Hector Hernandez and Tremonti, Christy A. and Troup, Nicholas W. and Tuttle, Sarah and Unda-Sanzana, Eduardo and Valentini, Marica and Vargas-Gonz{\'{a}}lez, Jaime and Vargas-Maga{\~{n}}a, Mariana and V{\'{a}}zquez-Mata, Jose Antonio and Vivek, M. and Wake, David and Wang, Yuting and Weaver, Benjamin Alan and Weijmans, Anne-Marie and Wild, Vivienne and Wilson, John C. and Wilson, Robert F. and Wolthuis, Nathan and Wood-Vasey, W. M. and Yan, Renbin and Yang, Meng and Y{\`{e}}che, Christophe and Zamora, Olga and Zarrouk, Pauline and Zasowski, Gail and Zhang, Kai and Zhao, Cheng and Zhao, Gongbo and Zheng, Zheng and Zheng, Zheng and Zhu, Guangtun and Zou, Hu},
    number = {1},
    month = {6},
    pages = {3},
    volume = {249},
    publisher = {IOP Publishing},
    url = {https://doi.org/10.3847/1538-4365/ab929e},
    doi = {10.3847/1538-4365/AB929E},
    issn = {0067-0049},
    arxivId = {1912.02905}
}

@article{ACSGC,
    title = {{The Advanced Camera for Surveys General Catalog: Structural Parameters for Approximately Half a Million Galaxies}},
    year = {2012},
    journal = {The Astrophysical Journal Supplement Series},
    author = {Griffith, Roger L. and Cooper, Michael C. and Newman, Jeffrey A. and Moustakas, Leonidas A. and Stern, Daniel and Comerford, Julia M. and Davis, Marc and Lotz, Jennifer M. and Barden, Marco and Conselice, Christopher J. and Capak, Peter L. and Faber, S. M. and Kirkpatrick, J. Davy and Koekemoer, Anton M. and Koo, David C. and Noeske, Kai G. and Scoville, Nick and Sheth, Kartik and Shopbell, Patrick and Willmer, Christopher N.A. and Weiner, Benjamin},
    number = {1},
    month = {5},
    pages = {9},
    volume = {200},
    publisher = {IOP Publishing},
    url = {https://doi.org/10.1088/0067-0049/200/1/9},
    doi = {10.1088/0067-0049/200/1/9},
    issn = {0067-0049},
    arxivId = {1203.1651}
}

@article{AEGIS,
    title = {{The All-Wavelength Extended Groth Strip International Survey (AEGIS) Data Sets}},
    year = {2007},
    journal = {The Astrophysical Journal},
    author = {Davis, M. and Guhathakurta, P. and Konidaris, N. P. and Newman, J. A. and Ashby, M. L. N. and Biggs, A. D. and Barmby, P. and Bundy, K. and Chapman, S. C. and Coil, A. L. and Conselice, C. J. and Cooper, M. C. and Croton, D. J. and Eisenhardt, P. R. M. and Ellis, R. S. and Faber, S. M. and Fang, T. and Fazio, G. G. and Georgakakis, A. and Gerke, B. F. and Goss, W. M. and Gwyn, S. and Harker, J. and Hopkins, A. M. and Huang, J.-S. and Ivison, R. J. and Kassin, S. A. and Kirby, E. N. and Koekemoer, A. M. and Koo, D. C. and Laird, E. S. and Le Floc'h, E. and Lin, L. and Lotz, J. M. and Marshall, P. J. and Martin, D. C. and Metevier, A. J. and Moustakas, L. A. and Nandra, K. and Noeske, K. G. and Papovich, C. and Phillips, A. C. and Rich, R. M. and Rieke, G. H. and Rigopoulou, D. and Salim, S. and Schiminovich, D. and Simard, L. and Smail, I. and Small, T. A. and Weiner, B. J. and Willmer, C. N. A. and Willner, S. P. and Wilson, G. and Wright, E. L. and Yan, R.},
    number = {1},
    month = {5},
    pages = {L1-L6},
    volume = {660},
    publisher = {American Astronomical Society},
    url = {https://doi.org/10.1086/517931},
    doi = {10.1086/517931/FULLTEXT/},
    issn = {0004-637X},
    arxivId = {astro-ph/0607355}
}

@article{Montage,
    title = {{The Application of the Montage Image Mosaic Engine To The Visualization Of Astronomical Images}},
    year = {2017},
    journal = {Publications of the Astronomical Society of the Pacific},
    author = {Berriman, G. Bruce and Good, J. C.},
    number = {975},
    month = {2},
    pages = {058006},
    volume = {129},
    url = {http://doi.org/10.1088/1538-3873/aa5456},
    doi = {10.1088/1538-3873/aa5456},
    issn = {0004-6280},
    arxivId = {1702.02593}
}

@article{LTSFit,
    title = {{The ATLAS3D project – XV. Benchmark for early-type galaxies scaling relations from 260 dynamical models: mass-to-light ratio, dark matter, Fundamental Plane and Mass Plane}},
    year = {2013},
    journal = {Monthly Notices of the Royal Astronomical Society},
    author = {Cappellari, Michele and Scott, Nicholas and Alatalo, Katherine and Blitz, Leo and Bois, Maxime and Bournaud, Frédéric and Bureau, M. and Crocker, Alison F. and Davies, Roger L. and Davis, Timothy A. and de Zeeuw, P. T. and Duc, Pierre Alain and Emsellem, Eric and Khochfar, Sadegh and Krajnovi{\'{c}}, Davor and Kuntschner, Harald and McDermid, Richard M. and Morganti, Raffaella and Naab, Thorsten and Oosterloo, Tom and Sarzi, Marc and Serra, Paolo and Weijmans, Anne Marie and Young, Lisa M.},
    number = {3},
    month = {7},
    pages = {1709--1741},
    volume = {432},
    publisher = {Oxford Academic},
    url = {https://doi.org/10.1093/mnras/stt562},
    doi = {10.1093/MNRAS/STT562},
    issn = {0035-8711},
    arxivId = {1208.3522}
}

@article{Behroozi2013,
    title = {{The Average Star Formation Histories of Galaxies in Dark Matter Halos from z = 0--8}},
    year = {2013},
    journal = {The Astrophysical Journal},
    author = {Behroozi, Peter S. and Wechsler, Risa H. and Conroy, Charlie},
    number = {1},
    month = {5},
    pages = {57},
    volume = {770},
    publisher = {IOP Publishing},
    url = {https://doi.org/10.1088/0004-637X/770/1/57},
    doi = {10.1088/0004-637X/770/1/57},
    issn = {0004-637X},
    arxivId = {1207.6105}
}

@article{ShannonICS,
    title = {{The commensal real-time ASKAP fast transient incoherent-sum survey}},
    year = {2025},
    journal = {Publications of the Astronomical Society of Australia},
    author = {Shannon, Ryan M and Bannister, Keith W and Bera, Apurba and Bhandari, Shivani and Day, Cherie K and Deller, Adam T and Dial, Tyson and Dobie, Dougal and Ekers, Ron D and Fong, Wen-fai and Glowacki, Marcin and Gordon, Alexa C and Gourdji, Kelly and Jaini, Akhil and James, Clancy W and Kumar, Pravir and Mahony, Elizabeth K and Marnoch, Lachlan and Muller, August R and Prochaska, Xavier and Qiu, Hao and Ryder, Stuart D and Sadler, Elaine M and Scott, Danica R and Tejos, N and Uttarkar, Pavan A and Wang, Yuanming and Ryan, Shannon M and Keith, Bannister W and Apurba, Bera and Shivani, Bhandari and Cherie, Day K and Adam, Deller T and Tyson, Dial and Dougal, Dobie and Ron, Ekers D and Wen-fai, Fong and Marcin, Glowacki and Alexa, Gordon C and Kelly, Gourdji and Akhil, Jaini and Clancy, James W and Pravir, Kumar and Elizabeth, Mahony K and Lachlan, Marnoch and August, Muller R and Xavier, Prochaska and Hao, Qiu and Stuart, Ryder D and Elaine, Sadler M and Danica, Scott R and Pavan, Uttarkar A and Yuanming, Wang},
    month = {8},
    pages = {e036},
    volume = {42},
    publisher = {Cambridge University Press},
    url = {https://doi.org/10.1017/PASA.2025.8},
    doi = {10.1017/PASA.2025.8},
    issn = {1323-3580},
    arxivId = {2408.02083}
}

@article{CRAFT,
    title = {{The Commensal Real-time ASKAP Fast Transients (CRAFT) survey}},
    year = {2010},
    journal = {Publications of the Astronomical Society of Australia},
    author = {Macquart, Jean-Pierre and Bailes, M. and Bhat, N. D. R. and Bower, G. C. and Bunton, J. D. and Chatterjee, S. and Colegate, T. and Cordes, J. M. and D'Addario, L. and Deller, A. and Dodson, R. and Fender, R. and Haines, K. and Hall, P. and Harris, C. and Hotan, A. and Johnston, S. and Jones, D. L. and Keith, M. and Koay, J. Y. and Lazio, T. J. W. and Majid, W. and Murphy, T. and Navarro, R. and Phillips, C. and Quinn, P. and Preston, R. A. and Stansby, B. and Stairs, I. and Stappers, B. and Staveley-Smith, L. and Tingay, S. and Thompson, D. and van Straten, W. and Wagstaff, K. and Warren, M. and Wayth, R. and Wen, L.},
    number = {3},
    month = {1},
    pages = {272--282},
    volume = {27},
    url = {https://doi.org/10.1071/AS09082},
    doi = {10.1071/AS09082},
    issn = {1323-3580},
    arxivId = {1001.2958}
}

@article{Ioka2003,
    title = {{The Cosmic Dispersion Measure from Gamma-Ray Burst Afterglows: Probing the Reionization History and the Burst Environment}},
    year = {2003},
    journal = {The Astrophysical Journal},
    author = {Ioka, Kunihito},
    number = {2},
    month = {12},
    pages = {L79-L82},
    volume = {598},
    publisher = {American Astronomical Society},
    url = {https://doi.org/10.1086/380598},
    doi = {10.1086/380598},
    issn = {0004-637X},
    arxivId = {astro-ph/0309200}
}

@article{CRACO1,
    title = {{The CRAFT coherent (CRACO) upgrade I: System description and results of the 110-ms radio transient pilot survey}},
    year = {2025},
    journal = {Publications of the Astronomical Society of Australia},
    author = {Wang, Z. and Bannister, K. W. and Gupta, V. and Deng, X. and Pilawa, M. and Tuthill, J. and Bunton, J. D. and Flynn, C. and Glowacki, M. and Jaini, A. and Lee, Y. W.J. and Lenc, E. and Lucero, J. and Paek, A. and Radhakrishnan, R. and Thyagarajan, N. and Uttarkar, P. and Wang, Y. and Bhat, N. D.R. and James, C. W. and Moss, V. A. and Murphy, Tara and Reynolds, J. E. and Shannon, R. M. and Spitler, L. G. and Tzioumis, A. and Caleb, M. and Deller, A. T. and Gordon, A. C. and Marnoch, L. and Ryder, S. D. and Simha, S. and Anderson, C. S. and Ball, L. and Brodrick, D. and Cooray, F. R. and Gupta, N. and Hayman, D. B. and Ng, A. and Pearce, S. E. and Phillips, C. and Voronkov, M. A. and Westmeier, T.},
    month = {1},
    pages = {e005},
    volume = {42},
    publisher = {Cambridge University Press},
    url = {https://doi.org/10.1017/pasa.2024.107},
    doi = {10.1017/PASA.2024.107},
    issn = {1323-3580}
}

@article{GordonProspector,
    title = {{The Demographics, Stellar Populations, and Star Formation Histories of Fast Radio Burst Host Galaxies: Implications for the Progenitors}},
    year = {2023},
    journal = {The Astrophysical Journal},
    author = {Gordon, Alexa C. and Fong, Wen-fai and Kilpatrick, Charles D. and Eftekhari, Tarraneh and Leja, Joel and Prochaska, J. Xavier and Nugent, Anya E. and Bhandari, Shivani and Blanchard, Peter K. and Caleb, Manisha and Day, Cherie K. and Deller, Adam T. and Dong, Yuxin and Glowacki, Marcin and Gourdji, Kelly and Mannings, Alexandra G. and Mahoney, Elizabeth K. and Marnoch, Lachlan and Miller, Adam A. and Paterson, Kerry and Rastinejad, Jillian C. and Ryder, Stuart D. and Sadler, Elaine M. and Scott, Danica R. and Sears, Huei and Shannon, Ryan M. and Simha, Sunil and Stappers, Benjamin W. and Tejos, Nicolas},
    number = {1},
    month = {8},
    pages = {80},
    volume = {954},
    publisher = {IOP Publishing},
    url = {https://doi.org/10.3847/1538-4357/ace5aa},
    doi = {10.3847/1538-4357/ACE5AA},
    issn = {0004-637X}
}

@article{James2022B,
    title = {{The fast radio burst population evolves, consistent with the star formation rate}},
    year = {2022},
    journal = {Monthly Notices of the Royal Astronomical Society},
    author = {James, C W and Prochaska, J X and Macquart, J.-P and North-Hickey, F O and Bannister, K W and Dunning, A},
    number = {1},
    month = {1},
    pages = {L18-L23},
    volume = {510},
    publisher = {Oxford Academic},
    url = {https://doi.org/10.1093/MNRASL/SLAB117},
    doi = {10.1093/MNRASL/SLAB117},
    issn = {1745-3925},
    arxivId = {2101.07998}
}

@article{GOODS,
    title = {{The Great Observatories Origins Deep Survey: Initial Results from Optical and Near-Infrared Imaging}},
    year = {2004},
    journal = {The Astrophysical Journal},
    author = {Giavalisco, M. and Ferguson, H. C. and Koekemoer, A. M. and Dickinson, M. and Alexander, D. M. and Bauer, F. E. and Bergeron, J. and Biagetti, C. and Brandt, W. N. and Casertano, S. and Cesarsky, C. and Chatzichristou, E. and Conselice, C. and Cristiani, S. and Da Costa, L. and Dahlen, T. and de Mello, D. and Eisenhardt, P. and Erben, T. and Fall, S. M. and Fassnacht, C. and Fosbury, R. and Fruchter, A. and Gardner, J. P. and Grogin, N. and Hook, R. N. and Hornschemeier, A. E. and Idzi, R. and Jogee, S. and Kretchmer, C. and Laidler, V. and Lee, K. S. and Livio, M. and Lucas, R. and Madau, P. and Mobasher, B. and Moustakas, L. A. and Nonino, M. and Padovani, P. and Papovich, C. and Park, Y. and Ravindranath, S. and Renzini, A. and Richardson, M. and Riess, A. and Rosati, P. and Schirmer, M. and Schreier, E. and Somerville, R. S. and Spinrad, H. and Stern, D. and Stiavelli, M. and Strolger, L. and Urry, C. M. and Vandame, B. and Williams, R. and Wolf, C.},
    number = {2},
    month = {1},
    pages = {L93-L98},
    volume = {600},
    publisher = {American Astronomical Society},
    url = {https://doi.org/article/10.1086/379232},
    doi = {10.1086/379232/FULLTEXT/},
    issn = {0004-637X},
    arxivId = {astro-ph/0309105}
}

@article{HorowiczMargalit2025,
    title = {{The Host Galaxies of Fast Radio Bursts Track a Combination of Stellar Mass and Star Formation, Similar to Type Ia Supernovae}},
    year = {2025},
    journal = {preprint (arXiv:2504.08038)},
    author = {Horowicz, Asaf and Margalit, Ben},
    month = {4},
    url = {https://arxiv.org/pdf/2504.08038},
    arxivId = {2504.08038}
}

@article{TendulkarR1,
    title = {{The Host Galaxy and Redshift of the Repeating Fast Radio Burst FRB 121102}},
    year = {2017},
    journal = {The Astrophysical Journal Letters},
    author = {Tendulkar, S. P. and Bassa, C. G. and Cordes, J. M. and Bower, G. C. and Law, C. J. and Chatterjee, S. and Adams, E. A. K. and Bogdanov, S. and Burke-Spolaor, S. and Butler, B. J. and Demorest, P. and Hessels, J. W. T. and Kaspi, V. M. and Lazio, T. J. W. and Maddox, N. and Marcote, B. and McLaughlin, M. A. and Paragi, Z. and Ransom, S. M. and Scholz, P. and Seymour, A. and Spitler, L. G. and Langevelde, H. J. van and Wharton, R. S.},
    number = {2},
    month = {1},
    pages = {L7},
    volume = {834},
    publisher = {IOP Publishing},
    url = {https://doi.org/10.3847/2041-8213/834/2/L7},
    doi = {10.3847/2041-8213/834/2/L7},
    issn = {2041-8205},
    arxivId = {1701.01100}
}

@article{FongBerger2013,
    title = {{The Locations of Short Gamma-Ray Bursts as Evidence for Compact Object Binary Progenitors}},
    year = {2013},
    journal = {The Astrophysical Journal},
    author = {Fong, W. and Berger, E.},
    number = {1},
    month = {9},
    pages = {18},
    volume = {776},
    publisher = {IOP Publishing},
    url = {https://doi.org/10.1088/0004-637X/776/1/18},
    doi = {10.1088/0004-637X/776/1/18},
    issn = {0004-637X},
    arxivId = {1307.0819}
}

@article{ProchaskaFRB20181112A,
    title = {{The low density and magnetization of a massive galaxy halo exposed by a fast radio burst}},
    year = {2019},
    journal = {Science},
    author = {Prochaska, J. Xavier and Macquart, Jean-Pierre and McQuinn, Matthew and Simha, Sunil and Shannon, Ryan M. and Day, Cherie K. and Marnoch, Lachlan and Ryder, Stuart and Deller, Adam and Bannister, Keith W. and Bhandari, Shivani and Bordoloi, Rongmon and Bunton, John and Cho, Hyerin and Flynn, Chris and Mahony, Elizabeth K. and Phillips, Chris and Qiu, Hao and Tejos, Nicolas},
    number = {6462},
    month = {9},
    pages = {231--234},
    volume = {366},
    url = {https://doi.org/10.1126/science.aay0073},
    doi = {10.1126/science.aay0073},
    issn = {0036-8075}
}

@article{EftekhariFRB20240902A,
    title = {{The Massive and Quiescent Elliptical Host Galaxy of the Repeating Fast Radio Burst FRB 20240209A}},
    year = {2025},
    journal = {The Astrophysical Journal Letters},
    author = {Eftekhari, T. and Dong, Y. and Fong, W. and Shah, V. and Simha, S. and Andersen, B. C. and Andrew, S. and Bhardwaj, M. and Cassanelli, T. and Chatterjee, S. and Coulter, D. A. and Fonseca, E. and Gaensler, B. M. and Gordon, A. C. and Hessels, J. W. T. and Ibik, A. L. and Joseph, R. C. and Kahinga, L. A. and Kaspi, V. and Kharel, B. and Kilpatrick, C. D. and Lanman, A. E. and Lazda, M. and Leung, C. and Liu, C. and Mas-Ribas, L. and Masui, K. W. and Mckinven, R. and Mena-Parra, J. and Miller, A. A. and Nimmo, K. and Pandhi, A. and Patil, S. S. and Pearlman, A. B. and Pleunis, Z. and Prochaska, J. X. and Rafiei-Ravandi, M. and Sammons, M. and Scholz, P. and Shin, K. and Smith, K. and Stairs, I.},
    number = {2},
    month = {1},
    pages = {L22},
    volume = {979},
    publisher = {IOP Publishing},
    url = {https://doi.org/10.3847/2041-8213/ad9de2},
    doi = {10.3847/2041-8213/AD9DE2},
    issn = {2041-8205},
    arxivId = {2410.23336}
}

@article{MagnetarCat,
    title = {{The McGill Magnetar Catalog}},
    year = {2014},
    journal = {The Astrophysical Journal Supplement Series},
    author = {Olausen, S. A. and Kaspi, V. M.},
    number = {1},
    month = {4},
    pages = {6},
    volume = {212},
    publisher = {IOP Publishing},
    url = {https://doi.org/10.1088/0067-0049/212/1/6},
    isbn = {164710.2−455216},
    doi = {10.1088/0067-0049/212/1/6},
    issn = {0067-0049},
    arxivId = {1309.4167}
}

@article{numpy,
    title = {{The NumPy array: A structure for efficient numerical computation}},
    year = {2011},
    journal = {Computing in Science and Engineering},
    author = {Van Der Walt, Stéfan and Colbert, S. Chris and Varoquaux, Gaël},
    number = {2},
    month = {3},
    pages = {22--30},
    volume = {13},
    url = {https://doi.org/10.1109/MCSE.2011.37},
    doi = {10.1109/MCSE.2011.37},
    issn = {15219615},
    arxivId = {1102.1523}
}

@article{PanSTARRS,
    title = {{The Pan-STARRS1 Surveys}},
    year = {2016},
    journal = {preprint (1612.05560)},
    author = {Chambers, K. C. and Magnier, E. A. and Metcalfe, N. and Flewelling, H. A. and Huber, M. E. and Waters, C. Z. and Denneau, L. and Draper, P. W. and Farrow, D. and Finkbeiner, D. P. and Holmberg, C. and Koppenhoefer, J. and Price, P. A. and Rest, A. and Saglia, R. P. and Schlafly, E. F. and Smartt, S. J. and Sweeney, W. and Wainscoat, R. J. and Burgett, W. S. and Chastel, S. and Grav, T. and Heasley, J. N. and Hodapp, K. W. and Jedicke, R. and Kaiser, N. and Kudritzki, R. -P. and Luppino, G. A. and Lupton, R. H. and Monet, D. G. and Morgan, J. S. and Onaka, P. M. and Shiao, B. and Stubbs, C. W. and Tonry, J. L. and White, R. and Ba{\~{n}}ados, E. and Bell, E. F. and Bender, R. and Bernard, E. J. and Boegner, M. and Boffi, F. and Botticella, M. T. and Calamida, A. and Casertano, S. and Chen, W. -P. and Chen, X. and Cole, S. and Deacon, N. and Frenk, C. and Fitzsimmons, A. and Gezari, S. and Gibbs, V. and Goessl, C. and Goggia, T. and Gourgue, R. and Goldman, B. and Grant, P. and Grebel, E. K. and Hambly, N. C. and Hasinger, G. and Heavens, A. F. and Heckman, T. M. and Henderson, R. and Henning, T. and Holman, M. and Hopp, U. and Ip, W. -H. and Isani, S. and Jackson, M. and Keyes, C. D. and Koekemoer, A. M. and Kotak, R. and Le, D. and Liska, D. and Long, K. S. and Lucey, J. R. and Liu, M. and Martin, N. F. and Masci, G. and McLean, B. and Mindel, E. and Misra, P. and Morganson, E. and Murphy, D. N. A. and Obaika, A. and Narayan, G. and Nieto-Santisteban, M. A. and Norberg, P. and Peacock, J. A. and Pier, E. A. and Postman, M. and Primak, N. and Rae, C. and Rai, A. and Riess, A. and Riffeser, A. and Rix, H. W. and R{\"{o}}ser, S. and Russel, R. and Rutz, L. and Schilbach, E. and Schultz, A. S. B. and Scolnic, D. and Strolger, L. and Szalay, A. and Seitz, S. and Small, E. and Smith, K. W. and Soderblom, D. R. and Taylor, P. and Thomson, R. and Taylor, A. N. and Thakar, A. R. and Thiel, J. and Thilker, D. and Unger, D. and Urata, Y. and Valenti, J. and Wagner, J. and Walder, T. and Walter, F. and Watters, S. P. and Werner, S. and Wood-Vasey, W. M. and Wyse, R.},
    month = {12},
    url = {http://arxiv.org/abs/1612.05560},
    arxivId = {1612.05560}
}

@article{Spearman,
    title = {{The Proof and Measurement of Association between Two Things}},
    year = {1904},
    journal = {The American Journal of Psychology},
    author = {Spearman, C.},
    number = {1},
    month = {1},
    pages = {72},
    volume = {15},
    publisher = {University of Illinois Press},
    url = {https://doi.org/10.2307/1412159},
    doi = {10.2307/1412159},
    issn = {00029556}
}

@article{KStest,
    title = {{The significance probability of the Smirnov two-sample test}},
    year = {1958},
    journal = {Arkiv f{\"{u}} Matematik},
    author = {Hodges, J. L.},
    number = {5},
    month = {1},
    pages = {469--486},
    volume = {3},
    publisher = {Kluwer Academic Publishers},
    url = {https://doi.org/10.1007/BF02589501},
    doi = {10.1007/BF02589501},
    issn = {0004-2080}
}

@article{MarnochFRB20210912A,
    title = {{The unseen host galaxy and high dispersion measure of a precisely localized fast radio burst suggests a high-redshift origin}},
    year = {2023},
    journal = {Monthly Notices of the Royal Astronomical Society},
    author = {Marnoch, Lachlan and Ryder, Stuart D and James, Clancy W and Gordon, Alexa C and Sammons, Mawson W and Prochaska, J Xavier and Tejos, Nicolas and Deller, Adam T and Scott, Danica R and Bhandari, Shivani and Glowacki, Marcin and Mahony, Elizabeth K and McDermid, Richard M and Sadler, Elaine M and Shannon, Ryan M and Qiu, Hao},
    number = {1},
    month = {8},
    pages = {994--1007},
    volume = {525},
    publisher = {Oxford Academic},
    url = {https://doi.org/10.1093/mnras/stad2353},
    doi = {10.1093/MNRAS/STAD2353},
    issn = {0035-8711}
}

@article{James2022A,
    title = {{The z –DM distribution of fast radio bursts}},
    year = {2021},
    journal = {Monthly Notices of the Royal Astronomical Society},
    author = {James, C W and Prochaska, J X and Macquart, J-P and North-Hickey, F O and Bannister, K W and Dunning, A},
    number = {4},
    month = {12},
    pages = {4775--4802},
    volume = {509},
    publisher = {Oxford Academic},
    url = {https://doi.org/10.1093/mnras/stab3051},
    doi = {10.1093/mnras/stab3051},
    issn = {0035-8711},
    arxivId = {2101.08005}
}

@article{UltraFlatGalaxies,
    title = {{Ultra-flat galaxies selected from RFGC catalog. I. The sample properties}},
    year = {2016},
    journal = {Astrophysical Bulletin},
    author = {Karachentseva, V. E. and Kudrya, Yu N. and Karachentsev, I. D. and Makarov, D. I. and Melnyk, O. V.},
    number = {1},
    month = {1},
    pages = {1--13},
    volume = {71},
    publisher = {MAIK Nauka - Interperiodica},
    url = {https://doi.org/10.1134/S1990341316010016},
    doi = {10.1134/S1990341316010016},
    issn = {19903421},
    arxivId = {1602.05368}
}
